\documentclass[]{spie}  
\usepackage[]{graphicx}

\title{Precision near-infrared radial velocity instrumentation I: absorption gas cells} 

\author{Peter P. Plavchan\supit{a}, Anglada-Escude, G.\supit{b}, White, R.\supit{c}, Gao, P.\supit{d}, Davison, C.\supit{c}; Mills, S.\supit{e}; Beichman, C.\supit{a}; Brinkworth, C.\supit{a}; Johnson, J.\supit{f}; Bottom, M.\supit{d}; Ciardi, D.\supit{a}; Wallace, K.\supit{g}; Mennesson, B.\supit{g}; von Braun, K.\supit{h}; Vasisht, G.\supit{g}; Prato, L.\supit{i}; Kane, S.\supit{j}; Tanner, A.\supit{k}; Walp, B.\supit{l}; Crawford, S.\supit{g}; Lin, S.\supit{g},
\skiplinehalf
\supit{a}NASA Exoplanet Science Institute, California Institute of Technology, 770 S Wilson Ave, Pasadena, CA, USA; \\
\supit{b}University of Goettingen\\
\supit{c}Georgia State University\\
\supit{d}California Institute of Technology\\
\supit{e}University of Chicago\\
\supit{f}Harvard University\\
\supit{g}Jet Propulsion Laboratory\\
\supit{h}Max Planck Institut Astronomie, Heidelberg\\
\supit{i}Lowell Observatory\\
\supit{j}San Francisco State University\\
\supit{k}Mississippi State University\\
\supit{l}SOFIA
}

\authorinfo{Further author information: (Send correspondence to P.P.P.)\\P.P.P.: E-mail: plavchan@ipac.caltech.edu, Telephone: 1 626 234 1628}

\begin{document} 
  \maketitle 

\begin{abstract}
We have built and commissioned gas absorption cells for precision spectroscopic radial velocity measurements in the near-infrared in the H and K bands.   We describe the construction and installation of three such cells filled with 13CH4, 12CH3D, and 14NH3 for the CSHELL spectrograph at the NASA Infrared Telescope Facility (IRTF).  We have obtained their high-resolution laboratory Fourier Transform spectra, which can have other practical uses.   We summarize the practical details involved in the construction of the three cells, and the thermal and mechanical control. In all cases, the construction of the cells is very affordable.  We are carrying out a pilot survey with the 13CH4 methane gas cell on the CSHELL spectrograph at the IRTF to detect exoplanets around low mass and young stars.  We discuss the current status of our survey, with the aim of photon-noise limited radial velocity precision.  For adequately bright targets, we are able to probe a noise floor of ~7 m/s with the gas cell with CSHELL at cassegrain focus.  Our results demonstrate the feasibility of using a gas cell on the next generation of near-infrared spectrographs such as iSHELL on IRTF, iGRINS, and an upgraded NIRSPEC at Keck.
\end{abstract}



\section{INTRODUCTION}
\label{sec:intro}  

More than 500 of the nearly 900 known exoplanets have been discovered with the radial velocity method, enabling a new ``golden age'' of precision synoptic and spectroscopic astronomy (Fig 1)\cite{akeson13}.  The US National Academy 2010 Astronomy and Astrophysics Decadal Survey made the bold and controversial recommendation:  ``The first task on the ground is to improve the precision radial velocity method\dots Using existing large ground-based or new dedicated mid-size ground-based telescopes equipped with a new generation of high-resolution spectrometers in the optical and near-infrared, a velocity goal of 10 to 20 centimeters per second is realistic.''   The future potential of the precision radial velocity method is exemplified by the recently reported detection of an Earth-mass exoplanet (radial velocity semi-amplitude K=51 cm/s) with a 3.2-day orbit around $\alpha$ Cen B with the HARPS-South spectrometer \cite{dumusque12}.   

\begin{figure}[tb]
  \begin{center}
    \includegraphics[width=0.80\textwidth]{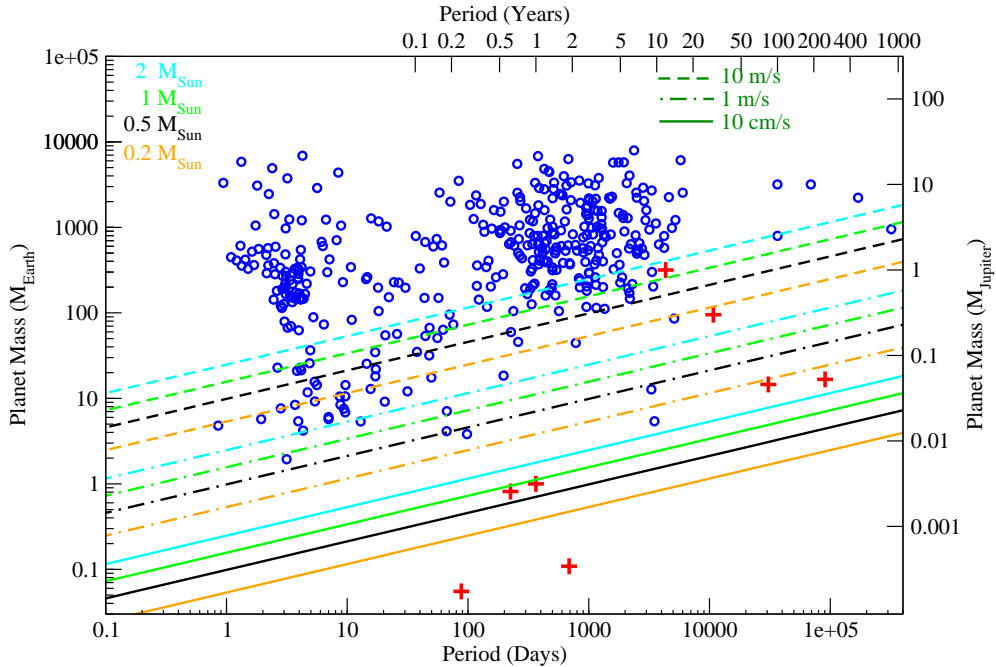}    
      \end{center}
      
  \caption{Confirmed exoplanets shown as blue circles from the NASA Exoplanet Archive \cite{akeson13} plotted as a function of orbital period on the horizontal axis, and planet mass on the vertical axis. Solar System planets are shown as red crosses.  The dashed, dot-dashed, and solid lines correspond to the sensitivity limits for radial velocity precisions of 10 m/s, 1m/s and 10 cm/s respectively, and the colors of these lines correspond to different stellar host masses of 2 M$_\odot$ (blue),  1 M$_\odot$ (green), 0.5 M$_\odot$ (black), and 0.2 M$_\odot$ (orange).  In the absence of red noise due to stellar activity or systematic noise sources, exoplanets above these lines can be detected for a given stellar mass and radial velocity precision. \label{fig:f1}}
\end{figure}

The vast majority of radial velocity exoplanet discoveries are made with visible wavelength spectrometers that achieve $\sim$1 m/s radial velocity precision.  By comparison, radial velocity measurements in the near-infrared would appear to be advantageous for a number of reasons, particularly for M dwarfs: 

\begin{itemize}
\item Lower mass M dwarfs are brightest in the near-infrared, with strong bands of sharp, deep absorption lines such as the CO band-head at 2.3 $\mu$m; 
\item M dwarfs have closer habitable zones and lower stellar masses, resulting in shorter habitable zone orbital periods and larger amplitude signals for the easier detection and characterization of terrestrial exoplanets; 
\item M dwarfs are more abundant than FGK stars comprising $\sim$75\% of the local stellar neighborhood \cite{howard12,henry06}; 
\item The Kepler mission has recently shown that terrestrial planets are more frequent around M dwarfs compared to solar-like FGK stars;
\item For solar-like FGK dwarfs, stellar ``jitter'' due to the rotational modulation of starspots is wavelength dependent and smaller in amplitude at longer (redder) wavelengths due to the lower flux contrast between star-spots and the stellar photosphere \cite{anglada12b}; 
\item Younger, more active photosphere, and/or embedded FGKM stars can be monitored (dust is more transparent in the infrared), for example to probe the dynamical oligarch epoch of planet formation;
\item The direct detection of radial velocity variations, (as opposed to stellar reflex motion) from Jovian planets with significant atmospheric CO absorption at 2.3 $\mu$m in the planet spectrum, orbiting featureless A stars such as Tau Boo\cite{rodler2012,snellen2012}.
\end{itemize}

However, radial velocity precision in the near-infrared historically lags the visible.   As of $\sim$2010, the most precise technique in the near-infrared uses telluric (atmospheric sky) lines for wavelength calibration to achieve a precision of ~50-100 m/s \cite{tanner12,crockett11,bailey11}.  Furthermore, instruments that require cryogenic cooling can often be a few times more expensive than their corresponding visible equivalents, and as a result there are currently relatively few operational high resolution near-infrared spectrographs, none of which are optimized for radial velocities -- CRIRES, CSHELL, Phoenix and NIRSPEC\cite{bean10,anglada12,bailey11}.  

Fortunately, a number of facilities are funded to become operational in the next five years, including the Habitable Zone Planet Finder, CARMENES, SPIRou, iGRINS, and iSHELL \cite{HZPFref,CARMENESref,iGRINSref,iSHELLref}.  The Habitable Zone Planet Finder, SPIRou and CARMENES are purposely built for precision radial velocity measurements.  Additionally, recent advances have pushed the achievable radial velocity precision in the near-infrared closer to parity with the visible.  These advances include the use of ammonia and isotopic methane absorption gas cells to achieve a precision of $\sim$5-10 m/s \cite{bean10,anglada12}, the recent commissioning of a near-infrared fiber scrambler prototype with non-circular core fibers on CSHELL at IRTF (NASA Infrared Telescope Facility) \cite{bottom13,plavchan13}, and the prototype for the fiber-fed Habitable Zone Planet Finder, Pathfinder, which made use of a novel laser frequency comb for wavelength calibration to achieve a precision of $\sim$15 m/s \cite{osterman10,PATHFINDERref}.   The next generation of near-infrared facilities hold the potential to achieve ~1-3 m/s single measurement precision, sufficient to detect terrestrial planets in the Habitable Zone of M dwarfs.  Active areas of technology development include Uranium-Neon emission lamps, stabilized Fabry-Perot etalons, and single-mode fiber-based spectrographs behind adaptive optics equipped telescopes.  However, much work is needed, including improvements in the analysis and removal of telluric lines and the generation of stellar templates\cite{seifahrt10}.

Given the costs of new high-resolution cryogenic spectrographs, we have taken the low-cost approach of modifying existing near-infrared spectrographs for enabling high precision radial velocity measurements through the use of absorption gas cells and fiber scramblers.  In this paper we present our instrumentation and survey efforts using isotopic and deuterated methane absorption gas cells at CSHELL on the IRTF \cite{tokunaga1990,greene1993}.  The first light paper with the isotopic methane absorption gas cell was presented in Anglada-Escude et al. 2012.  This paper provides more information on the instrumentation, as well as updates the current status of our survey data analysis and techniques.  A companion SPIE proceedings by Plavchan et al. 2013 will discuss first light observations with our prototype near-infrared fiber scrambler.  These instruments are available to the community to use on CSHELL, and interested parties are encouraged to contact the authors for more information.

\section{Design \& Construction}

In this section we provide a discussion of the detailed instrumental design of our absorption gas cell instrument.

\subsection{Spectrograph}

CSHELL is the ``host'' spectrograph for our ``symbiotic'' instrumentation.  The instrument is almost 20 years old, with (non-simultaneous) wavelength coverage from 1-5.5 $\mu$m, and with a Hughes SBRC 256x256 InSb CCD\cite{tokunaga1990,greene1993}.  The spectrograph is mounted at the Cassegrain focus of IRTF on the back of the primary mirror support structure.  CSHELL captures a single $\sim$5 nm order of spectra at 2.3 $\mu$m at a resolution of R$\sim$46,000 with a 0.''5 slit.  For comparison, the non-cross-dispersed CRIRES on the VLT had a spectral grasp of $\sim$50 nm of a single order that spans four CCDs, and has recently added a cross-disperser to increase the spectral grasp further.  Thus, the spectral grasp of CSHELL is limited compared to more modern spectrographs.   Since iSHELL is a funded replacement for CSHELL, we are graciously allowed by Dr. John Rayner and Dr. Alan Tokunaga to modify parts of the CSHELL fore-optics to accommodate our prototype instrumentation.  

Wavelengths are selected by adjusting the echelle grating central wavelength and a continuously variable filter for order selection that inadvertently introduces fringing at the $\sim$1-3\% flux level in our spectra.  We were unaware of the fringing at the time of commissioning our prototype, although this was a known limitation of CSHELL.  The fringing from the CVF filter adversely affected and delayed our data analysis pipeline development and resulting radial velocity precision.  Regardless, we were eventually able to incorporate a solution to the fringing into our analysis to obtain useful radial velocity measurements  (${\S}$3.2).

\subsection{Pre-Existing Calibration Unit}

The calibration unit of CSHELL is situated on top of the cryogenic entrance window (Fig 2). The calibration unit consists of a rectangular metal enclosure with a pass-through cavity approximately 12x6x7 inches.  The $\sim$2-inch diameter f/37.5 converging beam from the secondary mirror passes through this cavity along the longest dimension.  A fold mirror enters this cavity approximately half-way along the beam travel to inject calibration lamp light via an integrating sphere (flats, Xenon, Argon \& Krypton) that is mounted with the lamps on the other side of this cavity.  We have added an absorption gas cell into the space between this fold mirror and the calibration unit wall closest to the primary mirror, a volume approximately six inches per side.

\begin{figure}[tb]
  \begin{center}
    \includegraphics[width=0.30\textwidth]{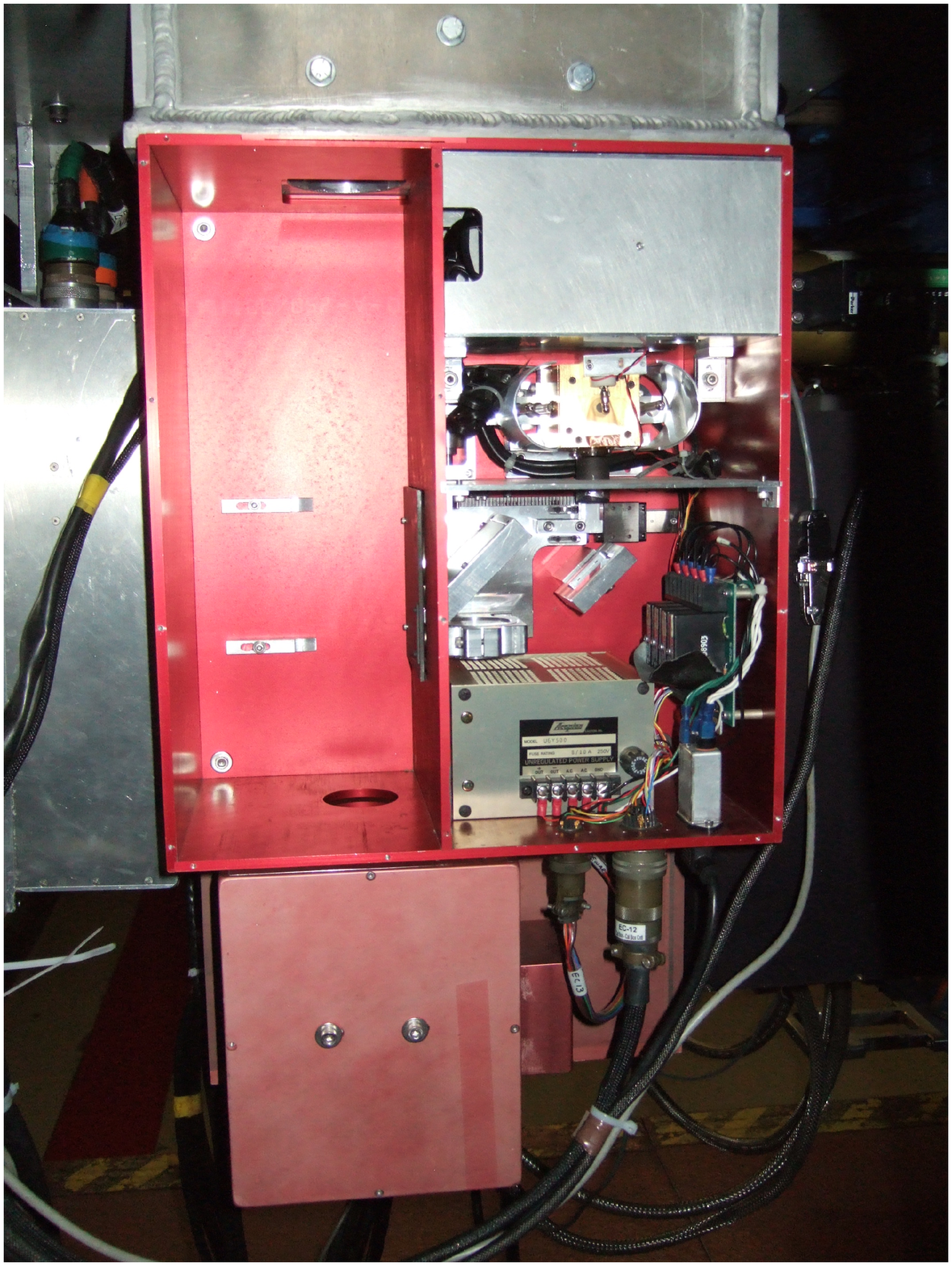}    
    \includegraphics[width=0.30\textwidth]{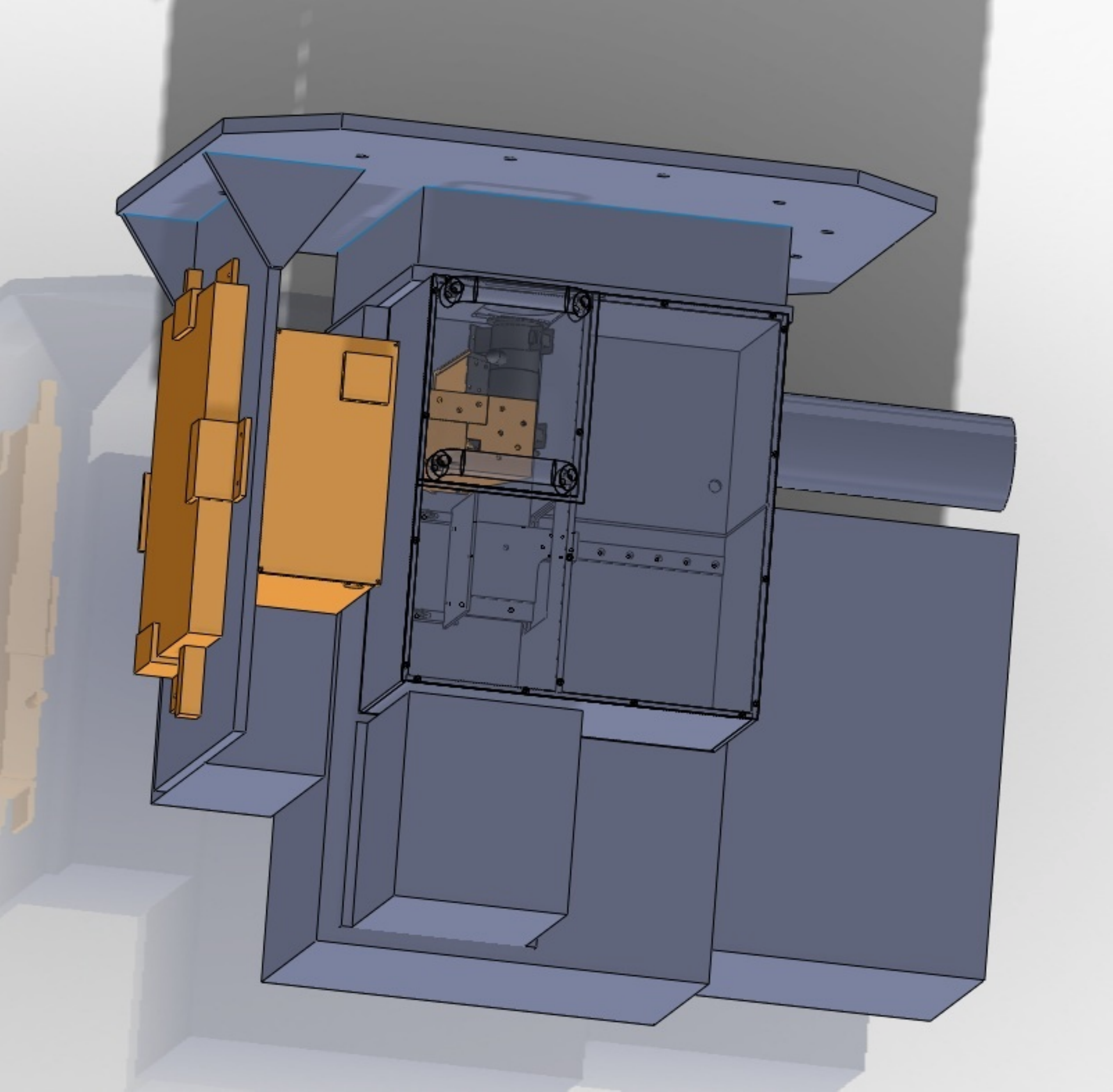}    
    \includegraphics[width=0.30\textwidth]{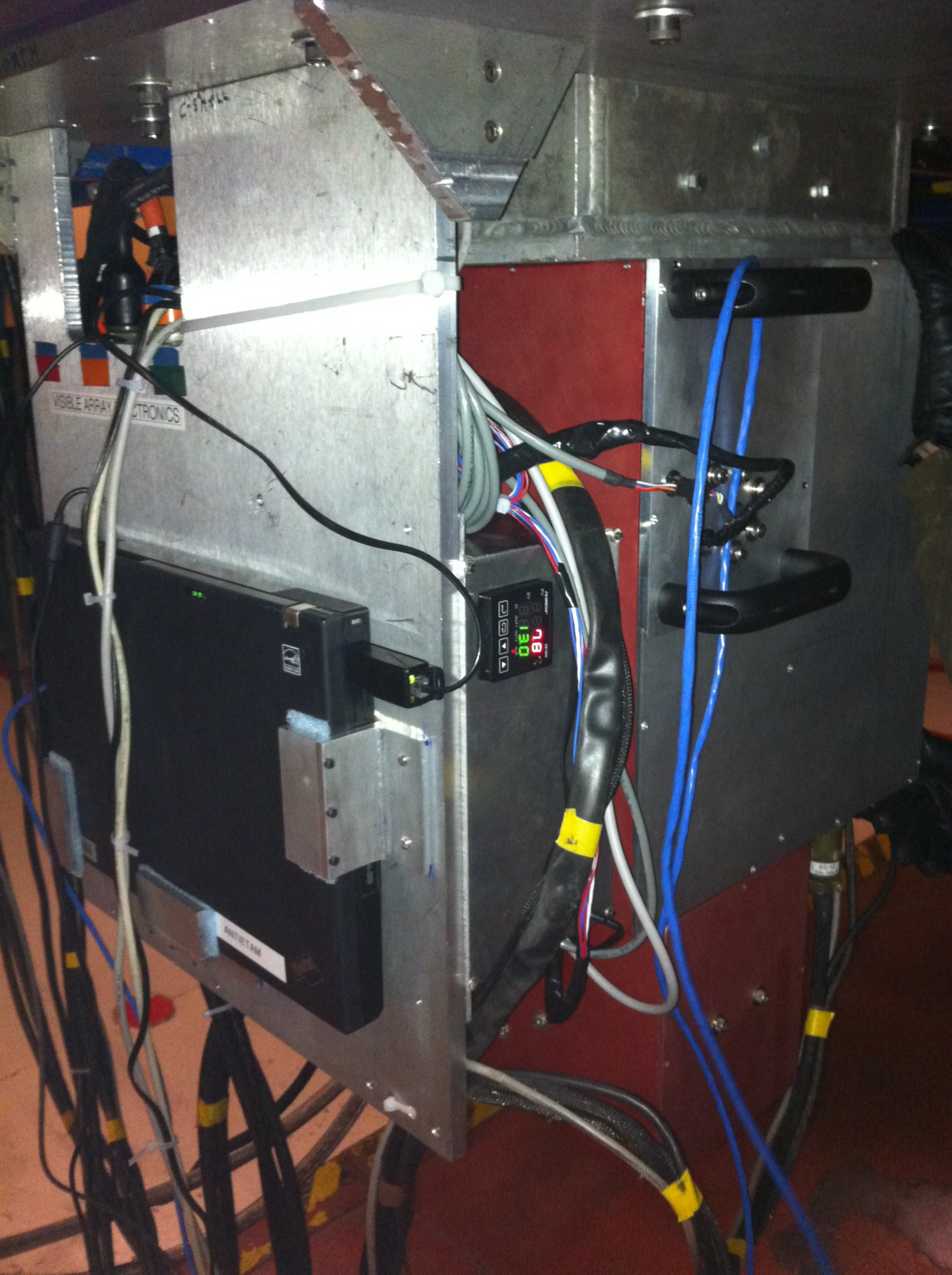}    
      \end{center}
      
  \caption{Left: CSHELL calibration unit with cover removed.  In this view, the primary mirror is towards the top of the image.  The beam converges from top to bottom on the left hand side of the calibration unit.  It is into this space we inserted the gas cell.  Middle: Solidworks model of the integrated gas cell  with the the CSHELL calibration unit.  The gas cell mount is shown in orange, along with the control electronics box and control laptop attached to a mounting plate to the left of the calibration unit.  Right: The completed instrument and control electronics box and laptop integrated with the CSHELL calibration unit, with a viewing angle rotated to the left of the view in the left panel. \label{fig:f2} }
\end{figure}

\subsection{Gas}

\begin{figure}[tb]
  \begin{center}
    \includegraphics[width=0.40\textwidth]{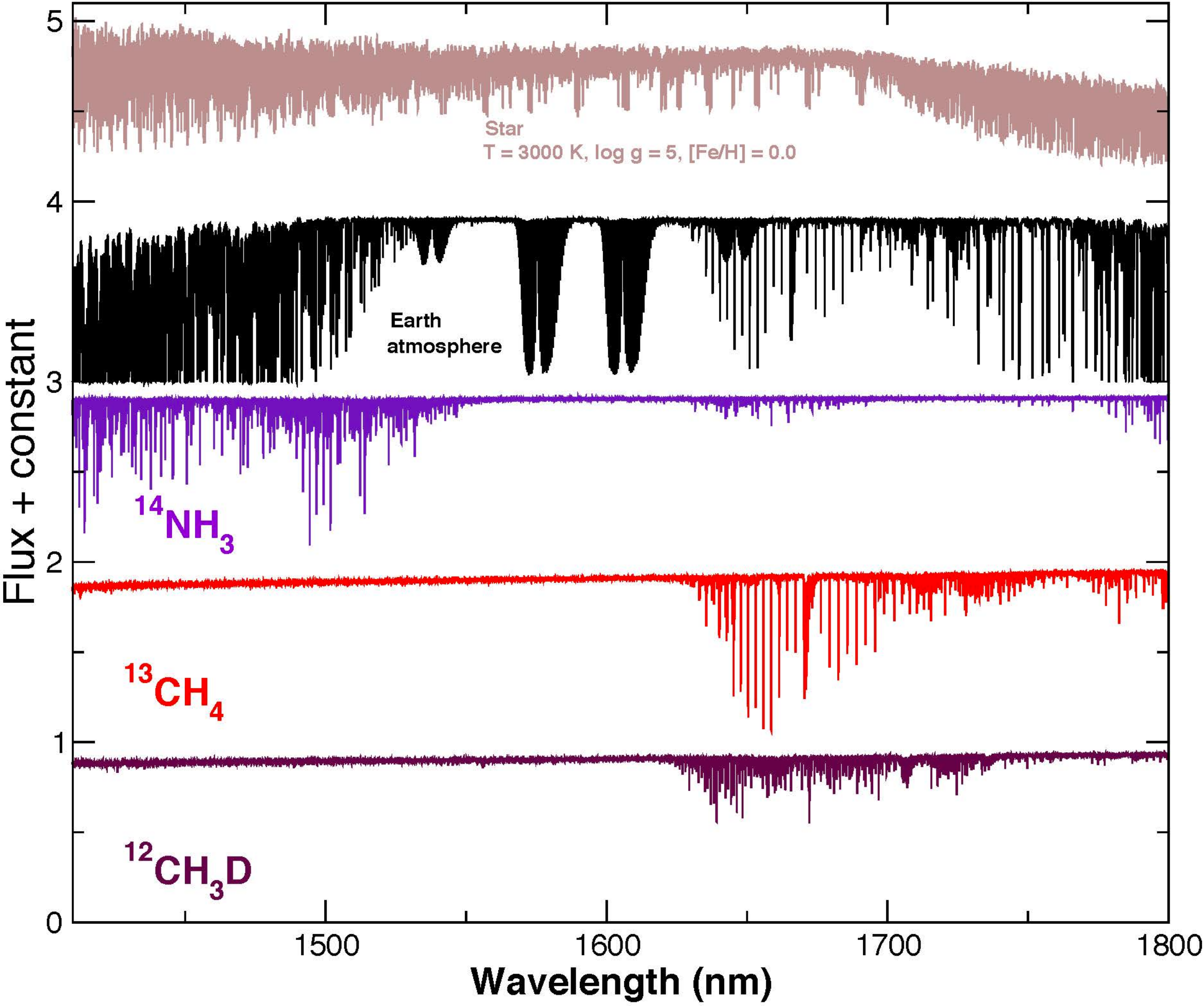}    
    \includegraphics[width=0.40\textwidth]{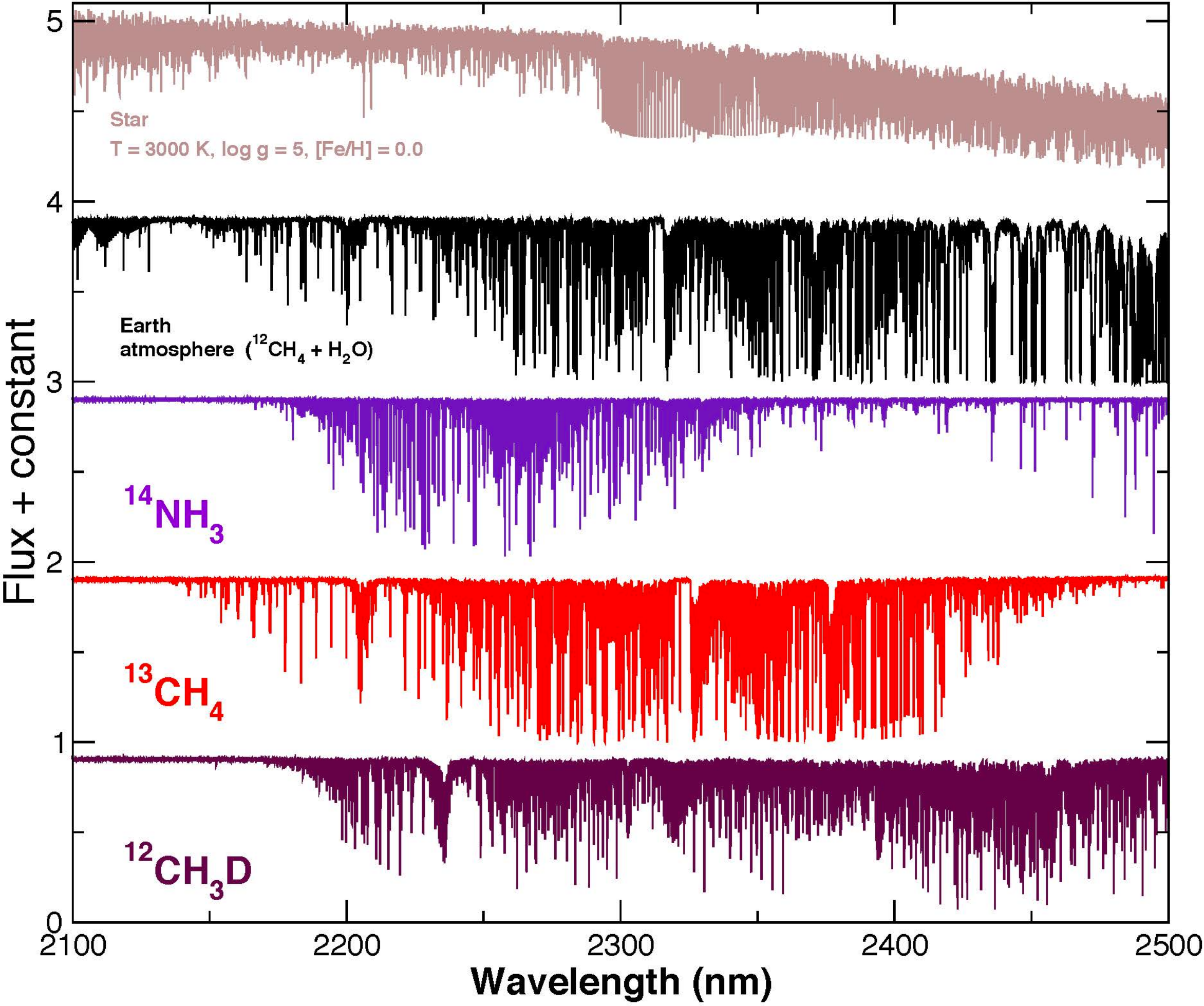}    
      \end{center}
      
  \caption{H and K band absorption spectra, normalized and offset flux as a function of wavelength, from top to bottom: synthetic stellar template in light brown, a telluric spectrum in black, ammonia ($^{14}$NH$_{3}$) in light purple, Isotopic methane ($^{13}$CH$_{4}$) in red, and deuterated methane ($^{12}$CH$_{3}$D) in maroon.  The gas cell spectra are obtained with a Fourier Transform Spectrometer, reproduced from Figures 6 and 5 respectively in Anglada-Escude et al. (2012).\label{fig:f3} }
\end{figure}

We built two methane cells containing isotopic methane ($^{13}$CH$_{4}$) and deuterated methane ($^{12}$CH$_{3}$D), as well as an ammonia ($^{14}$NH$_{3}$) cell for comparison.   In all cases, the construction of the cells is very affordable.    Pure anhydrous $^{14}$NH$_3$ ammonia was purchased from Matheson-trigas ($>$99.9\% quoted purity). The $^{13}$CH$_{4}$ isotopologue was purchased from SIGMA-ALDRICH (99\% quoted purity).  The more exotic $^{12}$CH$_{3}$D isotopologue was obtained from Cambridge Isotope Laboratories, Inc (98\% quoted purity). Absorption spectra of our three cells in the H and K bands are shown in Figure \ref{fig:f3}, demonstrating a large number of deep, narrow absorption lines suitable for precision radial velocity wavelength calibration across a broad spectral grasp.  

\begin{figure}[tb]
  \begin{center}
    \includegraphics[width=0.50\textwidth]{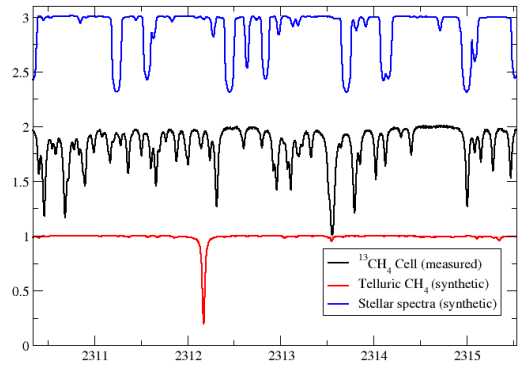}    
      \end{center}
      
  \caption{Spectra spanning the CSHELL order wavelength range: a synthetic stellar spectrum in blue, a synthetic telluric methane spectrum in red with only an isolated telluric line, and our measured isotopic methane spectrum in black.  \label{fig:f4} }
\end{figure}

Anglada-Escude et al. 2012 presents an analysis of the suitability of isotopic and deuterated methane gas for precision near-infrared radial velocity wavelength calibration in the H and K bands, and their radial velocity information content relative to ammonia.  Methane has been overlooked as a wavelength calibration source in the near-infrared due to the presence of atmospheric methane absorption lines in both the H and K wavelength bands.  A relatively stable wavelength reference (methane in a gas cell) aligned with a relatively unstable wavelength reference (telluric methane) is not very useful for precision radial velocity work.  The absorption lines produced in the near-infrared by methane are the result of ro-vibrational transitions whose wavelengths depend on the reduced mass of the molecule.  By using an isotope of methane, the reduced mass of the molecule changes, effectively shifting all of the lines by a large enough amount ($\sim$10 nm) to mis-align them from their telluric counterparts.  Additionally, deuterated methane,  $^{12}$CH$_{3}$D, breaks the rotational symmetry of regular methane, $^{12}$CH$_{4}$, effectively tripling the line density albeit at reduced optical depth.  In practice, our deuterated  methane cell did not provide adequate absorption line depth compared to the isotopic methane cell and the gas is more suitable for a higher resolution spectrograph.   Thus, we have only observed with the isotopic methane cell despite the lower line density. We chose to observe at $\lambda_{c}$ = 2312.5 nm, which is centered in a small window relatively free of telluric methane features, but rich in isotopic methane absorption lines (Figure \ref{fig:f4}).   At H band, we observe at $\lambda_{c}$ = 1671.7 nm.

We have obtained laboratory measurements of the three cells spectra at JPL using a Bruker IFS 125/HR spectrometer, whose instrumental setup can be found elsewhere \cite{sung08}. The FTIR spectra were taken at a resolution $R \approx 700 000$ at 2.0 $\mu$m and 298 K.  This resolution is much higher than CSHELL's resolution of $\sim$46,000 (with the 0.''5 slit), and allows for very precise resolution of the individual spectral absorption features of all three gases. In order to ensure we had complete coverage of the infrared bands we intended to use, a full scan from 1 to 5~$\mu$m wavelengths was performed.  The FTS system at JPL equipped with a temperature-stabilized He-Neon laser has enabled a frequency precision better than 0.0001 cm$^{-1}$ in the scanned region, where the units of frequency are wavenumbers per unit of length.  Since Ammonia gas is sticky and notorious in leaving permanent residues on the optical surfaces, we scanned the cell while the FTS was pumped down to 95 mb. This pressure is slightly higher than the cell pressure,  minimizing the risk of Ammonia leaking-out from the cell. For the methane isotopologues, the FTS was evacuated to better than 1 mb in pressure. For all three cases, we obtained the spectra without the cell at a pressure similar to that of each cell. In this way, the unwanted atmospheric residual features could be cancelled out by dividing the cell spectra by their \textit{no cell} counterparts.  These normalized FTIR spectra are an essential component of the radial velocity forward model pipeline discussed in Section 3. 

\subsection{Optical}

\begin{figure}[tb]
  \begin{center}
    \includegraphics[width=0.40\textwidth]{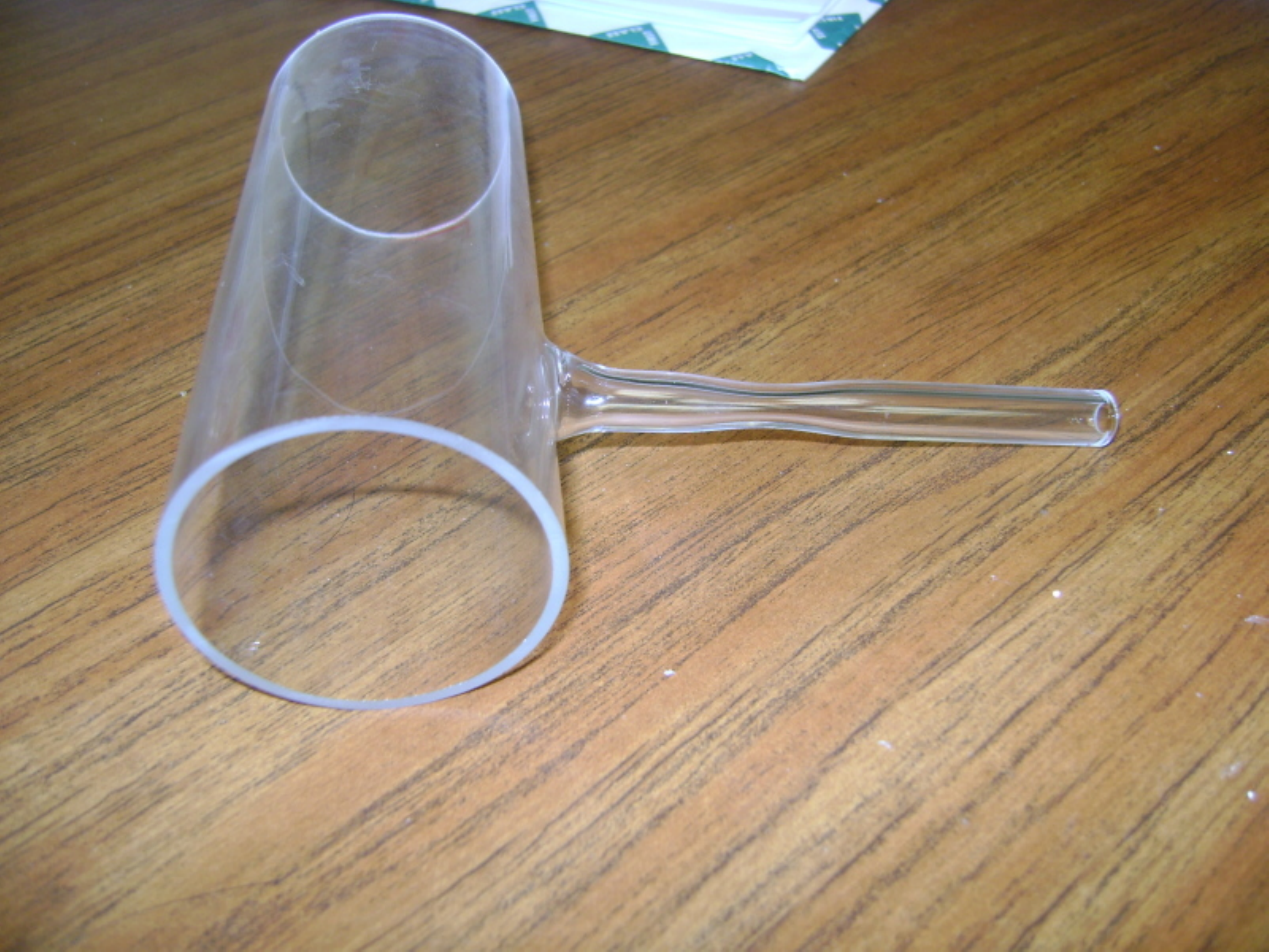}    
    \includegraphics[width=0.40\textwidth]{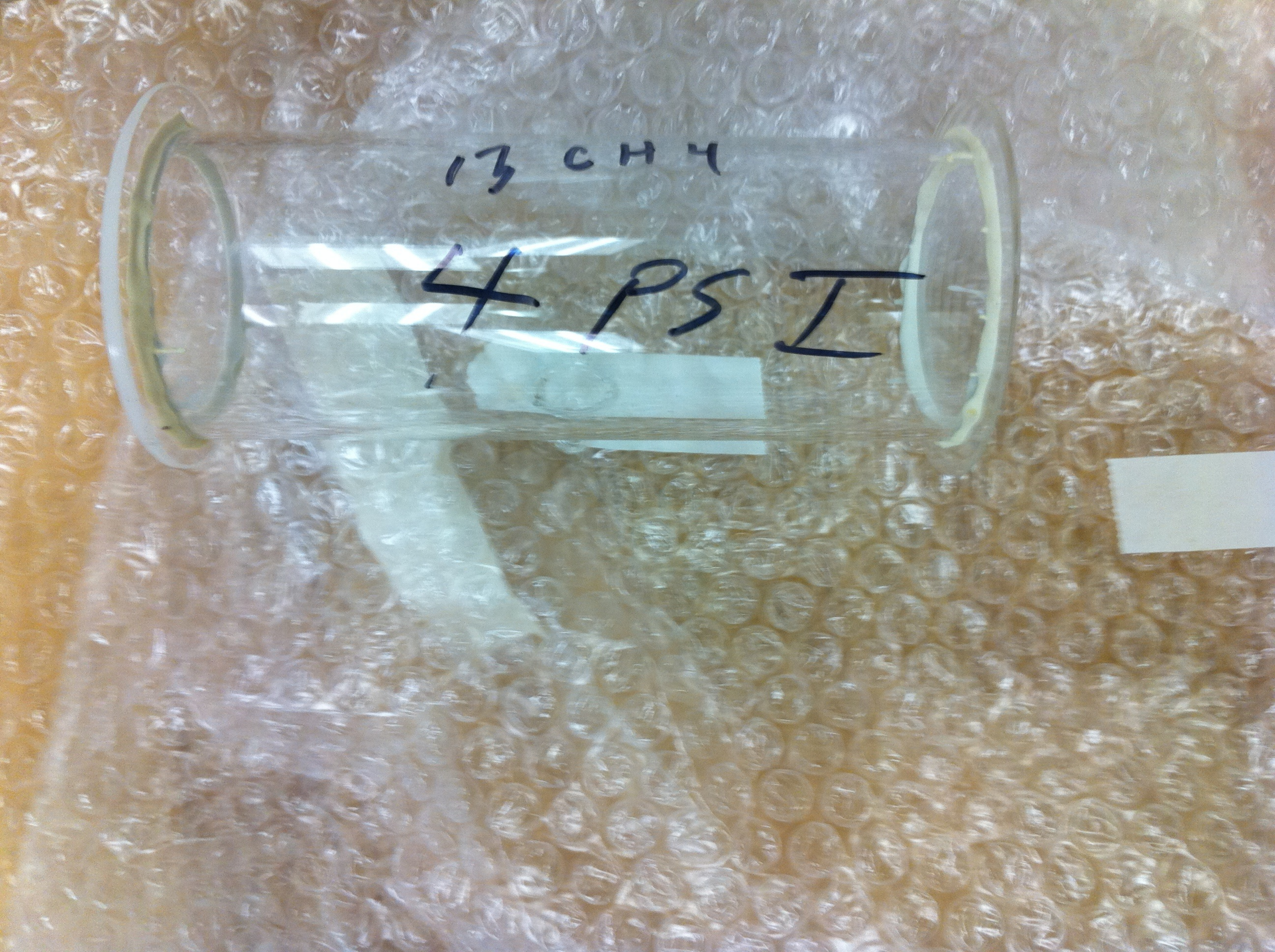}    
      \end{center}
      
  \caption{Gas cell body with inlet stem, before and after assembly.  The pinched off and sealed inlet stem can be seen on the bottom of the cell on the right hand side.\label{fig:f5} }
\end{figure}

Our gas cells generally follow the model of iodine and ammonia absorption gas cells, used respectively by the California Planet Search and other collaborations in the visible, and by Bean et al. 2010 in the near-infrared with CRIRES on the VLT.  Based on the spatial limitations imposed by the existing CSHELL calibration unit, the bodies of the cells were made using Pyrex tube 12.5 cm in length and 5.1 cm in diameter. A pyrex inlet tube for vacuum pumping and gas injection was annealed onto the gas cell body perpendicular to the body approximately one-third along it's length.  A ``pinch'' in the inlet tube was added to aid in the sealing of the cells (Figure \ref{fig:f5}).

Both ends of the cell bodies were capped with low-OH quartz windows over-sized at 7 cm in diameter which have excellent transmission in the NIR.  The windows were coated on one side with an infrared anti-reflective coating to further improve transmission.  The windows were initially incorrectly coated,  resulting in only a $\sim$50\% throughput verified by a quick FTIR scan, and had to be recoated.  The side of the windows exposed to the methane and ammonia gas were not anti-reflective coated to prevent any undesired chemical reaction that could destroy the coating and consequently the throughput.  Using the laboratory FTIR spectra of all the completed cells, we found that the overall transmission of the continuum (including the gases) is typically above 80\%.  

Each window is a wedge with an angle of 1.6$^{\circ}$ oriented 180$^{\circ}$ relative to the other window on the cell, corresponding to a 2 mm rise in window thickness over 7 cm, with a minimum window thickness at one end of 1 mm. This prevents ``ghost'' images from appearing on the stellar spectra within the slit, which arise from multiple reflections off the spectrograph's optics.  However, it also introduces a small beam displacement ($\sim$2'' on the sky)  that can be oriented along the slit direction.    The windows were sealed to the gas cell body using Varian Torr Seal, applied manually with a toothpick to the gas cell body to completely cover the circular edge leaving no glass-to-glass contact.  Torr Seal is a sealant specifically designed to connect glass and maintain a vacuum for a duration longer than a decade in nominal conditions.   The 180$^{\circ}$ window ``clocking'' was achieved to within 1$^{\circ}$ by hand given the quick dry time of the torr seal.  The torr seal was left to dry for 24 hours.

\subsection{Filling and sealing the cells}

All three cells were evacuated using a standard vacuum pump, and then filled with the gases via the use of a glass stopcock valve made custom for this project.  The stopcock was attached to the inlet tube of the cell via a vacuum o-ring connector from Thor Labs.  The stopcock had three lines -- one each to the cell, to the vacuum pump, and to the gas cylinder and pressure regulator.   As shown in Anglada-Escude et al. 2012, $^{14}$NH$_{3}$ and $^{13}$CH$_{4}$ cells were found to operate optimally at $\sim$75 mb and 275 mb respectively at 20 C (as a reference, 1013.25 mb is 1 atm or 1.013x$10^5$ Pa). Since we had no means to predict the optimal pressure for $^{12}$CH$_{3}$D at the K band, we filled the $^{12}$CH$_{3}$D cell at a higher pressure to increase the optical depth ($\sim$ 345 mb).  

In order to pinch and seal the inlet tube, the gases were then condensed to a liquid via immersion of the cell body in a liquid nitrogen bath.  The methane or ammonia liquid was then pooled at one end of the cell. This allowed the pinch point along the pyrex inlet tube to be locally heated to $\sim$ 1100 K without raising the interior gas temperature and pressure. If the gases had been left in the gas phase starting at room temperature and then heated, the internal pressure of the cell would have exceeded 1 atmosphere and the gas would have burst out of the hot malleable inlet tube at the pinch point. The first tests filling our gas cells with ammonia gas resulted in this outcome. With the gases properly condensed to a liquid inside the cell, the pinch in the gas cell inlet tube collapses upon itself due to the pressure differential between the atmosphere and the near-vacuum inside the cell.  The outer part of the inlet stem is then detached carefully to maintain the seal that is created.  The end result of the sealing process allows for the creation of an essentially permanent pyrex seal to the inlet such that the main body tube is a single piece of pyrex with a small nub of glass sticking out. We covered the gas cell nub in styrofoam and are careful in handling the cells to avoid breaking this nub and potentially releasing the gas.

\subsection{Mechanical}

The available physical space limitations for the cell ($\sim$ 6 $\times$6 $\times$7 in$^3$) place severe design constraints on the size of the cell and the motor mechanism to move the cell in and out of the telescope beam.  The length of the cell is limited on one end by the entrance window to the calibration unit, and on the other end by the descending fold mirror for the calibration lamps.  A Newmark rotary stage was chosen to move the cell in and out of the beam given that a suitable low-cost linear stage with adequate travel would not fit into the available space.  The rotary stage is mounted to a modified calibration unit cover.  The single-body cover was replaced with a two piece cover with handles to permit easy insertion and removal of the gas cell and stage as a unit.   The stage is held in place with a custom machined aluminum bracket attached to the cover adjustable via lock screws and slightly over-sized holes in the cover to permit changing the alignment of the stage.   

\begin{figure}[tb]
  \begin{center}
    \includegraphics[width=0.34\textwidth]{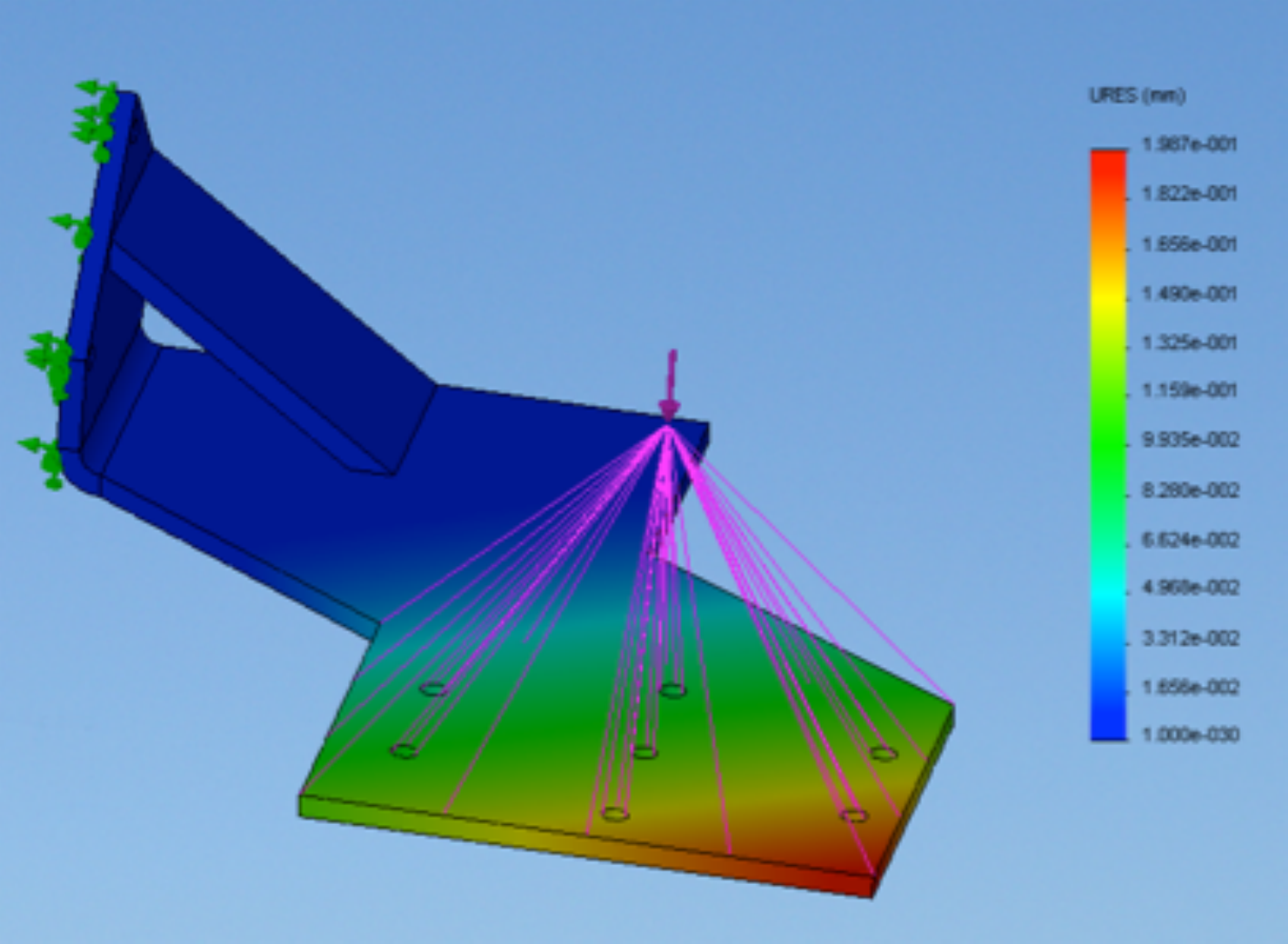}    
    \includegraphics[width=0.44\textwidth,clip=true,trim=3cm 0cm 0cm 0cm]{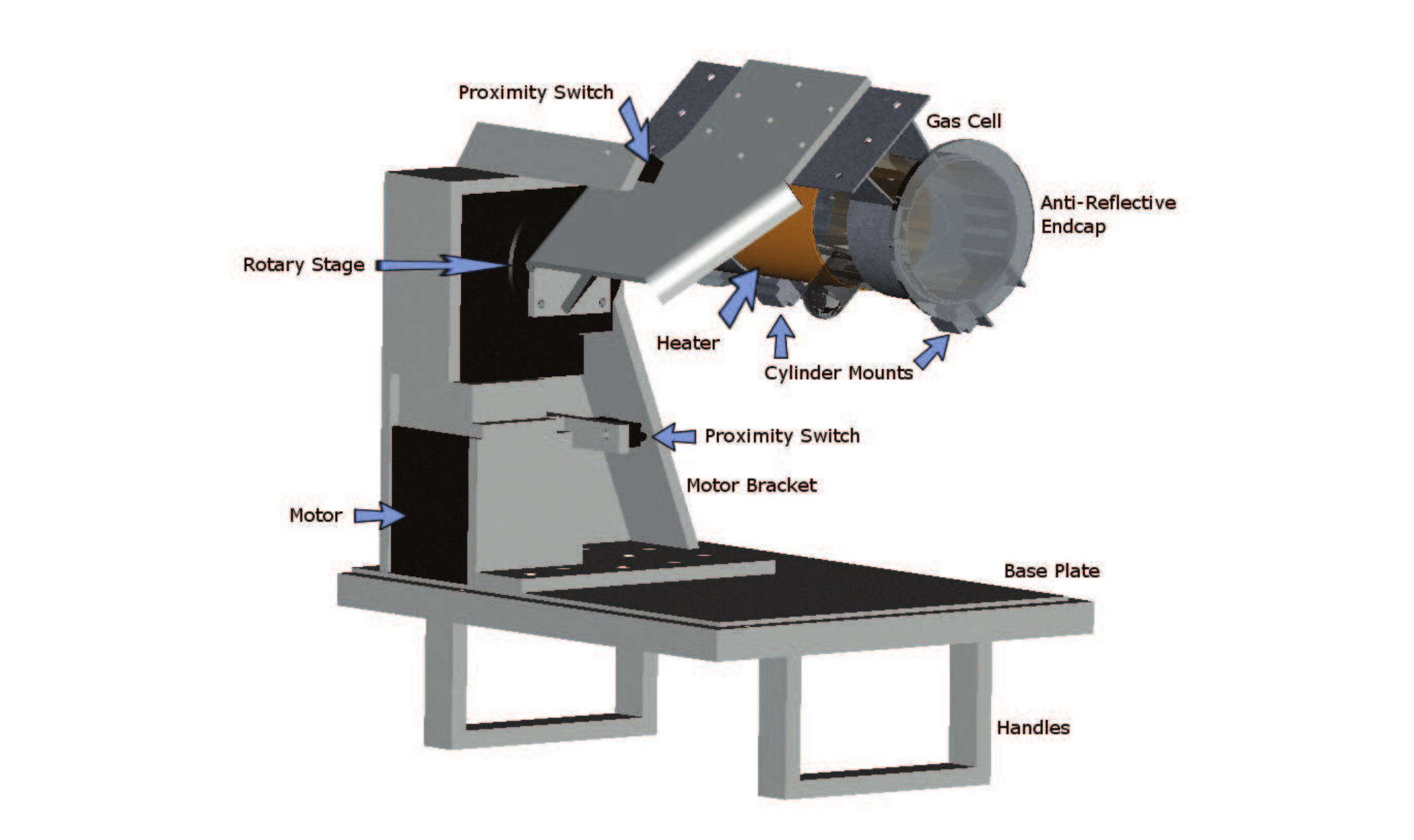}    
    \includegraphics[width=0.2\textwidth]{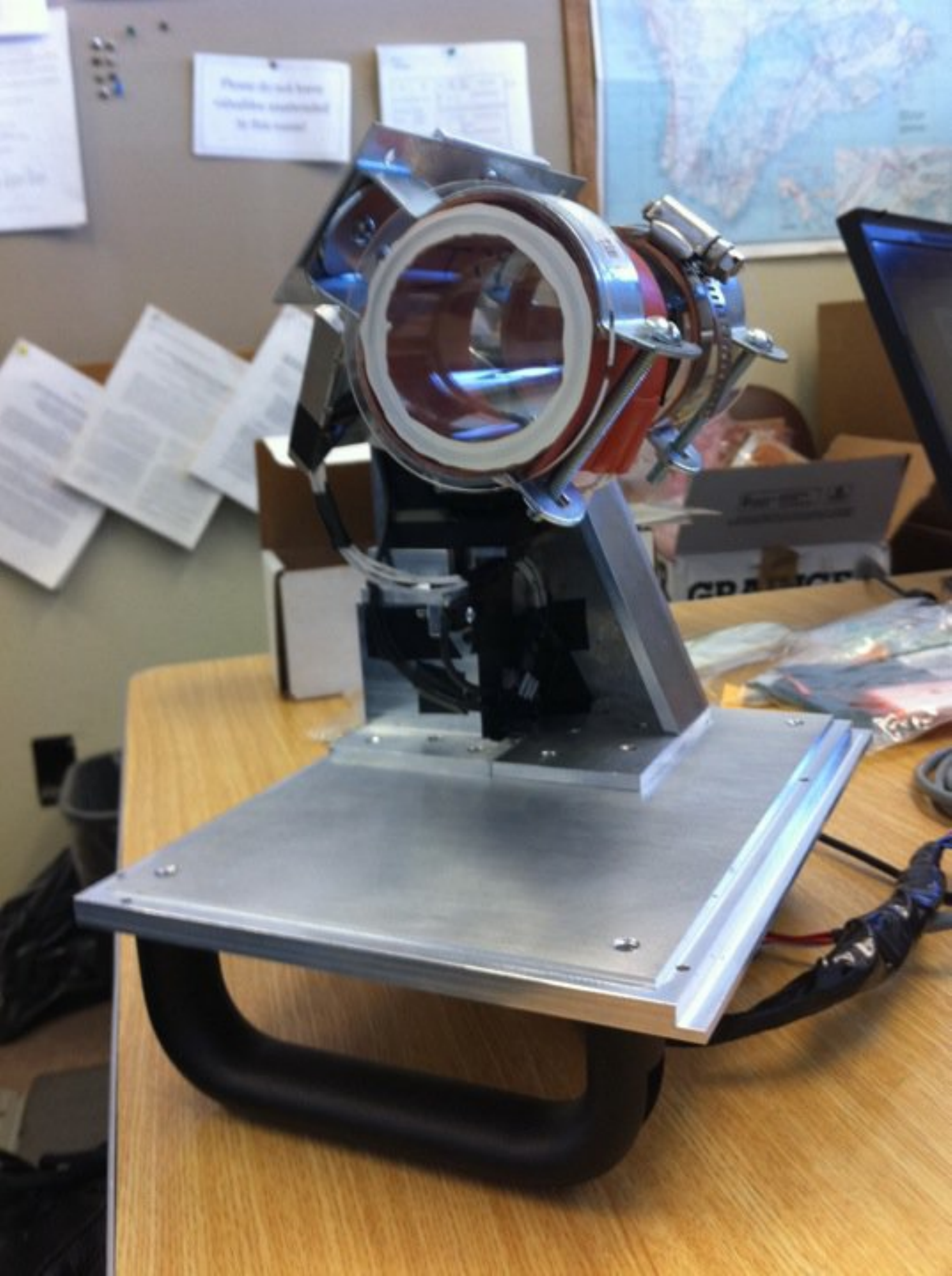}    
      \end{center}
      
  \caption{Left: Gas cell arm flexure modeling.  Middle: Solidworks model of the gas cell instrument with parts labeled.  Wires and bolts are omitted for visual clarity.  Right: Completed gas cell instrument. \label{fig:f6} }
\end{figure}

The gas cell glass body is held in place by fence clamps from Home Depot lined with rubber padding and shaped to the cell curvature.  The fence clamps are placed so as to avoid the ends of the cell, and avoid the remaining nub from the gas inlet tube.  The fence clamps were re-machined after the commissioning run to permit easier substitution of different gas cells.  The fence clamps are in turn affixed to a 2''x4'' aluminum join bracket made for joining wooden ``2x4s'', also obtained at Home Depot, which has suitable pre-machined holes for attaching the clamps. This bracket is in turn mounted to the stage via a custom machined aluminum arm.  Given the expected flexure from the instrument moving with the telescope at the cassegrain mount, the arm was designed and modeled to be thicker for minimal flexure deviation of the cell  (Figure \ref{fig:f6}).  Finally, we manually optimized the rotational angle of the stage and hence the gas cell alignment with the telescope beam by visual inspection by looking down on the calibration unit from above in a space in the primary mirror mount.  The optimal stage rotation angle deviated from our modeled estimate by a few degrees.

\subsection{Thermal}

\begin{figure}[tb]
  \begin{center}
    \includegraphics[width=0.40\textwidth]{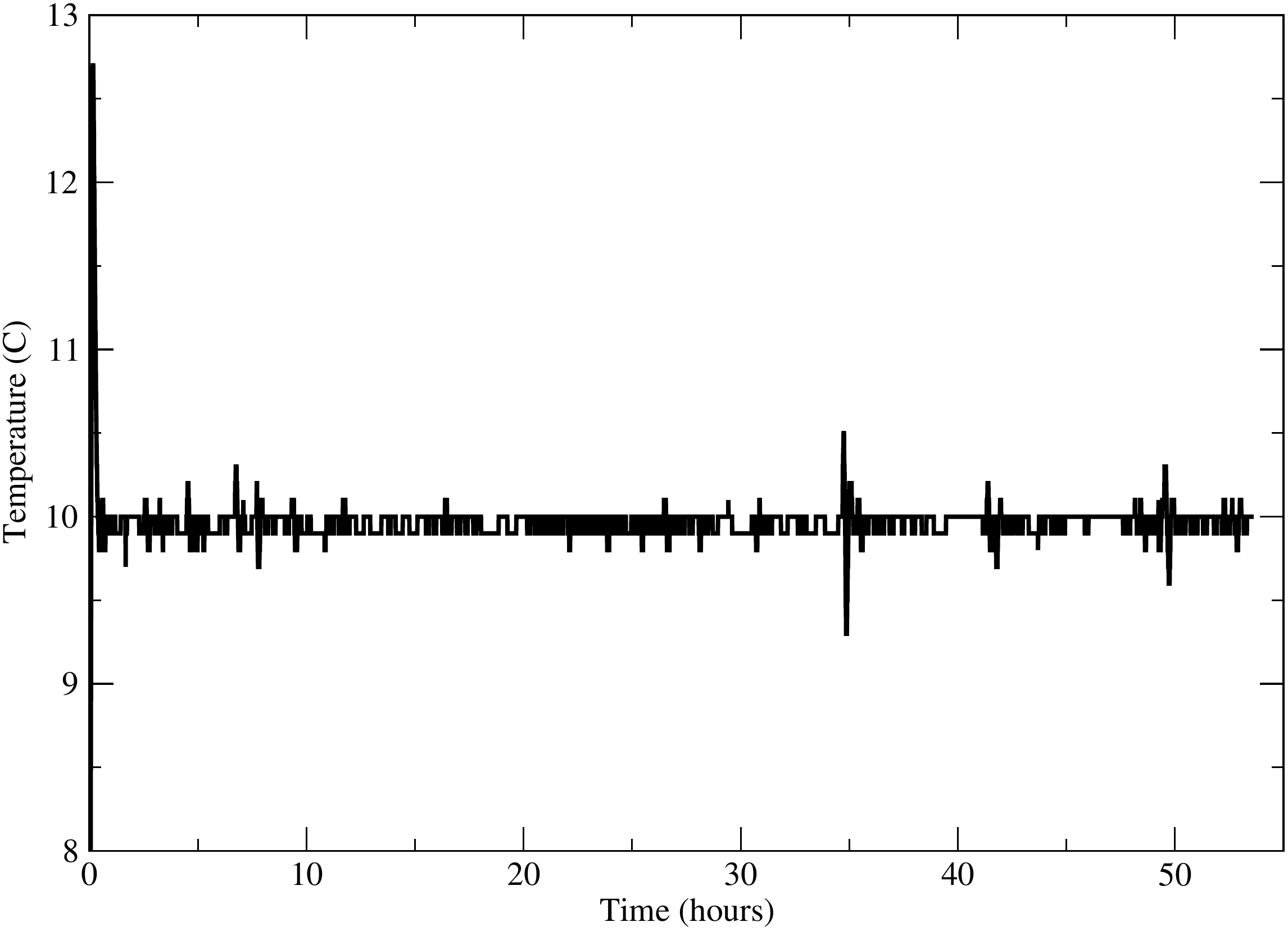}    
    \includegraphics[width=0.40\textwidth]{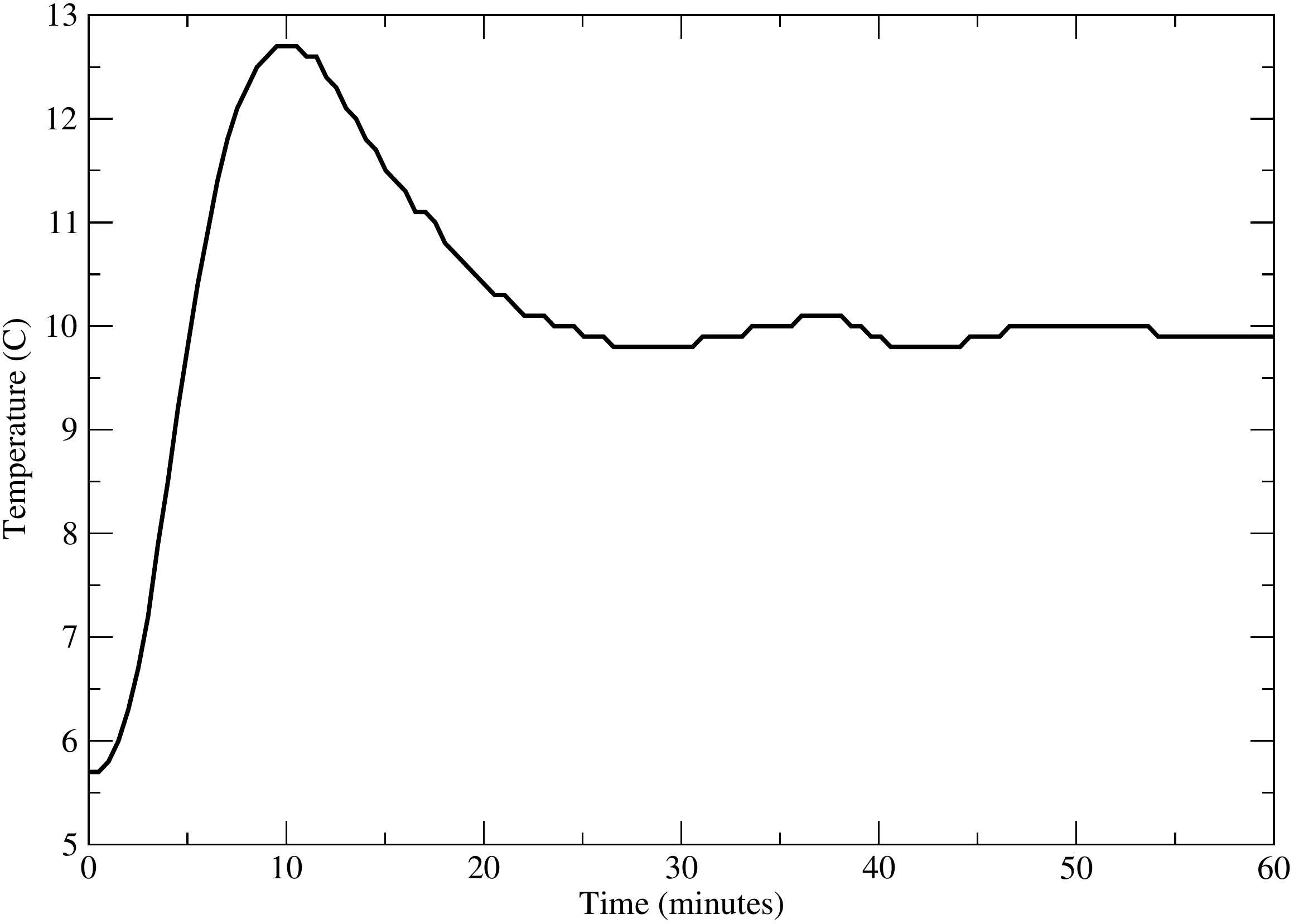}    
      \end{center}
      
 \caption{Reproduced from Anglada-Escude et al. 2012, Figure 4: Left: Gas cell temperature stability as a function of hours during first light.  Spikes in the temperature can be seen of up to 0.5$^\circ$ C, and are due to observations of very bright standards heating the gas cell.  Right:  Gas cell temperature as a function of minutes during first hour of turning heater on.  The gas cell stabilizes at the desired temperature within 1 hour. \label{fig:f7} }
\end{figure}

Methane and ammonia are in the gas phase at room temperature and approximately atmospheric pressures.  The CSHELL calibration unit entrance window is open to the ambient air of the IRTF telescope dome, which experiences temperature ranges of 276 to 284 K over the course of a year.  In order to mitigate velocity calibration errors due to temperature changes, we stabilize the gas cell temperature with a small silicon heater. For consistency, it is heated to 283 K (10$^\circ$ C), at the high end of dome temperatures experienced over the course of a year. Heating the cell had the effect of stabilizing the temperature within the entire calibration unit.  The cell has an RTD sensor attached giving temperature feedback to a temperature controller, which is expected to maintain the temperature to within $\pm$0.1 K. The temperature controller is a CN7823 model from Omega Engineering, and can be set and logged remotely to ensure stability. This should result in temperature-induced errors well below 1 m/s \cite{bean10}.  As a curiosity, we find that bright telluric standard stars (e.g. Sirius) heat the cell by up to $\sim$0.5$^\circ$ C before the temperature controller re-establishes an equilibrium gas cell temperature, a consequence of using near-infrared heat-absorbing greenhouse gases (Figure \ref{fig:f7}).

\subsection{Control Electronics and Software}

The remote operation of the gas cell and associated thermal and motor electronics is an essential design feature for ease of use and efficiency of observations, given that CSHELL mounts at the telescope cassegrain focus.   A small 1/2 inch hole in the calibration unit cover was made to pass the control cables through, including the motor control cable, two electromagnetic position switches for controlling the stage rotation, and thermal control lines, the latter two of which were integrated into a single 12-pin detachable connector.   The control lines enter a control electronics box mounted to the side of the calibration unit and shown in Figure 2.  This box contained the Omega thermal controller and power supply, as well as a RS-232 serial Newmark Systems NSC-1S Single Axis Stepper Motor Controller and power supply, an A/C D/C adapter, a KAX-10 fuse for the thermal controller, and the laptop power supply.  A commercial laptop is connected via two serial-to-USB converters for sending commands to the thermal and motor controllers, and is clamped to the side of the mounting plate opposite the electronics box and in close proximity to the CSHELL calibration unit.  Wiring diagrams are shown in Figure 8.  A small reset switch was added to the outside of the electronics box for the motor controller power supply.

\begin{figure}[tb]
  \begin{center}
    \includegraphics[width=0.40\textwidth]{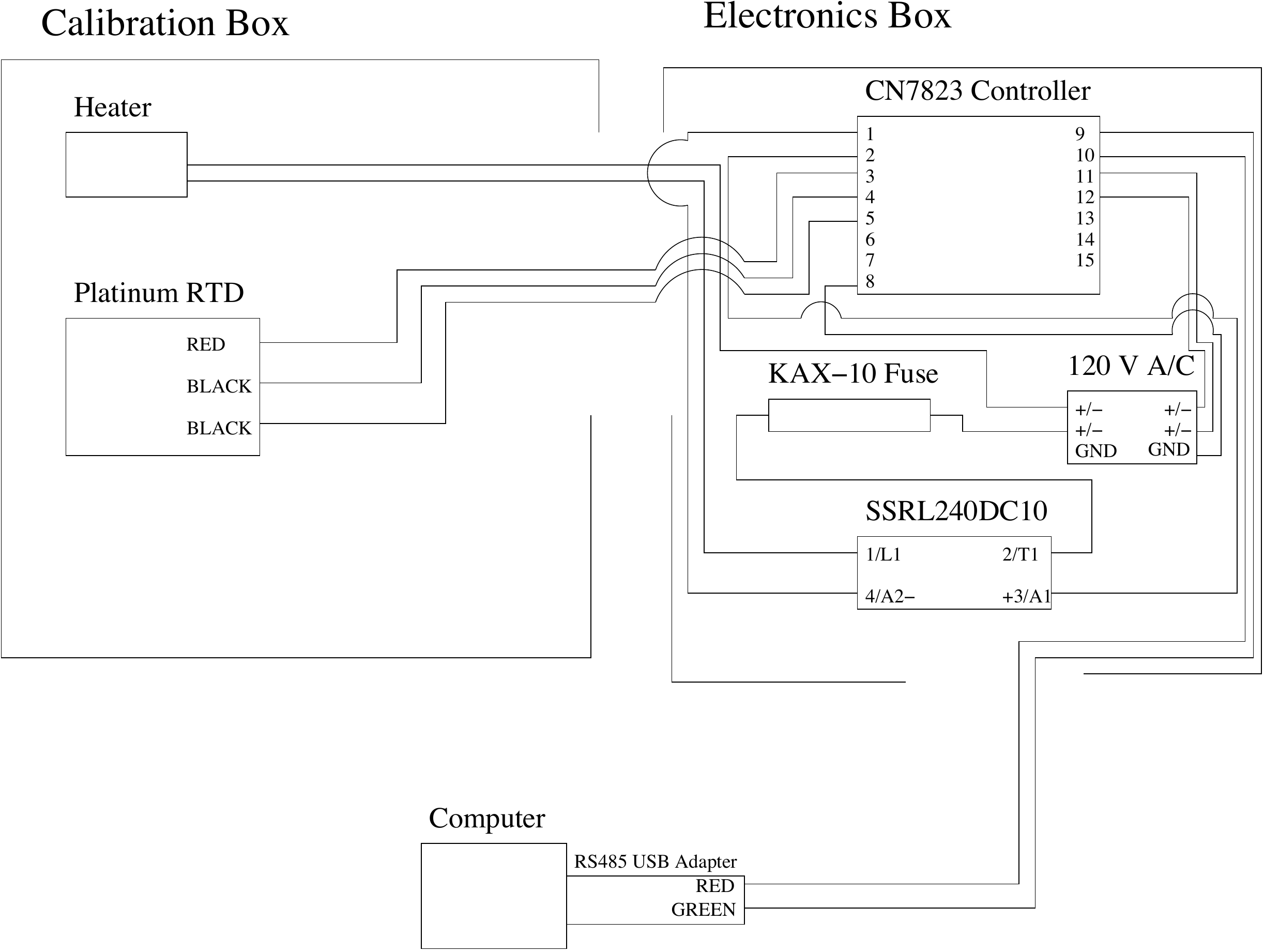}    
    \includegraphics[width=0.40\textwidth]{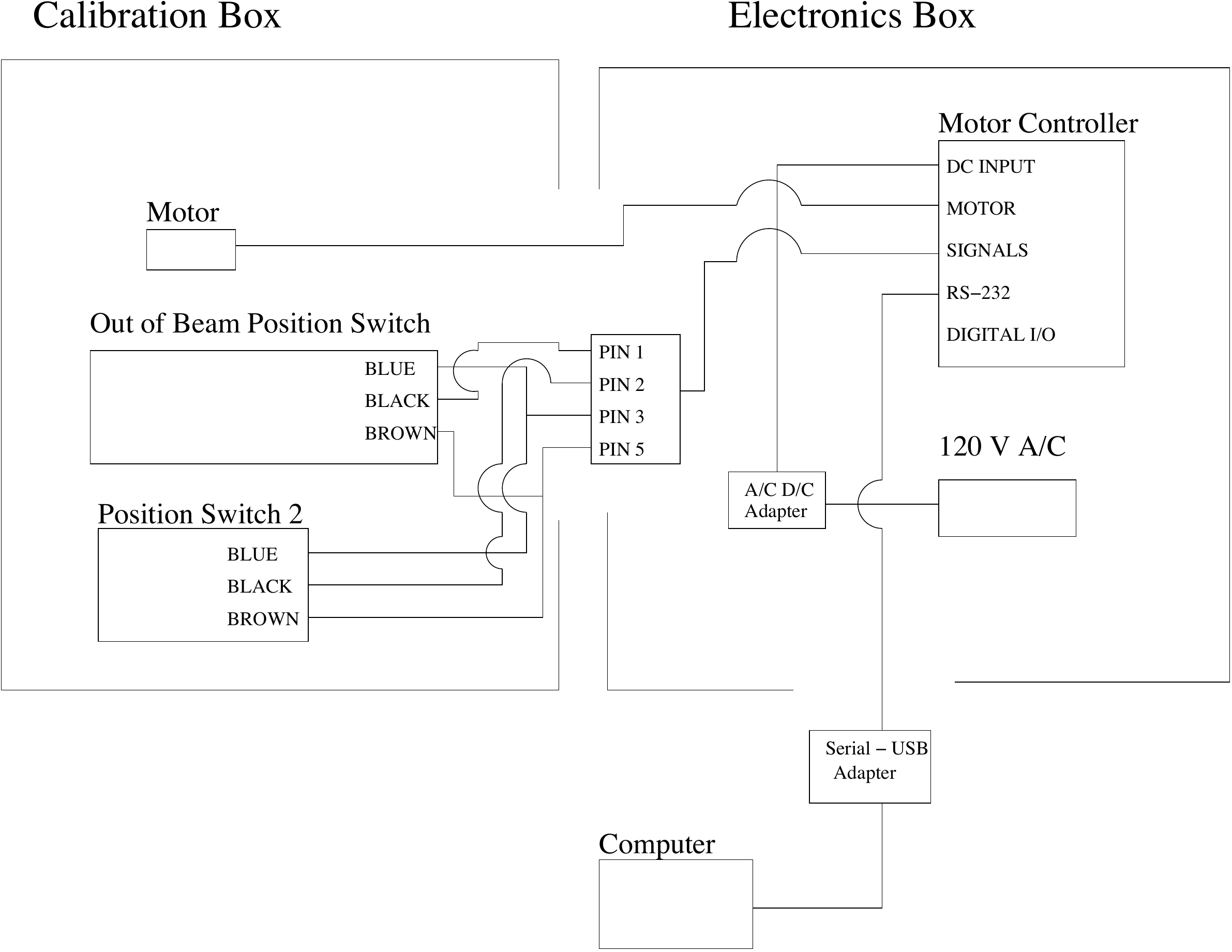}    
      \end{center}
      
 \caption{Thermal and motor wiring diagrams. \label{fig:f8} }
\end{figure}

We developed custom Python programs with command line interfaces to separately send commands to the thermal and motor controllers.   The motor control program is based upon the ``miniterm'' Python serial control program provided with the Newmark motor controller.  A low-level Microlynx language program is loaded into the motor controller containing a list of controller commands, whereas the temperature controller commands were written in binary directly to the serial port from within the Python program based upon the Omega controller user guide.  Commands for the motor controller include calibrating, rotating into and out of the beam, rotating by a specific angle, checking position switch status, and printing the current rotary stage angle.  Commands for the temperature controller include reading the current temperature, starting a controlled temperature feedback loop with logging to the screen or disk, and turning the heater on and off.  These Python programs are loaded onto the laptop, running Debian linux and configured to stay powered on when the laptop lid is closed, connected via ethernet to a hub on the telescope primary.  This allows for remote ssh connections to the laptop via an xterm from the instrument control computer in the telescope control room.  The laptop and control electronics are shut down between runs, requiring manual powering up at the telescope at the start of each observing run.  User guides for these Python programs are available for the community to use along with the gas cell hardware.

\section{Data Pipeline}

Observations have been obtained in a pilot survey of young, low mass stars with the gas cell on CSHELL at IRTF. Data have been collected and reduced from a number of observing runs from the fall of 2010 through the fall of 2012.  The detailed survey results will be presented in a future publication.  In this section, we present an updated overview of the process of starting from raw FITs images to arriving at our final radial velocity measurements, with some example radial velocity time-series and pipeline output.  A detailed presentation of the data pipeline will be presented in a future publication as well.  The pipeline is a significantly evolved version of the pipeline presented in Anglada-Escude et al. 2012, and has been developed over the past two years to overcome a number of significant data challenges identified since the preliminary analysis of first light data. 

\subsection{Optimal Spectral Extraction}

Spectra are extracted from the raw 2-D FITS images using a custom IDL pipeline to perform unweighted (standard) and flux-weighted (optimal) sums of counts over the spatial direction as a function of wavelength in the dispersion direction.  The software is derived from the software used in Bailey et al. 2012 and Tanner et al. 2012, and adapted for the nod-less observing strategy employed for the majority of our pilot survey.  The standard extraction is subtracted from the optimal extraction, and bad spectral pixels are flagged when the absolute deviation between these two spectra exceed a normalized threshold of 0.1 (Figure 9).  Despite the optimal extraction incorporating a bad pixel rejection, the substituted values for the bad pixels can still introduce systematic errors in the radial velocities.   Right ascension, declination, hour angle, date of observation, exposure time, and other telescope parameters are extracted from the FITS headers and combined with telescope latitude, longitude and elevation to generate barycenter radial velocity corrections from the motion of the Earth about the barycenter of the Solar System with a custom IDL routine \cite{stumpff1980}.

\begin{figure}[tb]
  \begin{center}
    \includegraphics[width=0.25\textwidth]{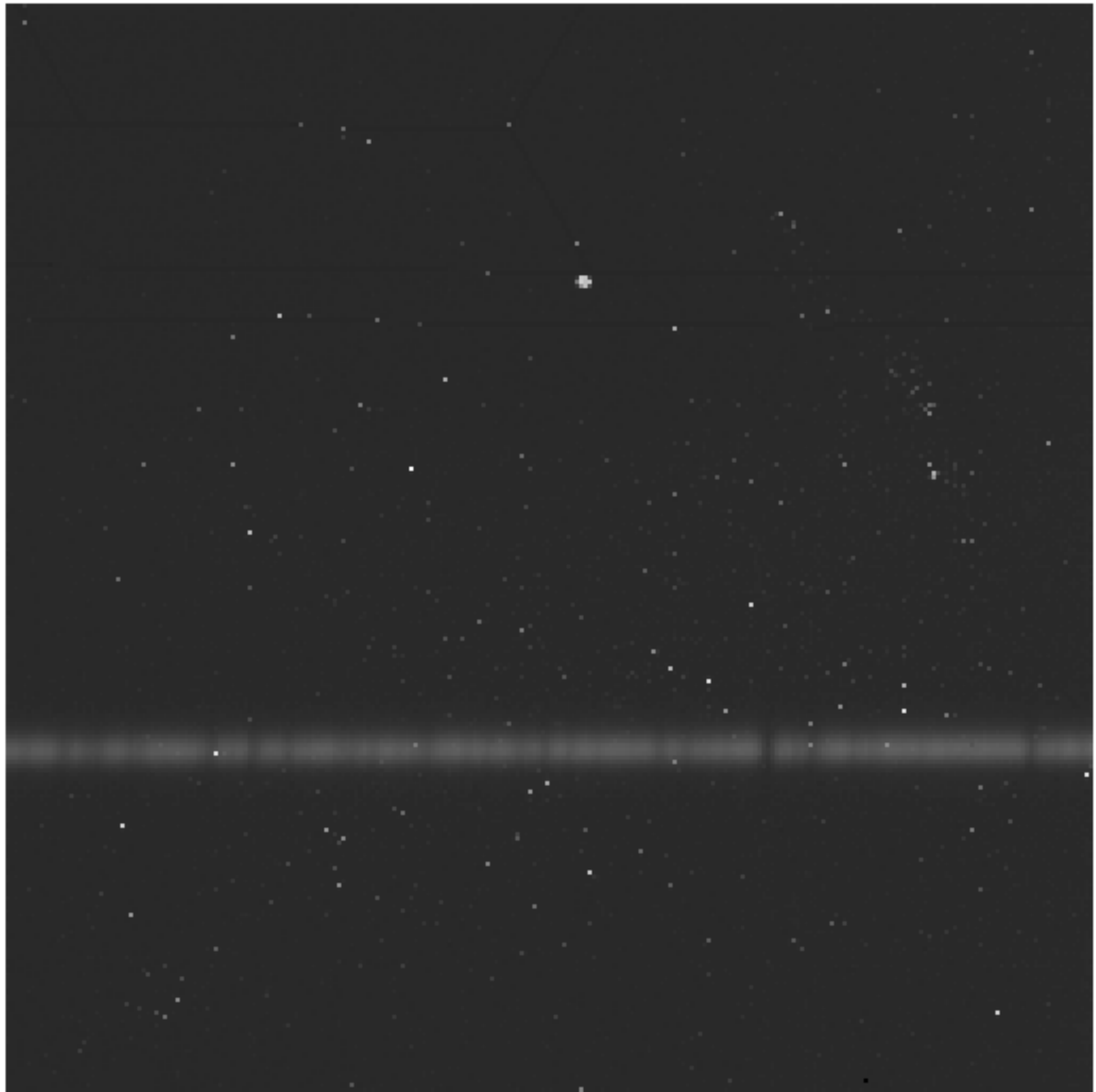}    
    \includegraphics[width=0.35\textwidth,clip=true,trim=0cm 0cm 0cm 0.7cm]{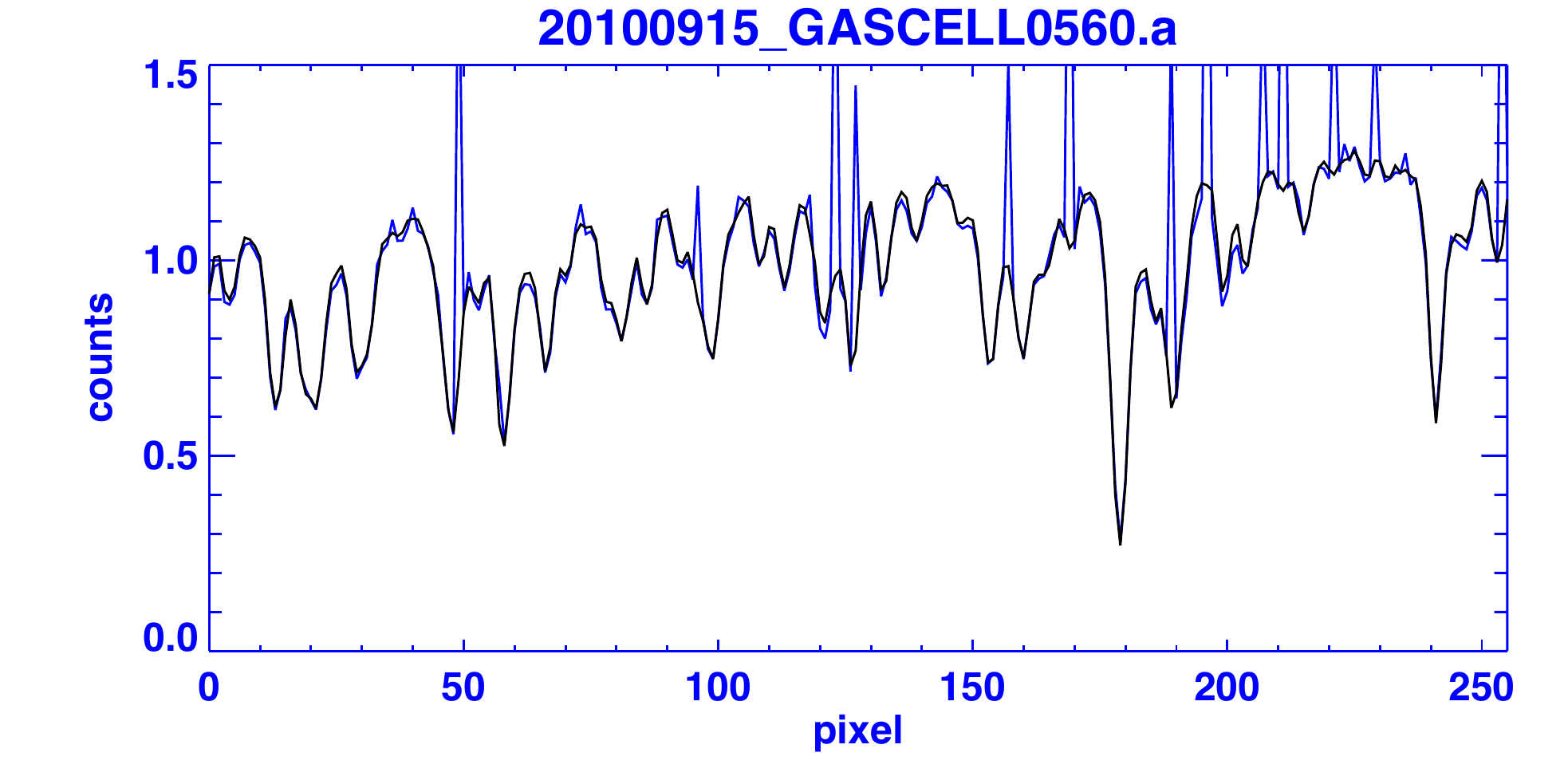}    
    \includegraphics[width=0.38\textwidth]{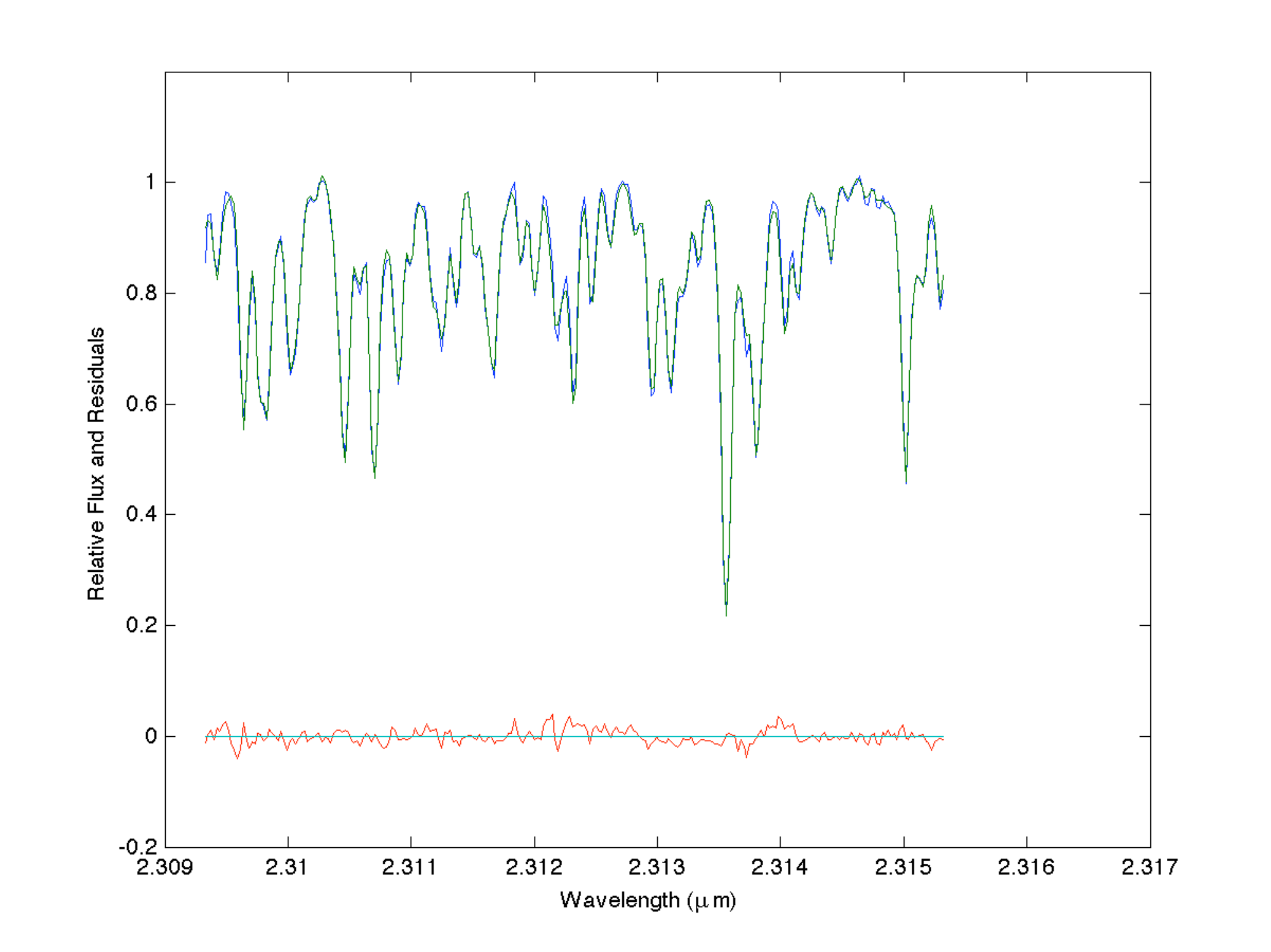}    
      \end{center}
      
 \caption{Left: Example FITS image of spectra of GJ 15A taken with the gas cell in the beam and obtained September 15th, 2010.  Middle: Example optimal (black) and standard (blue) extracted spectra for the same observation of GJ 15A.   Bad pixels in the FITS images result in large blue spikes in the standard extracted spectra.  Although bad pixels are removed by the optimal extraction process, they nevertheless negatively impact the radial velocities. Right: Optimal extracted spectra (blue) for the same observation of GJ 15A, with best fit forward model (green), with residuals (red) that include flagged bad pixels that are excluded in the minimization process.  \label{fig:f9} }
\end{figure}

\subsection{Forward Model Challenges}

Extracting precise radial velocities is challenging, with a 30 m/s shift corresponding to $\sim$1/100$^{th}$ of a detector pixel.  For example, Keck HIRES visible wavelength spectra require a multi-parameter forward model fit to the data via a least squares minimization\cite{Butler96}.  The data are first independently continuum normalized with a series of three 6th order polynomials applied to each order to correct for the spectrograph blaze function, and each order is then divided into 2 Angstrom snippets.  The model for a given snippet of spectra is constructed by multiplying wavelength-shifted ``infinite resolution'' templates for the stellar and gas cell spectra convolved with a variable approximation of the line-spread function for a given wavelength solution.   The radial velocities are determined by subtracting the best-fitting wavelength shifts for the gas cell and stellar template spectra and then subtracting the barycenter correction.

In the near-infrared, we started with a similar approach to that used for Keck HIRES, given its success.  However, our analysis required several increases in complexity, including the inclusion of a model for the telluric absorption in both the radial velocity extraction and the generation of the empirical stellar template.  Consequently, we developed a Matlab-based pipeline from scratch.  Readily apparent from Figure 9, the stellar and gas cell line density is so high (a desirable problem!) that a true continuum level cannot be determined in a straightforward fashion.  Thus, the correction for the spectrograph blaze function and the forward modeling of the observed spectra must be accomplished simultaneously.  This increase in the number of free parameters computationally forces us to use a more advanced technique for multi-dimensional parameter space exploration beyond a least-squares minimization.  We implemented a Nelder-Mead amoeba simplex search\cite{nelder}.  

We were initially unable to obtain adequate long term radial velocity precision, and we noticed an unidentified low frequency noise source in the Fourier transform of fit residuals despite the high S/N.  We discovered that the poor long-term radial velocity precision and low frequency residuals in the model fits were due to variable fringing present in the spectra at a flux amplitude of $\sim$1-4\%.  We were able to attribute the source of the fringing to  the continuously variable filter for the order selection (Figure 10).  Fringing is the wavelength-dependent interference of the starlight with itself, which produces an alternating constructive and destructive pattern as a function of wavelength that can be described by a sinusoid.  Due to the high line density in the survey spectra, this fringing was not apparent by eye until we intentionally obtained A-star spectra with no absorption lines and without the gas cell present in December 2012.  This was not part of our normal observing plan, since only variable telluric lines were present.  We implemented a fringing correction that adds a sinusoid multiplication step in our forward model generation, in addition to the other components described below.  The fringing contributes $\sim$1 km/s long-term RV errors, and correcting for the fringing in our forward model fortunately and dramatically improves our radial velocity precision.  Finally, despite our small spectral grasp, we uncovered a small but variable quadratic term in our wavelength solution.  Including a correction for this term as a free parameter enables us to obtain long-term RV precision of less than $\sim$100 m/s and approaches our expected photon noise limit of $\sim$40 m/s.  However, the variable wavelength solution over our small spectral grasp introduces a non-trivial source of radial velocity noise.

\begin{figure}[tb]
  \begin{center}
    \includegraphics[width=0.45\textwidth]{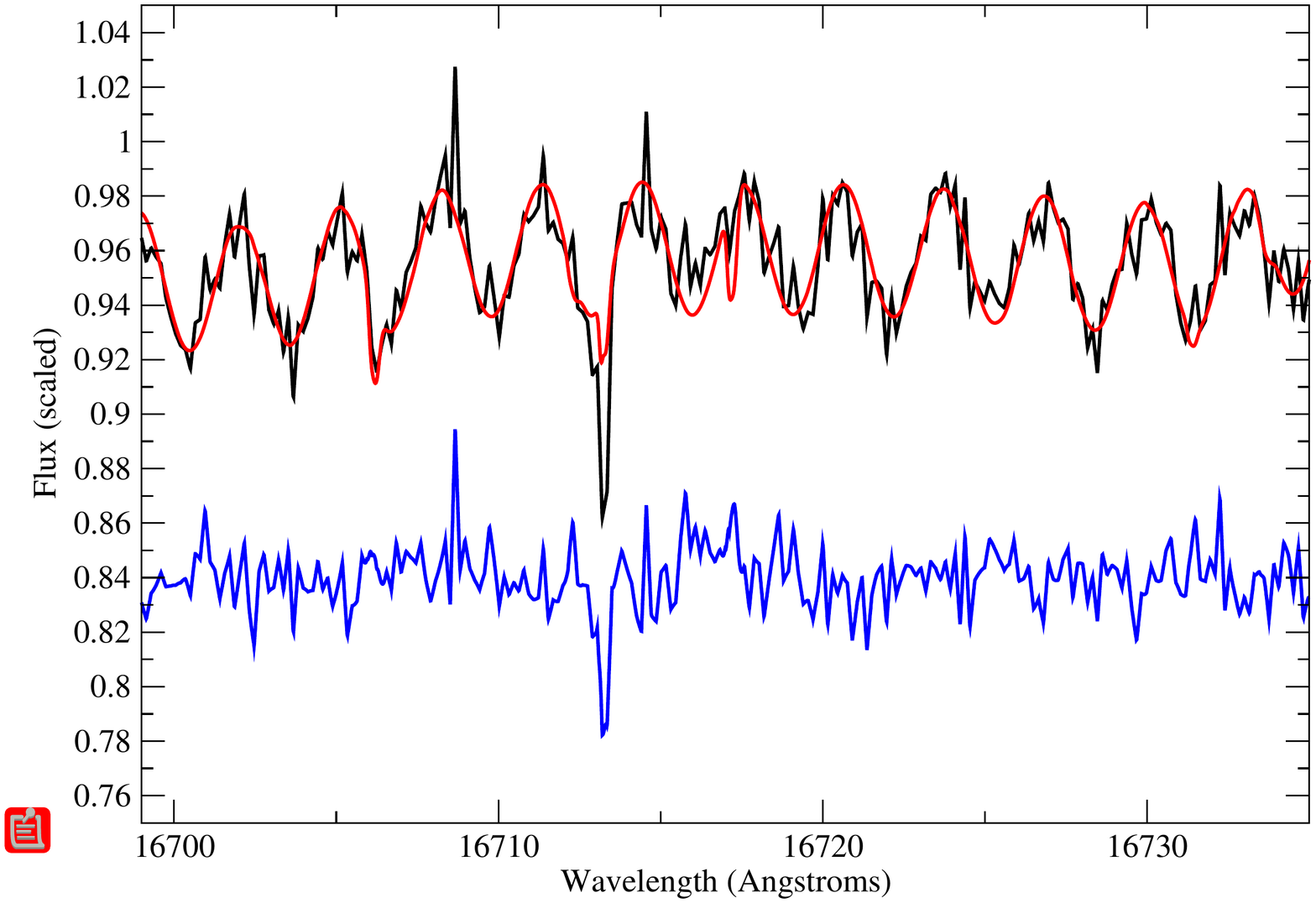}    
    \includegraphics[width=0.45\textwidth]{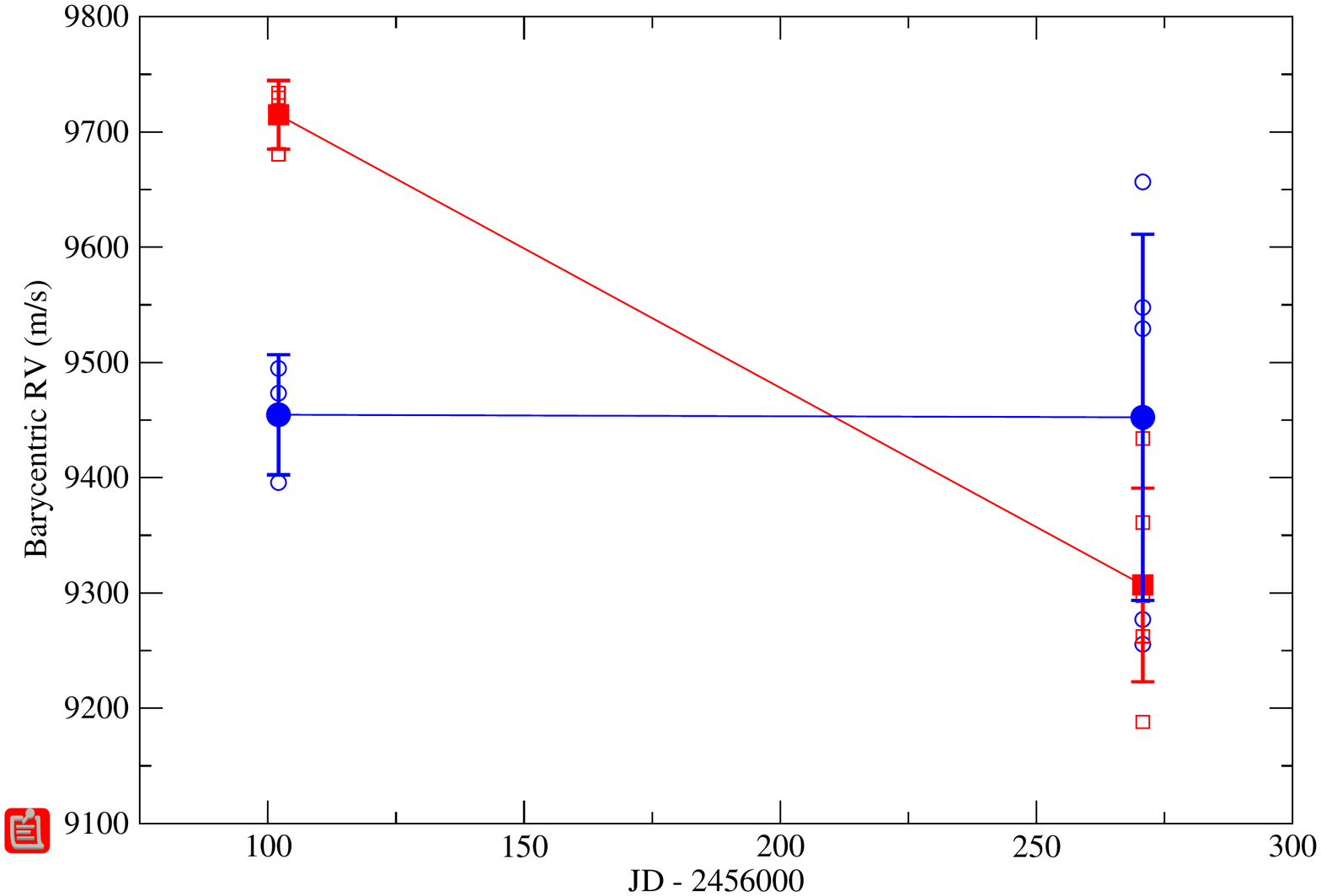}    
      \end{center}
      
 \caption{Left: December 2012 spectrum of 32 Peg (B9 star with one telluric line) shown in black with large amplitude fringing -- the systematic sinusoidal pattern as a function of wavelength -- clearly visible in this continuum source. Our model fit to the fringing is shown in red and the shifted residuals in blue. Removing the telluric line was not a goal of this particular analysis, but we demonstrate that the fringing in this mostly featureless continuum can be successfully removed. The remaining noise level is consistent with the photon, flatfield, sky and read noise sources, and two bad pixels (rms=1.02\%).   Right:  Red: Radial velocity measurements taken over 6 months at 1.6717 $\mu$m for SV Peg without fringing correction taken with the fiber scrambler and gas cell in 2012. Blue: Radial velocity measurements with the fringing correction included. The systematic radial velocity offset for this data set decreases from 437 to 2 m/s!}
\end{figure}

\subsection{Forward Model Pipeline Overview}

The pipeline is divided into two main portions: (1) the generation of an empirical stellar template, and (2) the calculation of the radial velocities. Both of these portions rely on the generation of a model to fit to the observations.   The model of the observations is constructed from the derived empirical stellar template spectrum multiplied by the experimental high resolution FTIR gas cell spectrum and multiplied by an observed telluric spectrum obtained at high resolution from a solar telescope.  Wavelength shifts are applied to all three spectra to simulate the relative systematic instrumental, and genuine astrophysical Doppler effects. The gas cell and telluric spectral lines are also augmented by associated optical depths. Due to the stability of the gas cell, we assume the gas cell optical depth is constant for every observation in a dataset. The combined spectrum is convolved with a model line spread function (LSF) consisting of a central Gaussian distribution surrounded on both sides by an equal number of satellite Gaussians at fixed distances away from the central Gaussian. Our default number of satellite Gaussians was four, and all the Gaussians are characterized purely by their relative maxima and standard deviations. In order to account for an uneven continuum in the data spectra blaze function, we divide out a linear polynomial from the spectra, and then multiply our model spectra with another linear polynomial. Higher order terms in the polynomial were excluded from our pipeline as they introduced extraneous correlations and radial velocity noise into our fits.  Finally, we account for fringing in CSHELLÕs CVF filter as previously mentioned by multiplying the model spectra with a sinusoid characterized by its amplitude, period, and phase. A summary of the model parameters are listed in Table 1.

\begin{table}[h]
\caption{Table of Forward Model Parameters} 
\label{tab:t1}
\begin{center}       
\begin{tabular}{|l|l|l|} 
\hline
\rule[-1ex]{0pt}{3.5ex}  Number & Parameter & Variable or Fixed? \\
\hline
\rule[-1ex]{0pt}{3.5ex}  1 & Standard Deviation of left-most Satellite Gaussian in LSF model & Bounded variable \\
\rule[-1ex]{0pt}{3.5ex}  2 & Standard Deviation of 2nd left-most Satellite Gaussian in LSF model & Bounded variable \\
\rule[-1ex]{0pt}{3.5ex}  3 & Standard Deviation of central satellite Gaussian in LSF model  & Bounded variable \\
\rule[-1ex]{0pt}{3.5ex}  4 & Standard Deviation of 2nd right-most Satellite Gaussian in LSF model & Bounded variable \\
\rule[-1ex]{0pt}{3.5ex}  5 & Standard Deviation of right-most satellite Gaussian in LSF model & Bounded variable\\
\rule[-1ex]{0pt}{3.5ex}  6 & Amplitude of left-most satellite Gaussian in LSF model & Bounded variable \\
\rule[-1ex]{0pt}{3.5ex}  7 & Amplitude of 2nd left-most satellite Gaussian in LSF model & Bounded variable\\
\rule[-1ex]{0pt}{3.5ex}  8 & Amplitude of Central satellite Gaussian in LSF model & Fixed = 1\\
\rule[-1ex]{0pt}{3.5ex}  9 & Amplitude of 2nd right-most satellite Gaussian in LSF model & Bounded variable \\
\rule[-1ex]{0pt}{3.5ex}  10 & Amplitude of right-most satellite Gaussian in LSF model & Bounded variable \\
\rule[-1ex]{0pt}{3.5ex}  11 & Normalization constant for model flux scale & Variable \\
\rule[-1ex]{0pt}{3.5ex}  12 & Normalization linear slope for model flux scale & Variable \\
\rule[-1ex]{0pt}{3.5ex}  13 & Wavelength shift of star template in Angstroms & Variable \\
\rule[-1ex]{0pt}{3.5ex}  14 & Wavelength shift of gas cell spectra in Angstroms & Variable \\
\rule[-1ex]{0pt}{3.5ex}  15 & Wavelength shift of telluric spectra in Angstroms & Variable \\
\rule[-1ex]{0pt}{3.5ex}  16 & Optical depth of gas cell spectra & Fixed \\
\rule[-1ex]{0pt}{3.5ex}  17 & Optical depth of telluric spectra & Variable \\
\rule[-1ex]{0pt}{3.5ex}  18 & Linear term in the stellar template wavelength scale & Fixed to optimized value \\
\rule[-1ex]{0pt}{3.5ex}  19 & Normalization constant for the star template flux scale & Fixed = 1 \\
\rule[-1ex]{0pt}{3.5ex}  20 & Amplitude of the data fringing sinusoid  & Variable \\
\rule[-1ex]{0pt}{3.5ex}  21 & 2*pi/period of the data fringing sinusoid & Variable \\
\rule[-1ex]{0pt}{3.5ex}  22 & Phase of the data fringing sinusoid  & Variable \\
\rule[-1ex]{0pt}{3.5ex}  23 & Linear term for the model wavelength scale & Variable \\
\rule[-1ex]{0pt}{3.5ex}  24 & A wavelength shift for the model wavelength scale in Angstroms & Variable \\
\rule[-1ex]{0pt}{3.5ex}  25 & Amplitude of the stellar template fringing sinusoid & Fixed = 0 \\
\rule[-1ex]{0pt}{3.5ex}  26 & 2*pi/period of the star template fringing sinusoid & Fixed = 0 \\
\rule[-1ex]{0pt}{3.5ex}  27 & Phase of the star template fringing sinusoid w/r/t the stellar template & Fixed = 0 \\
\rule[-1ex]{0pt}{3.5ex}  28 & Quadratic wavelength correction for the model wavelength scale & Variable \\
\rule[-1ex]{0pt}{3.5ex}  29 & Quadratic wavelength correction for the stellar template wavelength scale & Fixed to optimized value \\

\hline
\end{tabular}
\end{center}
\end{table}

The wavelength solution of the experimental gas cell spectrum and the observed telluric spectrum are assumed to be ``perfect'', while that of the derived stellar template and the data have quadratic wavelength solutions that can deviate from the initial estimated linear wavelength solution and are also variable. Therefore, we apply a quadratic wavelength solution correction to both the stellar template and the data. The stellar template wavelength solution is assumed to be fixed for all spectra in a dataset. Conversely, the wavelength solution is different for every single data spectrum, and thus it is treated as free parameters in the fits. 

The model is compared to the data after the former is down-sampled (binned) from 4096 ($\lambda$,flux) pairs to the 256 pixels of the data and their associated normalized counts. Comparison of the model and the data is done via calculating the RMS of the residuals (data minus model). In order to account for edge effects, we taper the last five pixels of the residuals on either end by multiplying by linearly decreasing weights. We also ignore any Òbad pixelsÓ from the RMS calculation, as discussed earlier. 

The above processes are used in both the stellar template generation and the radial velocity calculation portions of the pipeline. The template is derived by initially fitting a model that uses a guess template (usually a flat line) to data taken without the gas cell (e.g. the model is made up of only the guess template and the telluric spectrum), then adding the averaged residuals back into said template to improve its agreement with the data. The wavelength solution corrections are ÒreversedÓ during the addition in order to maintain the original template wavelength solution. This procedure is repeated until the RMS of the fits between the model and the sans gas cell observations converge. This template is then used in conjunction with the gas cell and the telluric spectra in the radial velocity calculations, where all parameters are fit simultaneously. These parameters include the relative maxima and standard deviations of the Gaussians in the LSF; the coefficients of the linear polynomial; the wavelength shifts of the stellar template, gas cell, and telluric spectra; the optical depths of the gas cell and telluric spectra; the wavelength corrections to the stellar template and data wavelength solutions; and the amplitude, period, and phase of the fringing-compensating sinusoid. The barycenter-corrected radial velocities are calculated from the difference between the best-fit stellar and gas cell wavelength shifts divided by the central wavelength of the data wavelength solution, multiplied by the speed of light, and with the barycenter velocities subtracted off. Finally, these RVs are optimized by fixing the gas cell optical depth and stellar template wavelength solution to specific values for all observations in a dataset such that the RV scatter is minimized. 

\section{Results}

Here we discuss the current radial velocity results from our pilot survey.

\subsection{Short Term Noise Floor -- SV Peg}

\begin{figure}[tb]
  \begin{center}
    \includegraphics[width=0.45\textwidth]{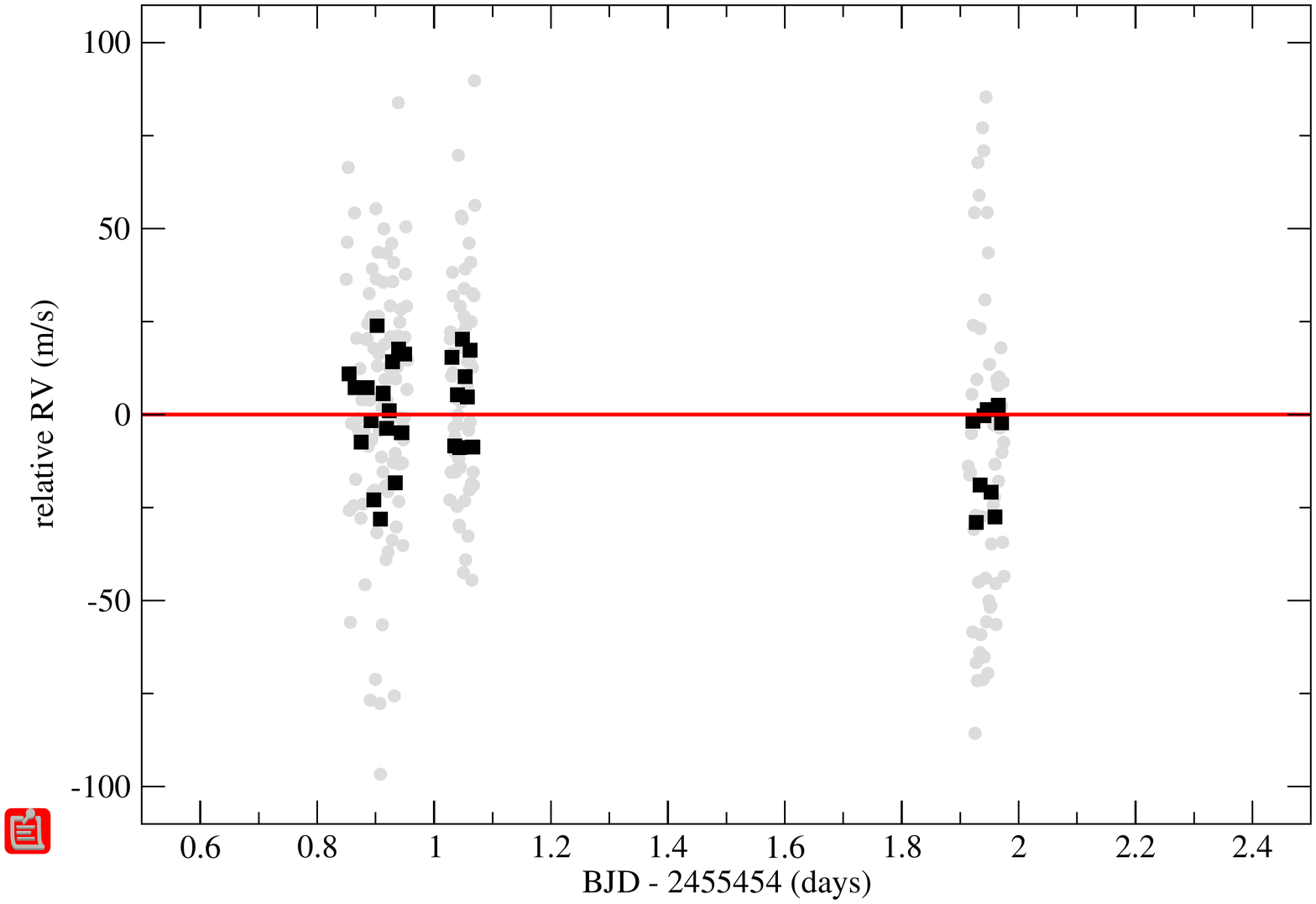}    
    \includegraphics[width=0.45\textwidth]{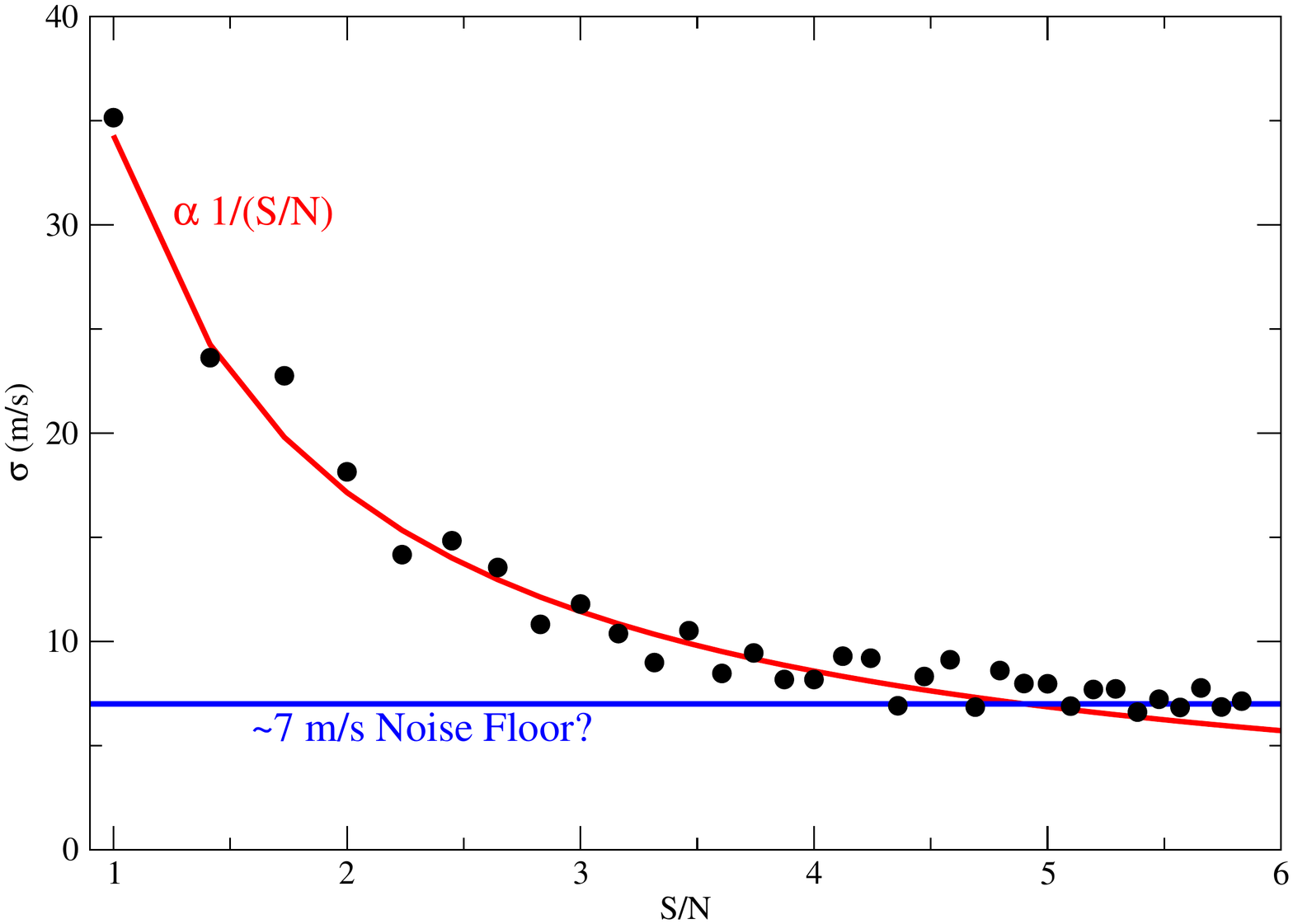}    
      \end{center}
      
 \caption{Left: Relative radial velocity measurements of the super giant star SV Peg (K = -0.4 mag) obtained during the first two nights. Each spectra has a formal S/N of $\sim$100. SV Peg is a supergiant star  and, as such, should show some RV variability in time-scales of  a few days.  The gray points have an RMS of 35 m/s.  The black points result from co-adding every 6 gray points and have a resulting precision of 14.8 m/s.   Right: Coadding the radial velocity measurements and re-computing the radial velocity precision improves the radial velocity precision, following a 1/(S/N) ``white'' photon noise curve down to $\sim$7 m/s.  The horizontal axis is the (S/N)/100, and the vertical axis is the corresponding radial velocity precision.  The theoretical ``white'' photon noise curve is shown in red, with measurements as black circles, and a possible noise floor shown as a blue horizontal line at 7 m/s, corresponding to where the measured precision deviates from the photon noise curve.\label{fig:svpeg_rv}}
\end{figure}

Several hundred high S/N spectra (S/N = 100, one spectra every 30 seconds) of the K=-0.4 mag supergiant M7 star SV Peg were obtained (Figure \ref{fig:svpeg_rv}).  From the radial velocity measurements of SV Peg using an older version of the pipeline (without the fringing correction), we can co-add measurements to probe the short-term noise floor with the gas cell as a function of S/N.  From a precision of $\sim$35 m/s for a S/N$\sim$100, the RV precision follows an expected 1/(S/N) curve down to $\sim$7 m/s for S/N$>$400.   A re-analysis of this data is underway with the updated pipeline, but the impact is expected to be minimal.

\subsection{Long Term Precision -- GJ 15A}

We have applied the new pipeline to the K=4.02 mag M2V dwarf and radial velocity standard GJ 15A, known from visible radial velocity surveys to be stable to $<$3 m/s and thus lacking any companion that would produce a signal $>$3 m/s.  We have observed GJ 15A over the past two years with an expected photon noise limited precision of $\sim$40 m/s per night, after co-adding measurements taken within a night to achieve a S/N$\sim$100 per night.  Our single night measurement uncertainties have a median value of 37 m/s.  Our single night measurement uncertainties are calculated by taking the radial velocity standard deviation of individual co-adds taken within a night, and dividing by the square root of the number of measurements.  Thus, the single night uncertainties are consistent with the expected photon noise precision.  

\begin{figure}[tb]
  \begin{center}
    \includegraphics[width=0.48\textwidth]{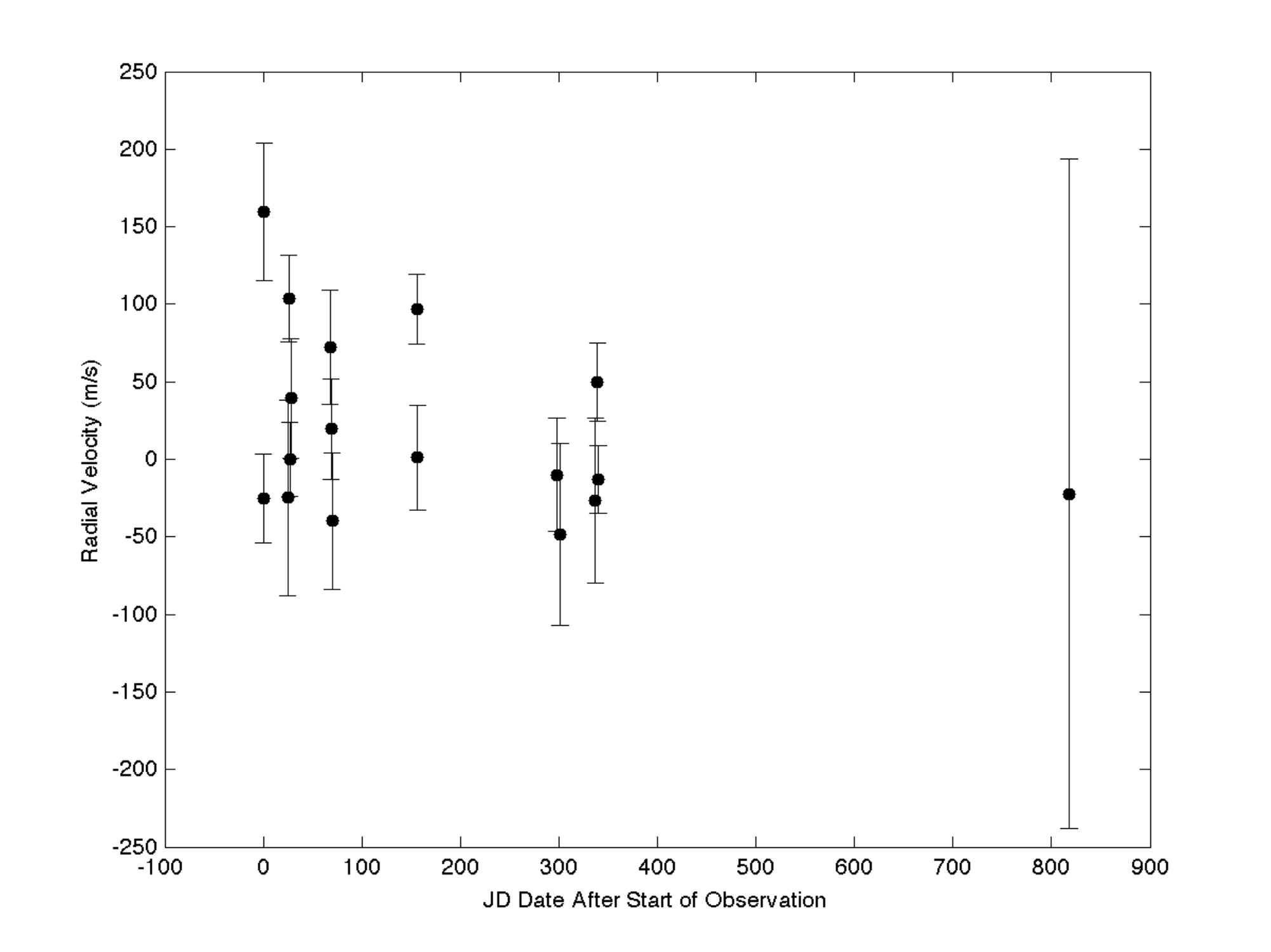}    
    \includegraphics[width=0.48\textwidth]{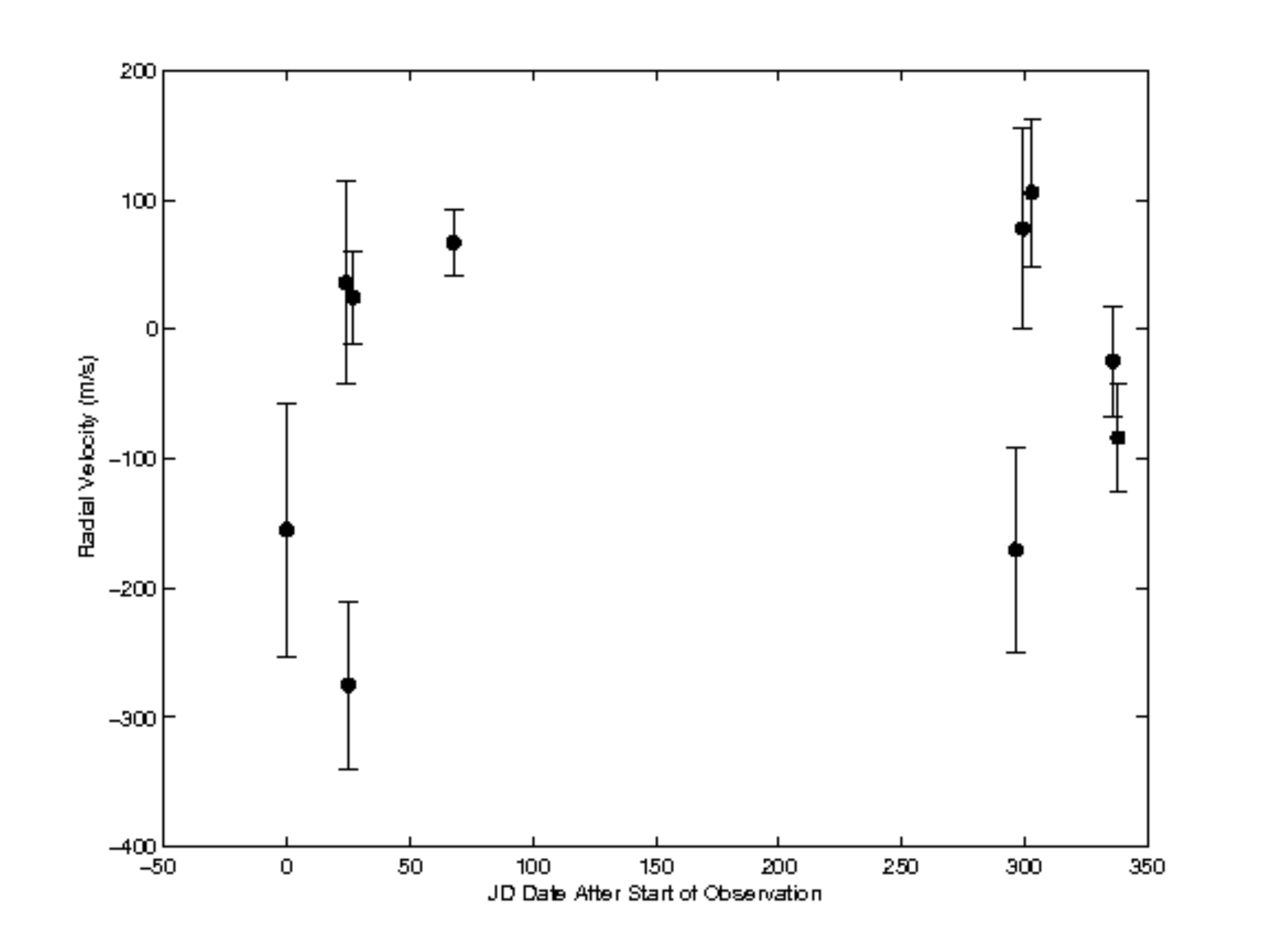}    
      \end{center}      
 \caption{Left: Radial velocity time-series spanning 800 days for GJ 15A taken with the gas cell on CSHELL at IRTF from September 2010 through December 2012.  The rms radial velocity precision is 58 m/s, 47 m/s after discounting a data point from the commissioning run.  Right: Radial velocity time-series for AU Mic, where the precision is 126 m/s.  We confirm the result in Bailey et al. (2012) that AU Mic is a radial velocity variable. \label{fig:f12}}
\end{figure}

However, we have measured a long term radial velocity precision of 58 m/s, implying a $\chi^2=2.64$, shown in Figure \ref{fig:f12}.  The final precision is worse than the expected photon noise limited precision.  Removing a measurement from our first commissioning run, when we had not yet chosen an optimal detector row on which to place the star every time, the long-term radial velocity precision drops to 48 m/s, much closer to the expected photon noise performance.  Regardless, this indicates a long-term systematic noise source is present in our data, which we attribute to a number of possible factors.  For one, systematic noise in the detector from bad pixels, even after they are flagged, adversely affect our long term RV precision since they change between runs.  Second, the variable fringing, and variable quadratic wavelength solution for the wavelength scale of the detector is expected to introduce systematic night-to-night radial velocity errors on the order of $\sim$25 m/s each that are not entirely accounted by the differential shifts between the gas cell and stellar lines (25 m/s $\sim$1 pixel / 256 pixels $\times$ c / R).  These two noise sources alone are adequate to explain the difference between our expected photon-noise limited precision and the observed.  This result suggests that future efforts with a modern detector (e.g. a H2RG) with no fringing will lead to excellent long term radial velocity precision, such as is observed with CRIRES on the VLT with an ammonia gas cell in Bean et al. 2010.

\subsection{Model Parameter Stability}

In addition to the radial velocity measurements from our forward model pipeline, another useful output is the stability of the spectrograph as inferred from the forward model parameters, which are listed in Table 1.  In Figures 13 \& 14, we present a panel of the dependence of the radial velocities on each of the forward model parameters.  As can be seen, the radial velocities do not depend on any of our model parameters, indicating we have optimized the model.  Other tests involving higher order polynomial models for the continuum introduced radial velocity trends correlated with the magnitudes of those parameters.  Similarly, we find that the gas cell does serve as a better wavelength calibration source that the telluric lines, as evidenced by the strength (scatter) of the correlation of the gas cell and telluric model shifts with the stellar shift (not shown).  Finally, we have not yet had an opportunity to optimize the number of satellite Gaussians nor the spacing thereof in our LSF model, which will be the subject of a future analysis.

\begin{figure}[tb]
  \begin{center}
    \includegraphics[width=0.3\textwidth]{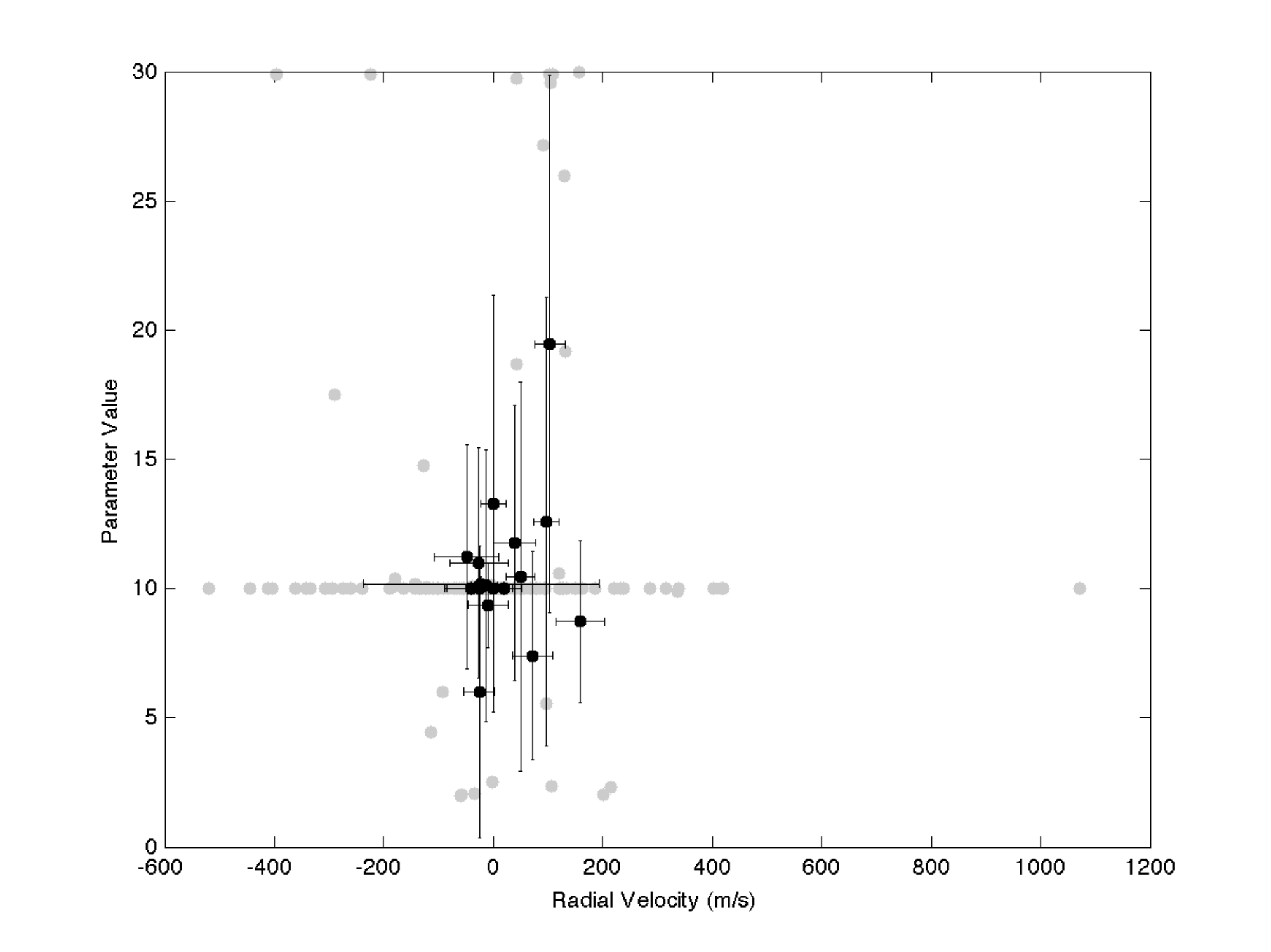}    
    \includegraphics[width=0.3\textwidth]{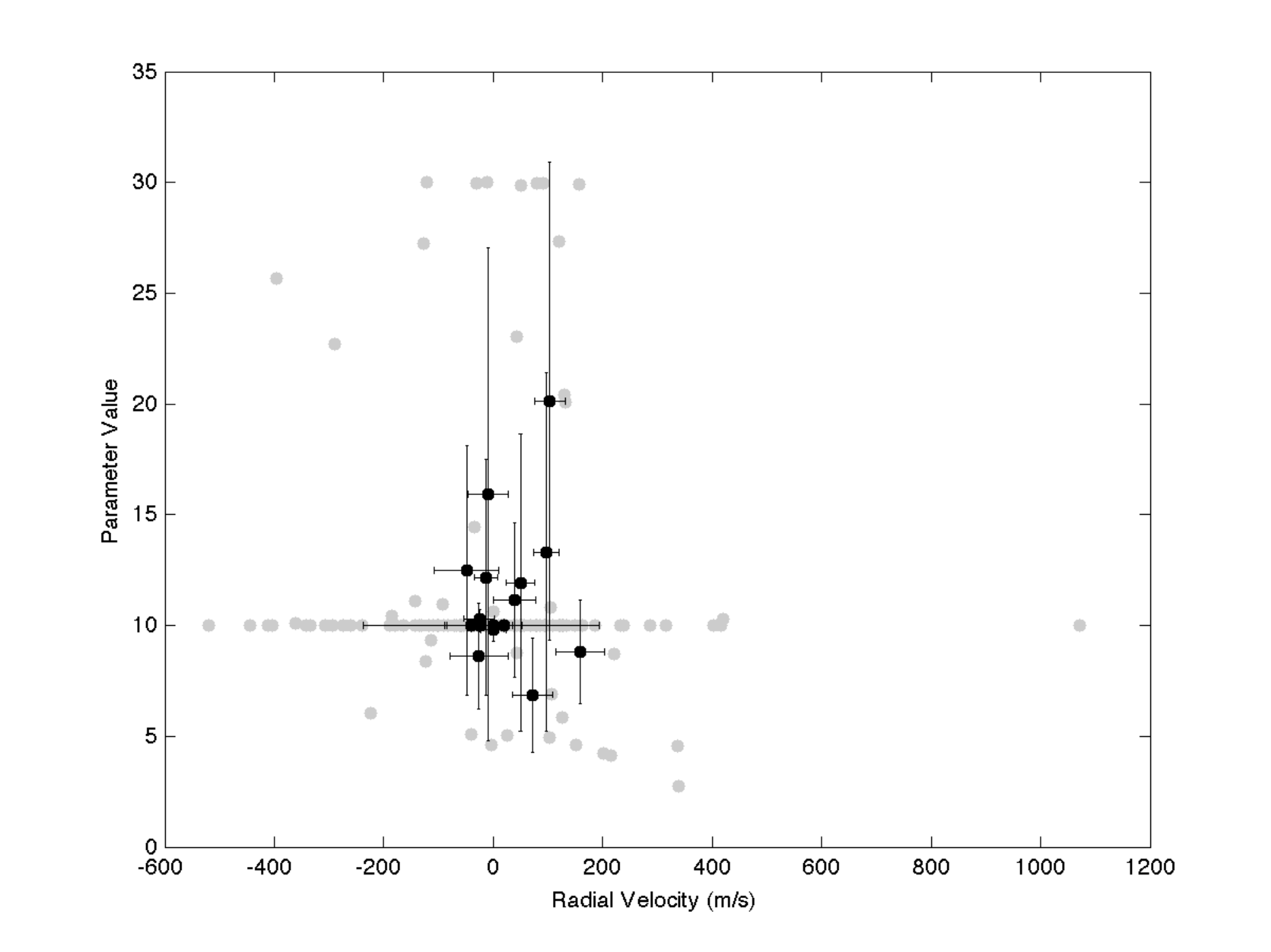}    
    \includegraphics[width=0.3\textwidth]{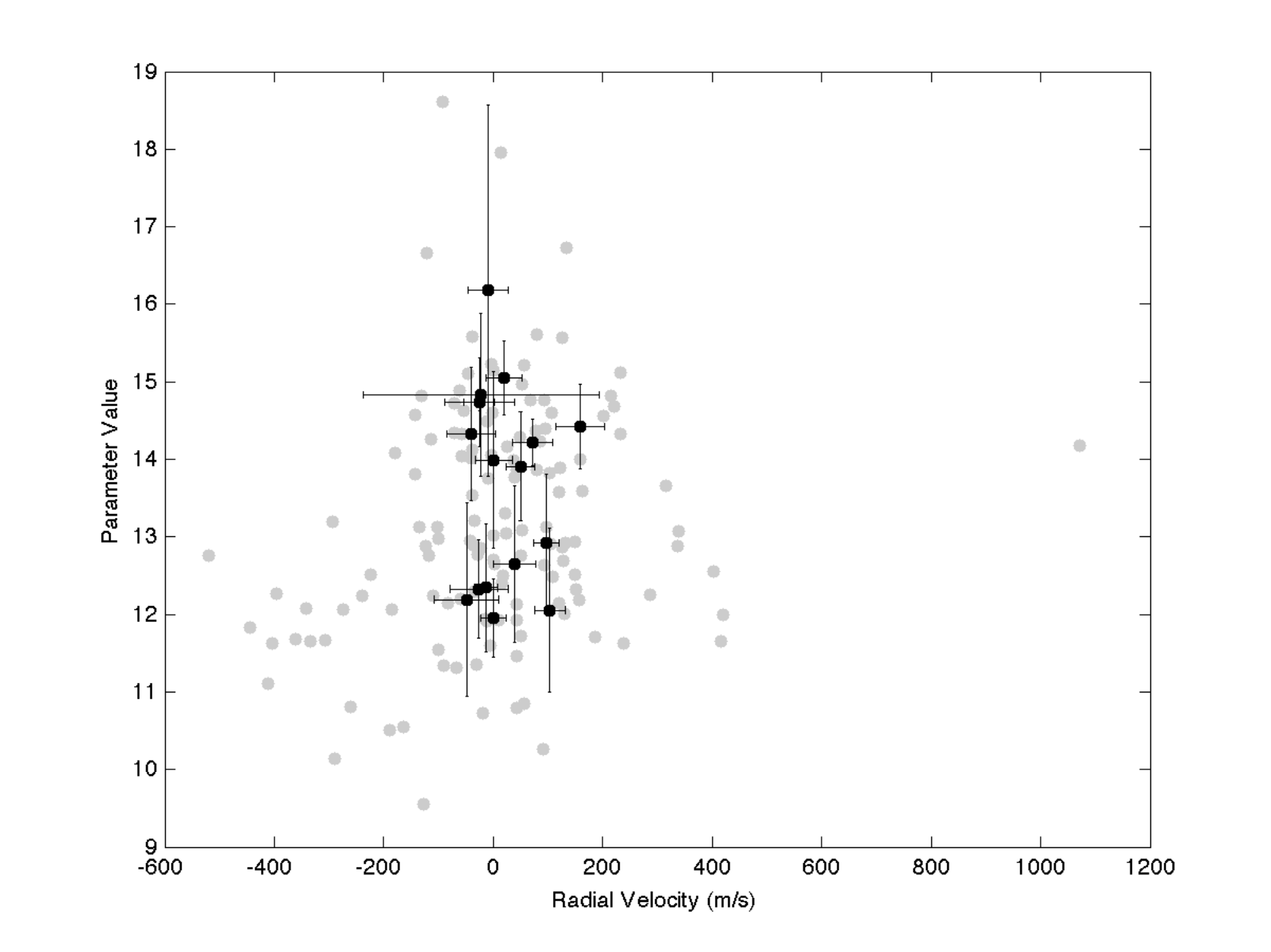}    
    \includegraphics[width=0.3\textwidth]{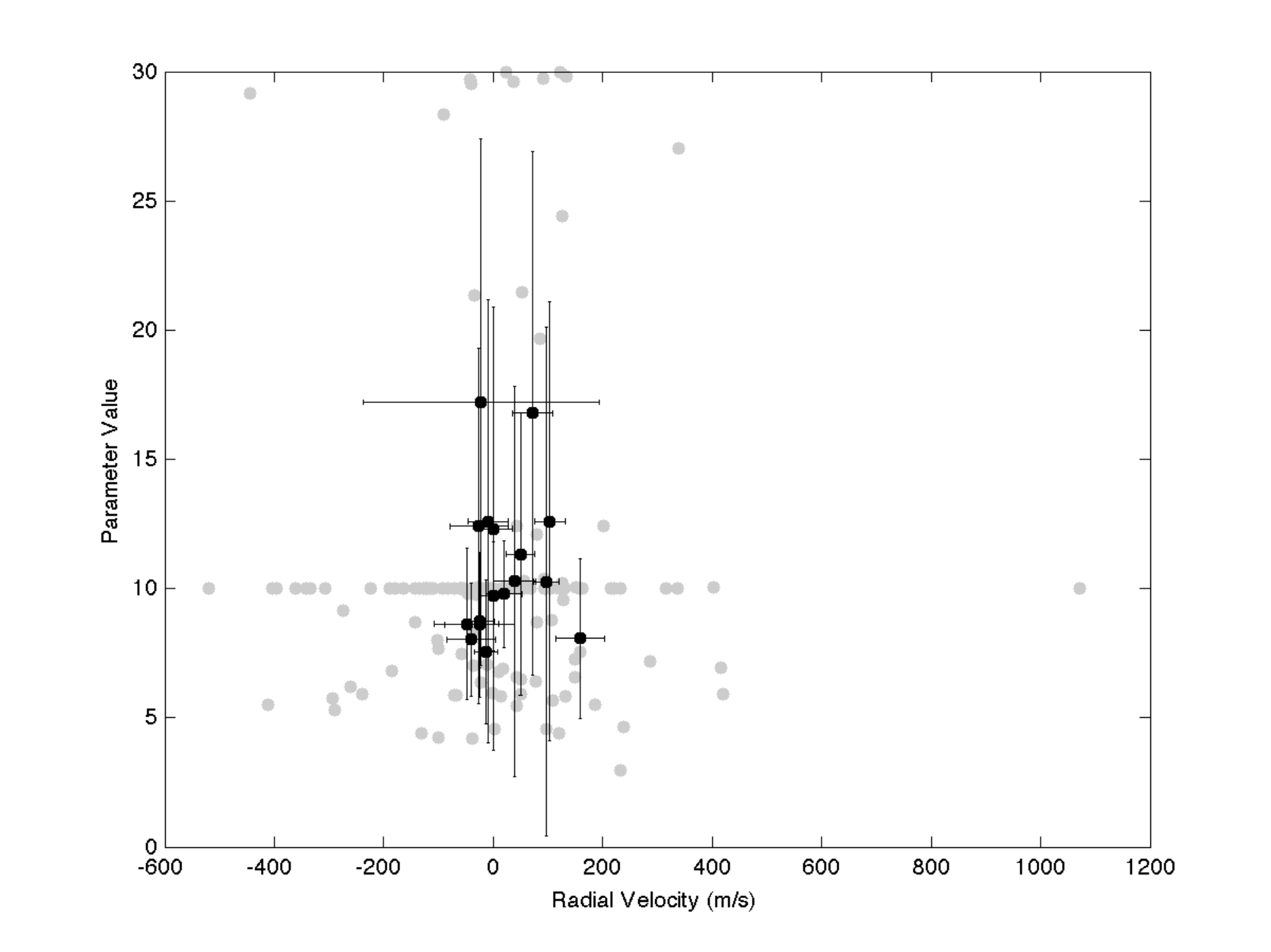}    
    \includegraphics[width=0.3\textwidth]{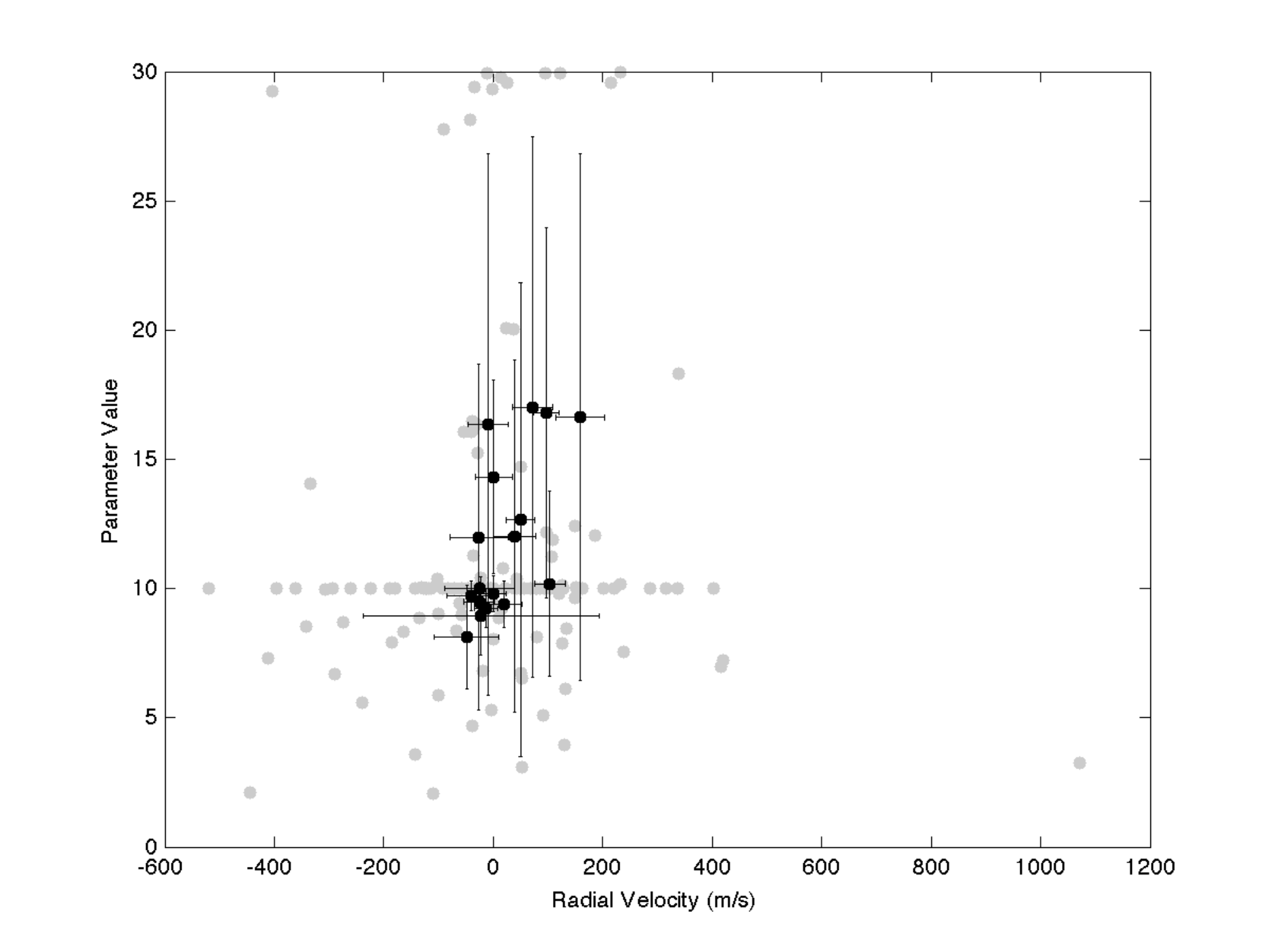}    
    \includegraphics[width=0.3\textwidth]{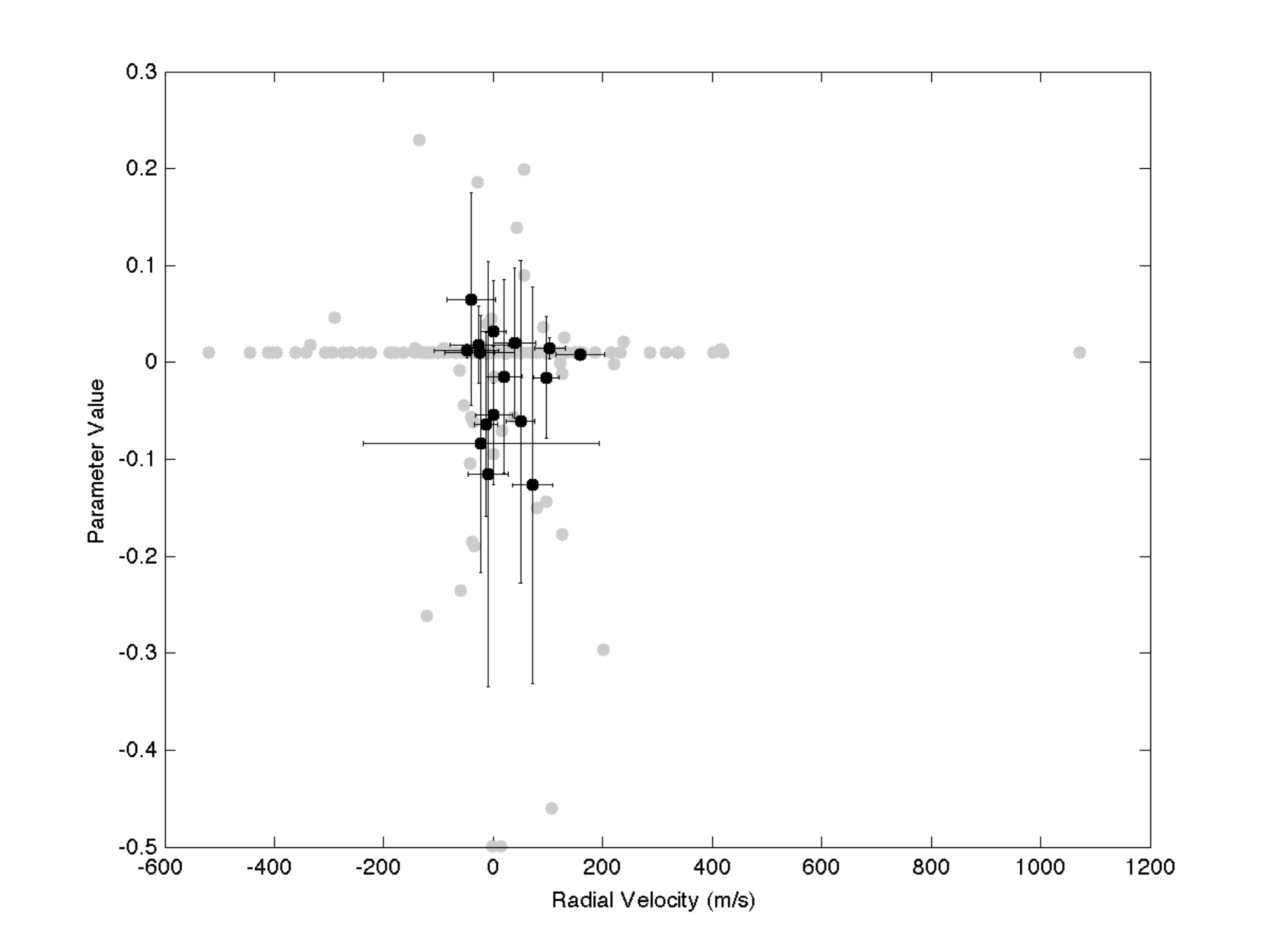}    
    \includegraphics[width=0.3\textwidth]{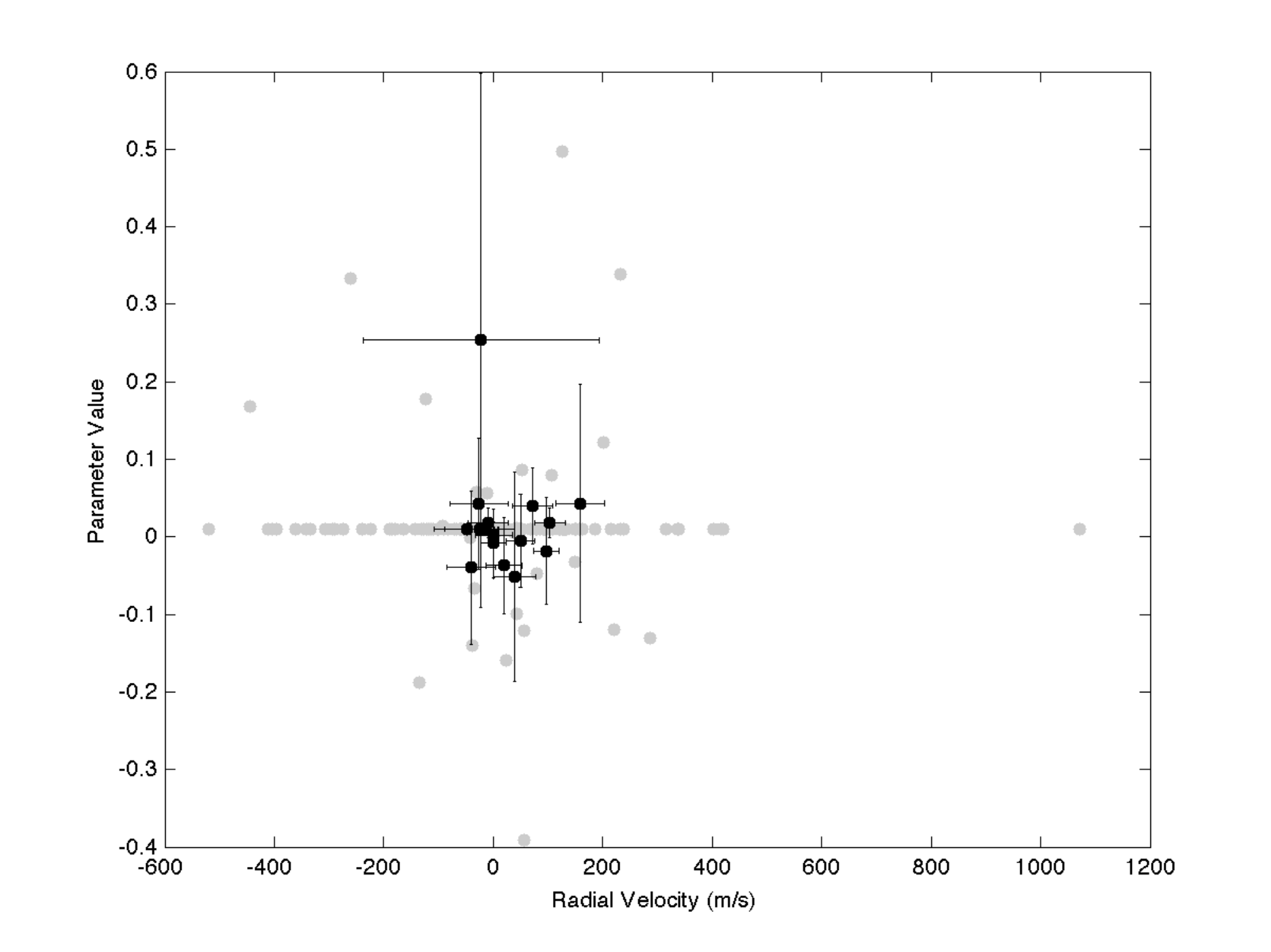}    
    \includegraphics[width=0.3\textwidth]{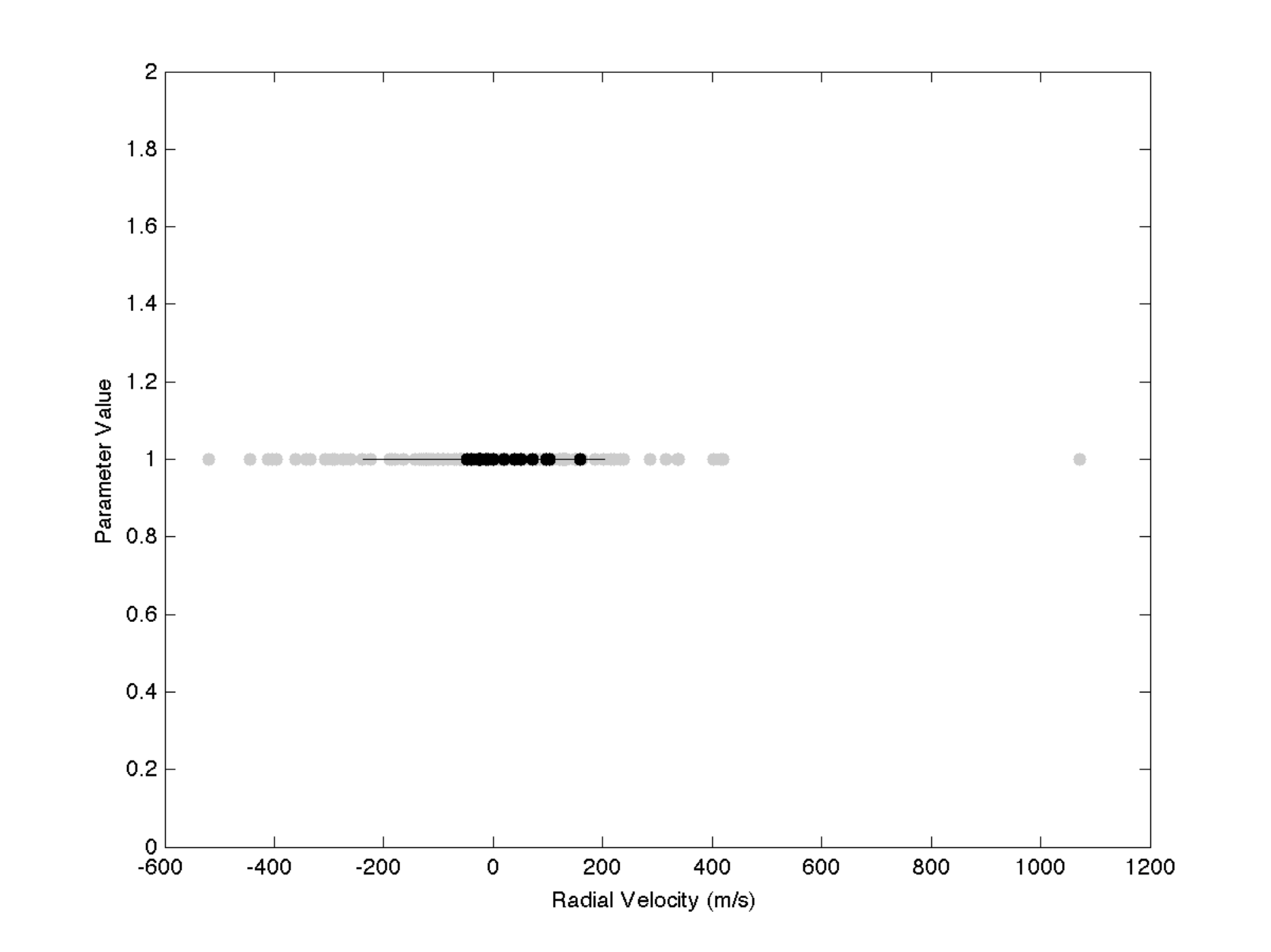}    
    \includegraphics[width=0.3\textwidth]{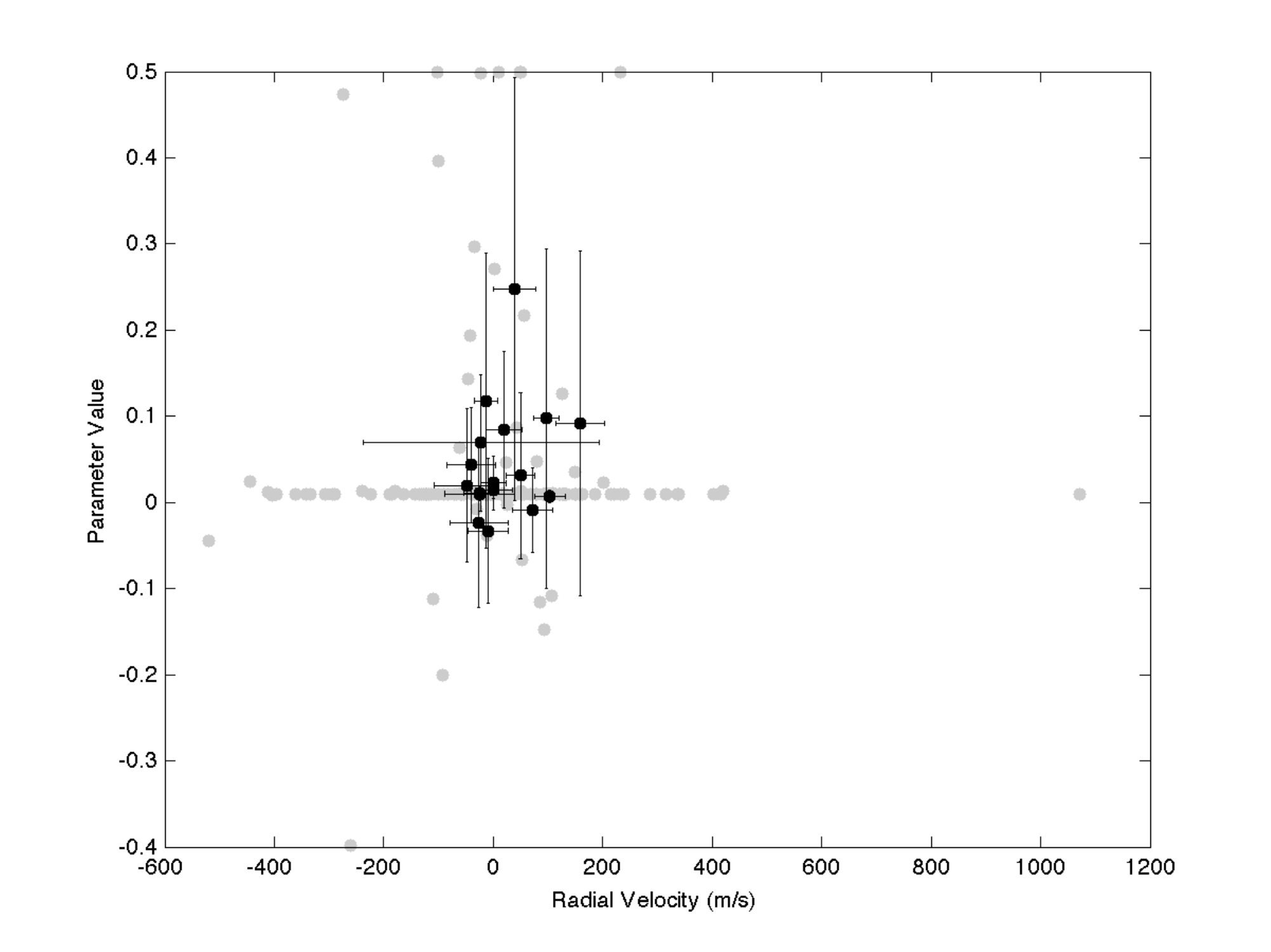}    
    \includegraphics[width=0.3\textwidth]{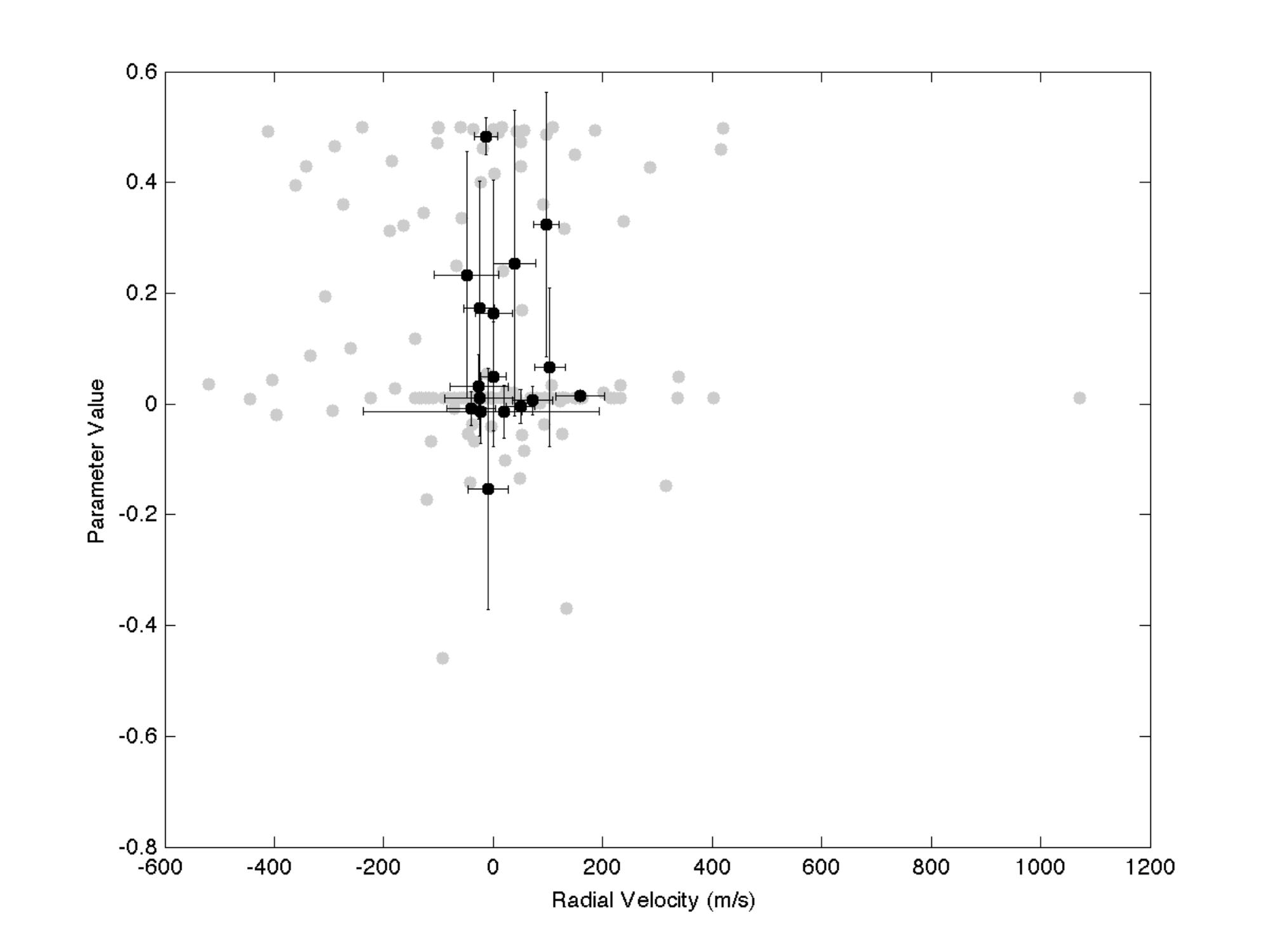}    
    \includegraphics[width=0.3\textwidth]{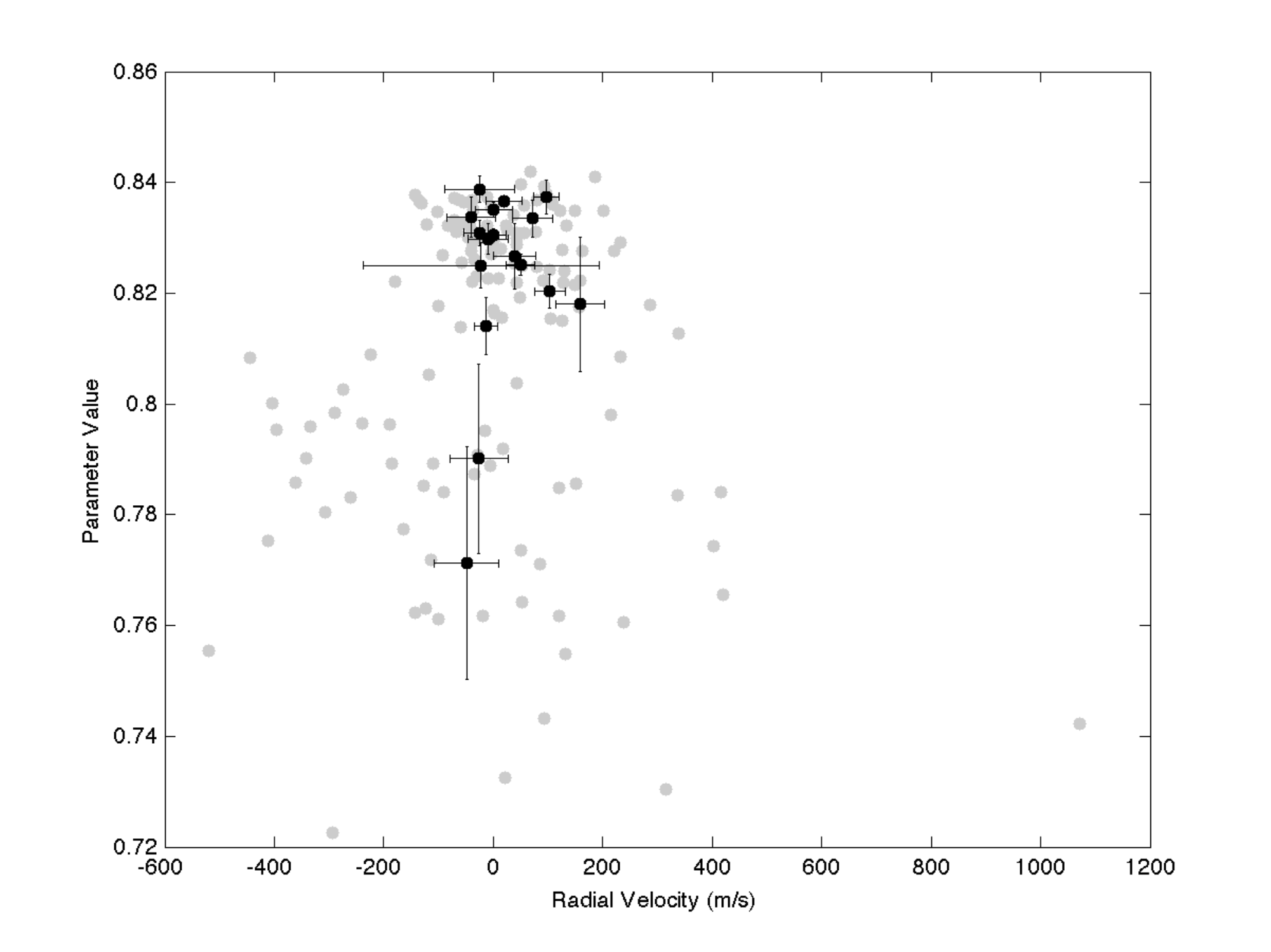}    
    \includegraphics[width=0.3\textwidth]{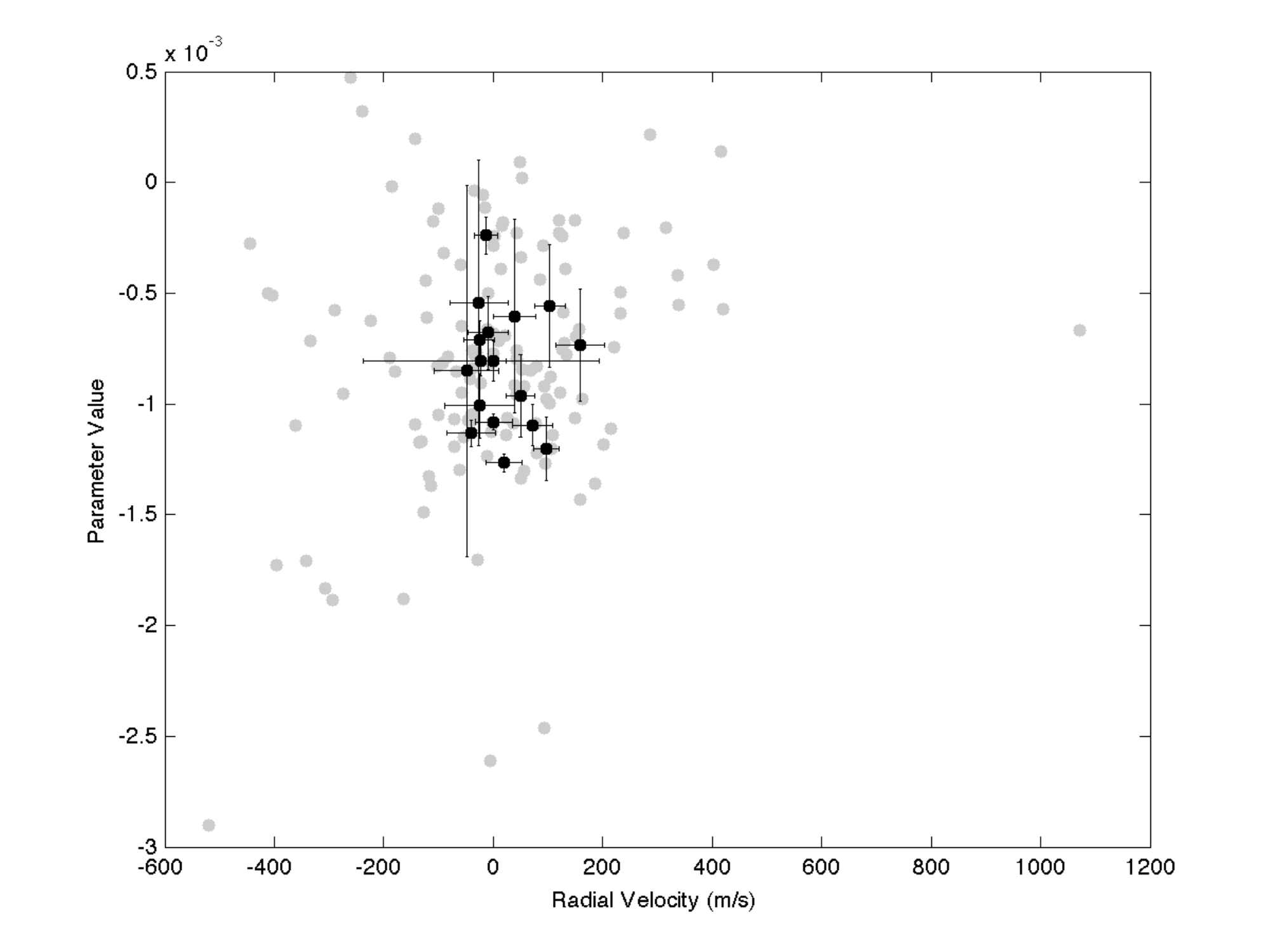}    
    \includegraphics[width=0.3\textwidth]{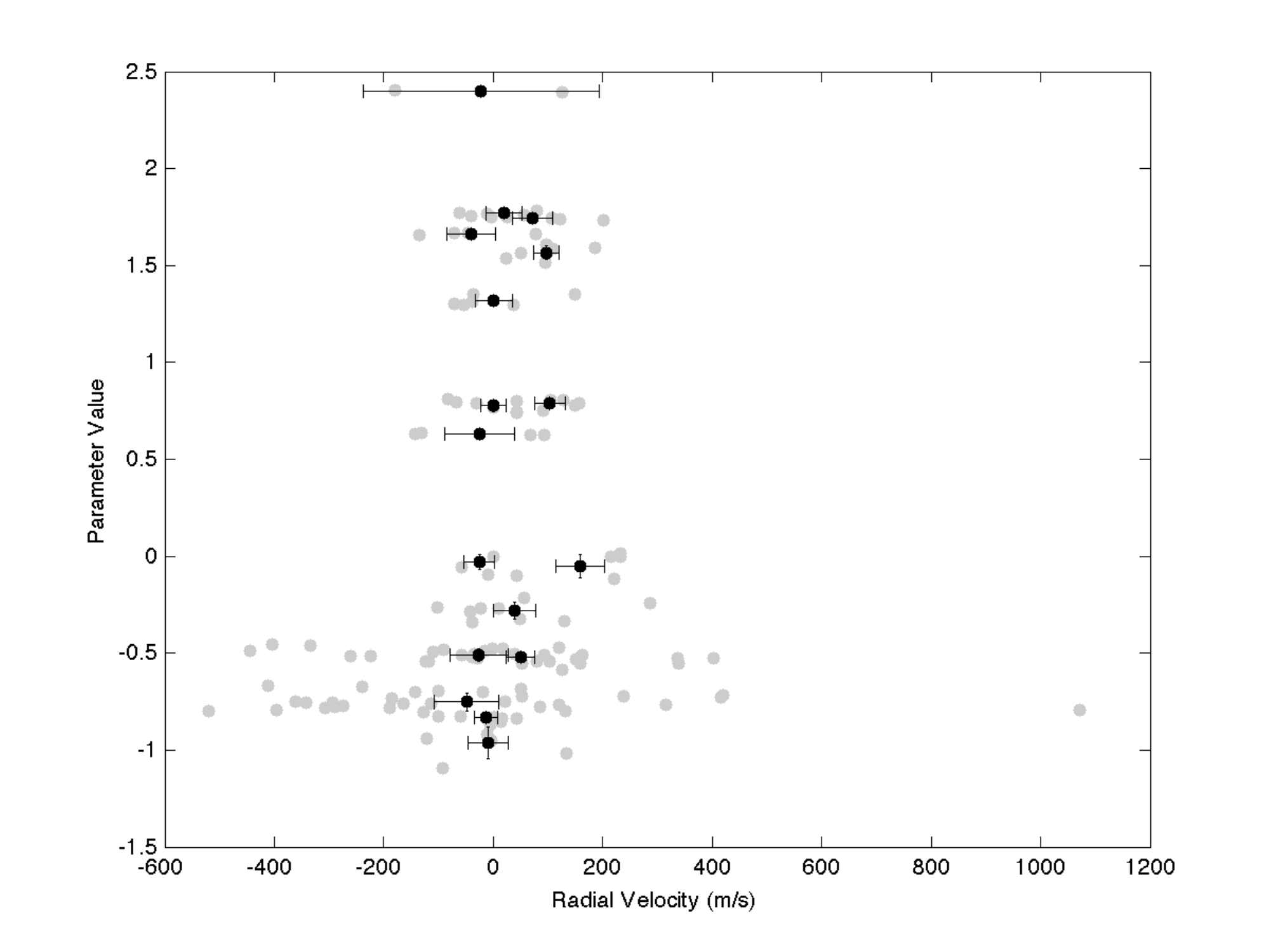}    
    \includegraphics[width=0.3\textwidth]{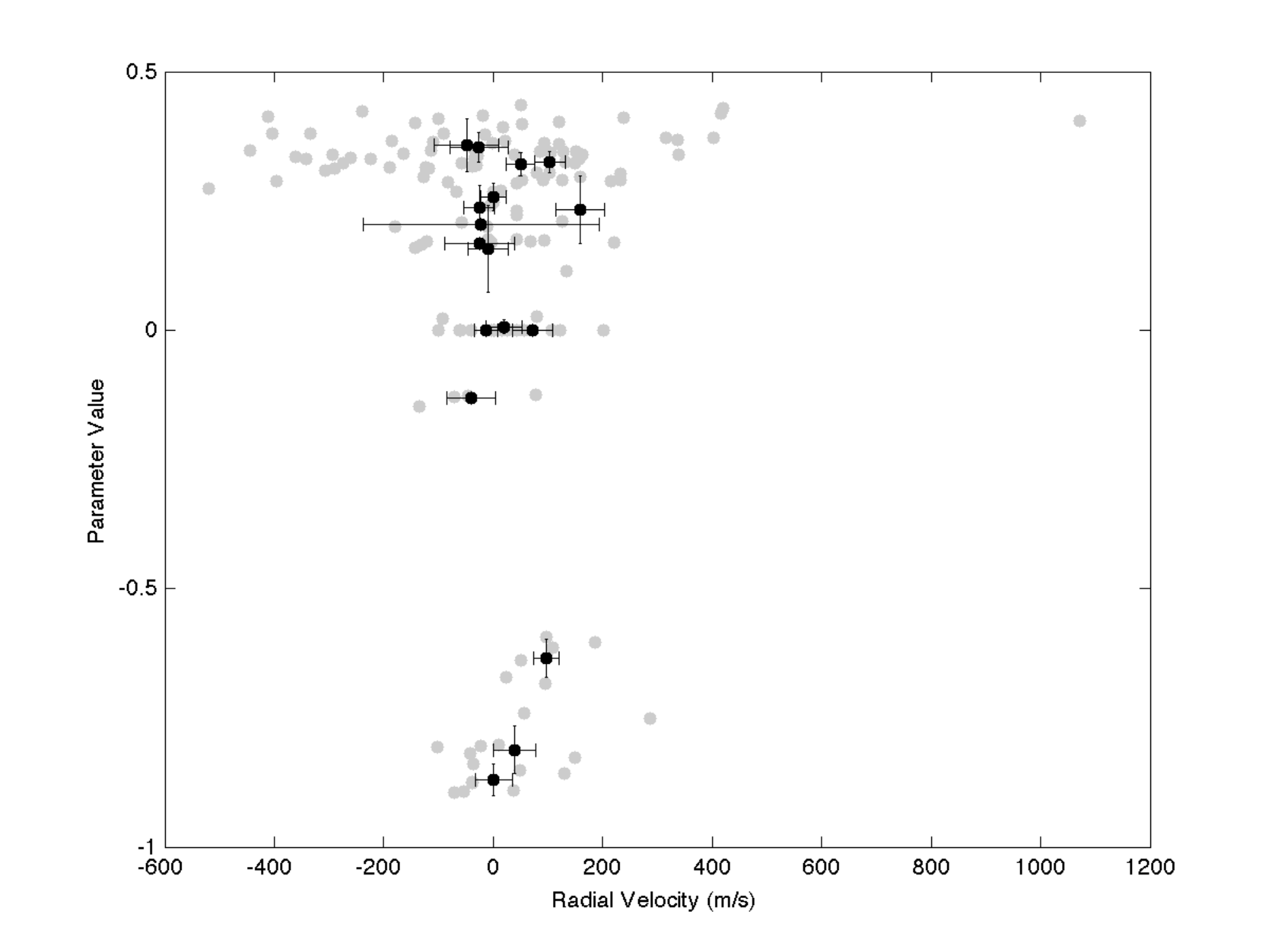}    
    \includegraphics[width=0.3\textwidth]{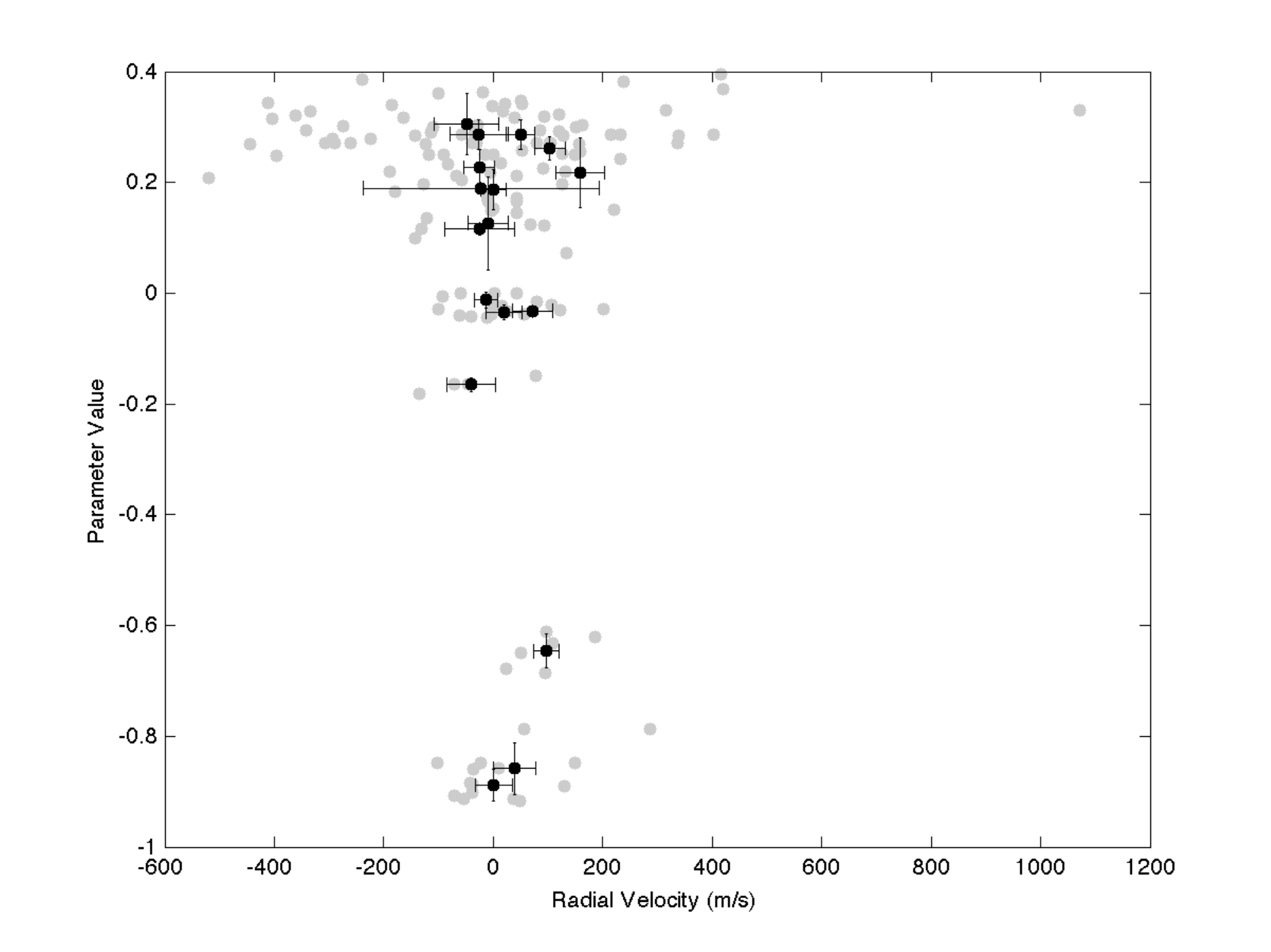}    
      \end{center}     
 \caption{The dependence of the radial velocity measurements for GJ 15A on parameters 1-15 as listed in Table 1, and run in the same order from top left to top right to the second row left to the second row right, etc.  The gray points correspond to the individual coadd observation radial velocities (fit independently), and the black points correspond to the means of each night.  The error bars correspond to the standard deviation of the gray points within a given night.  \label{fig:f13}}
\end{figure}

\begin{figure}[tb]
  \begin{center}
    \includegraphics[width=0.3\textwidth]{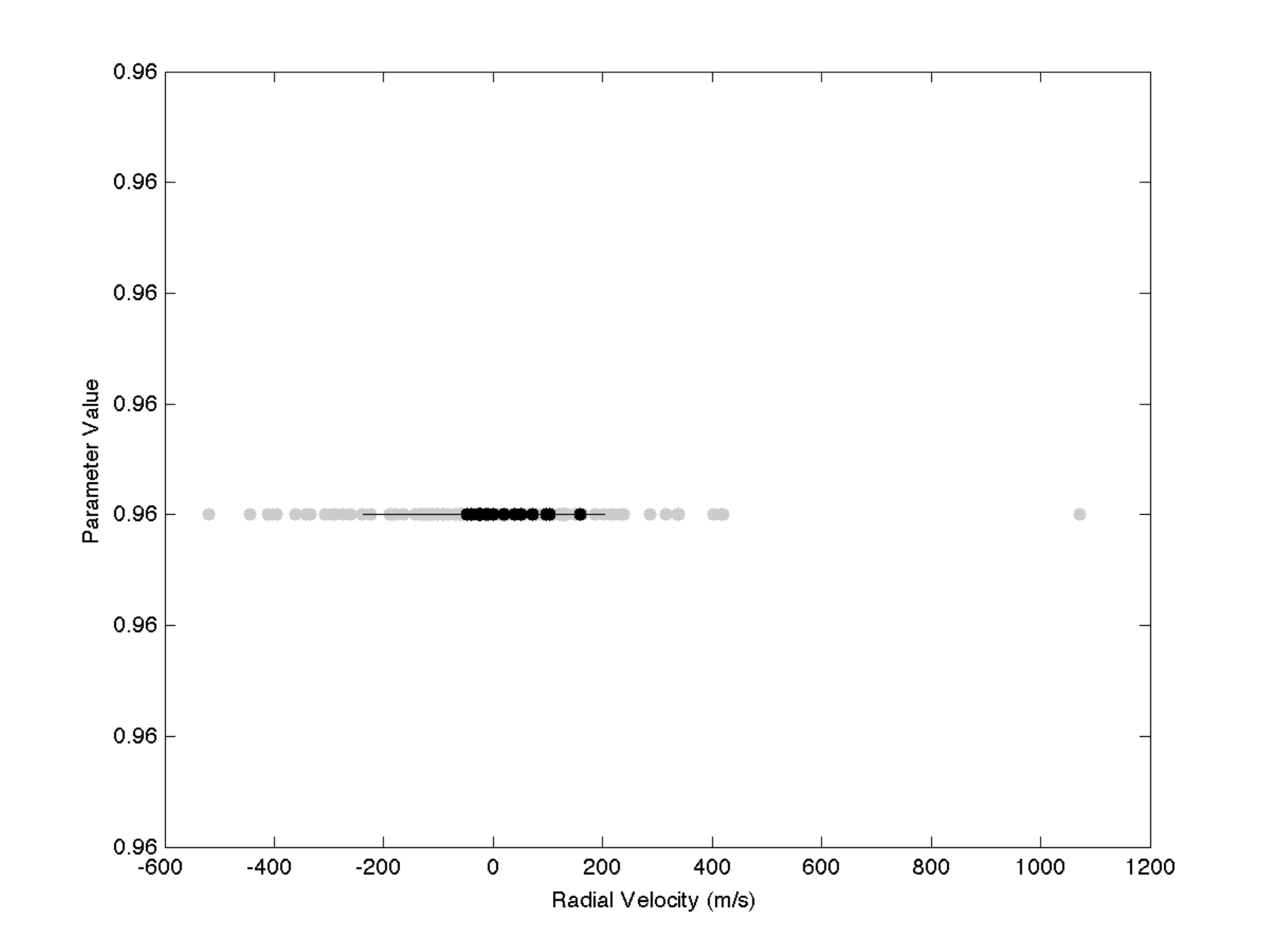}    
    \includegraphics[width=0.3\textwidth]{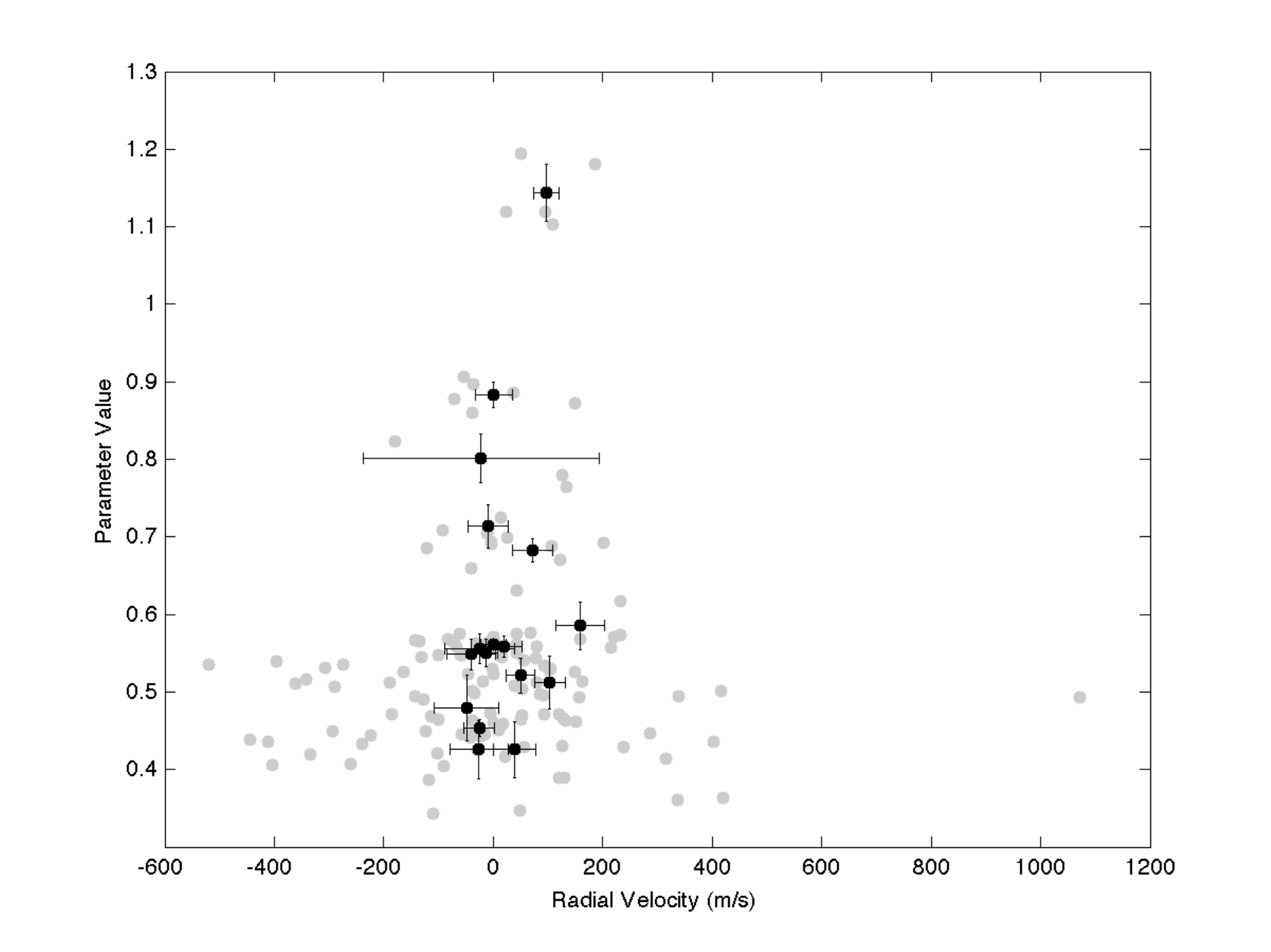}    
    \includegraphics[width=0.3\textwidth]{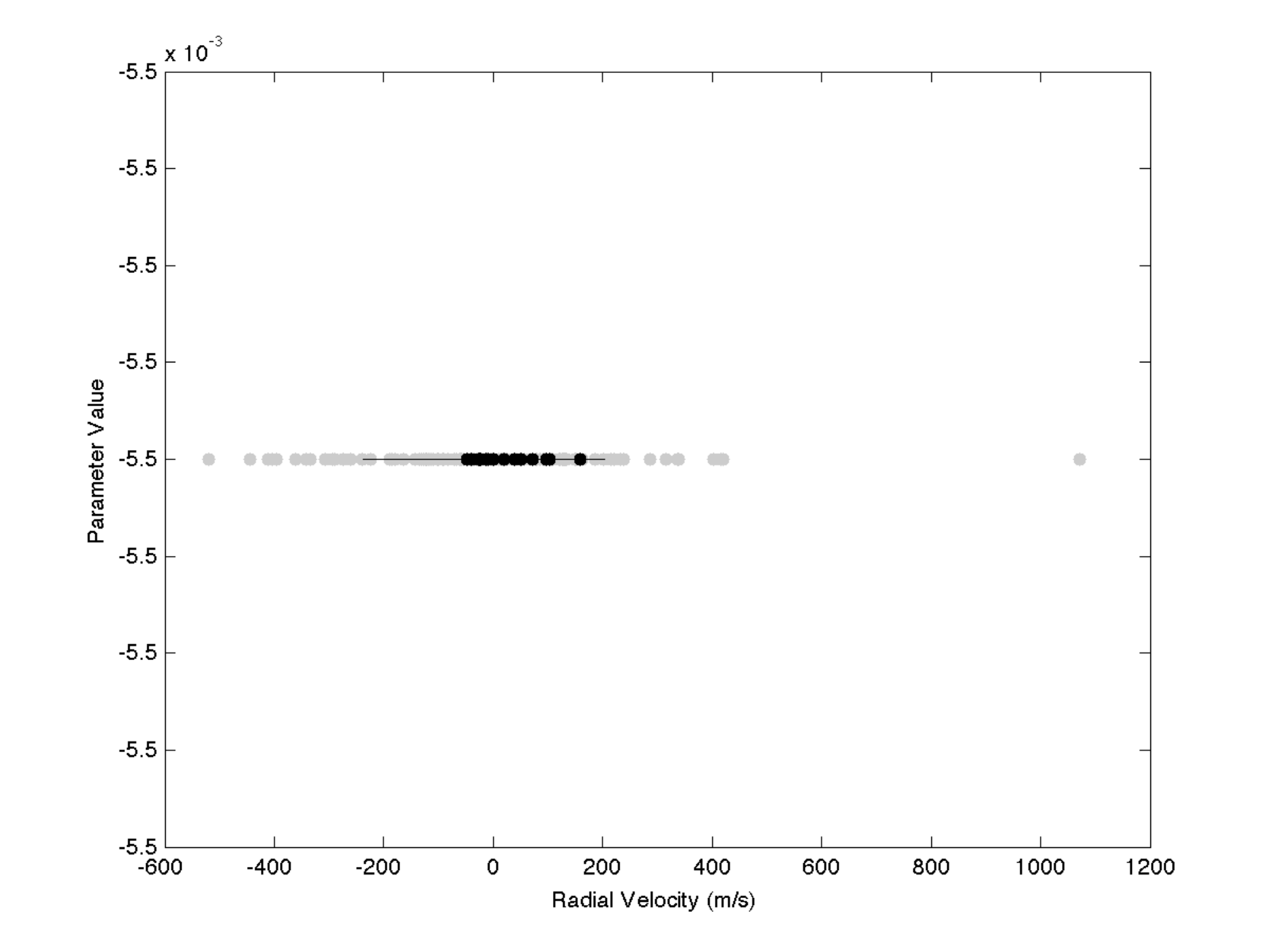}    
    \includegraphics[width=0.3\textwidth]{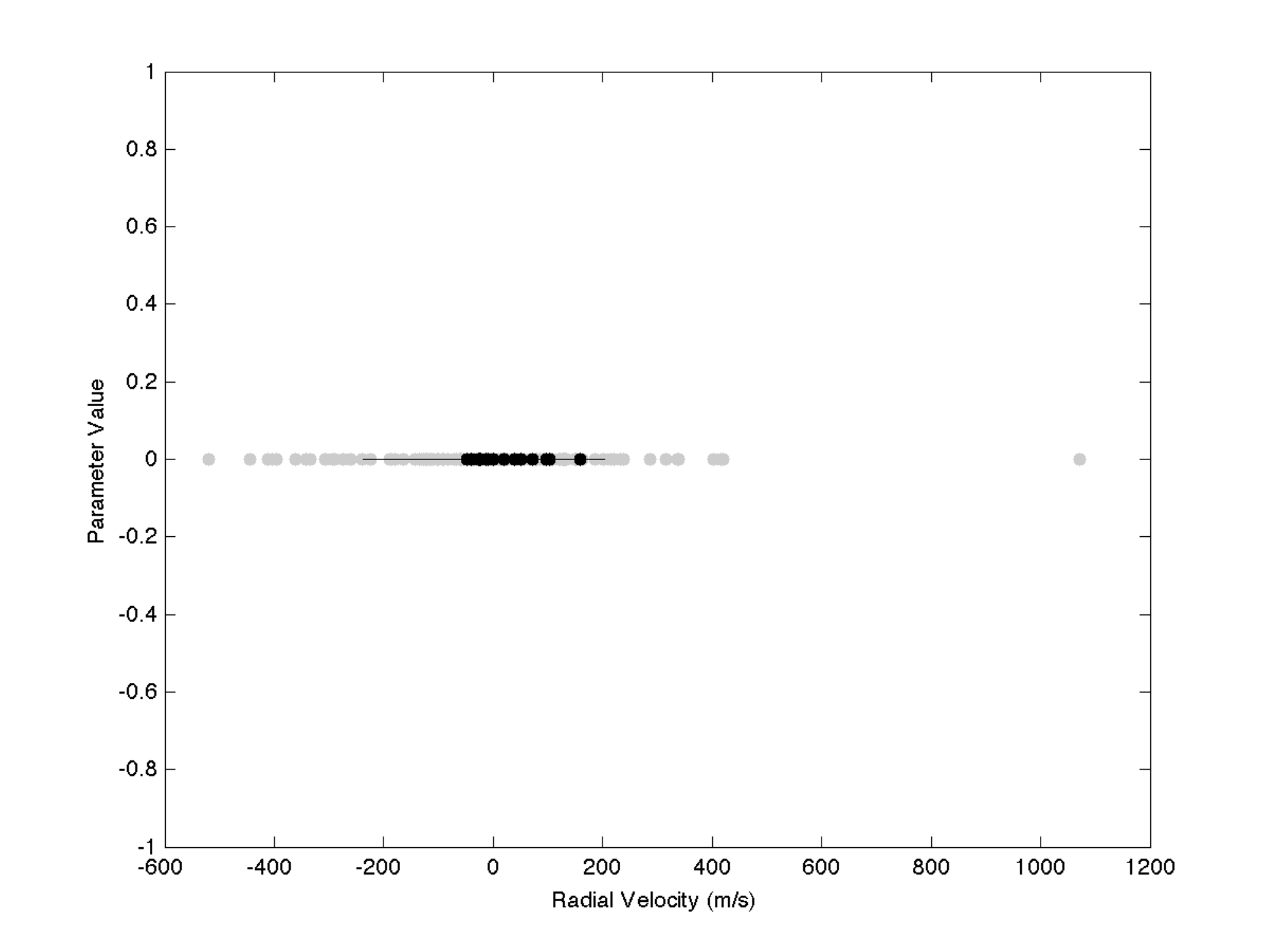}    
    \includegraphics[width=0.3\textwidth]{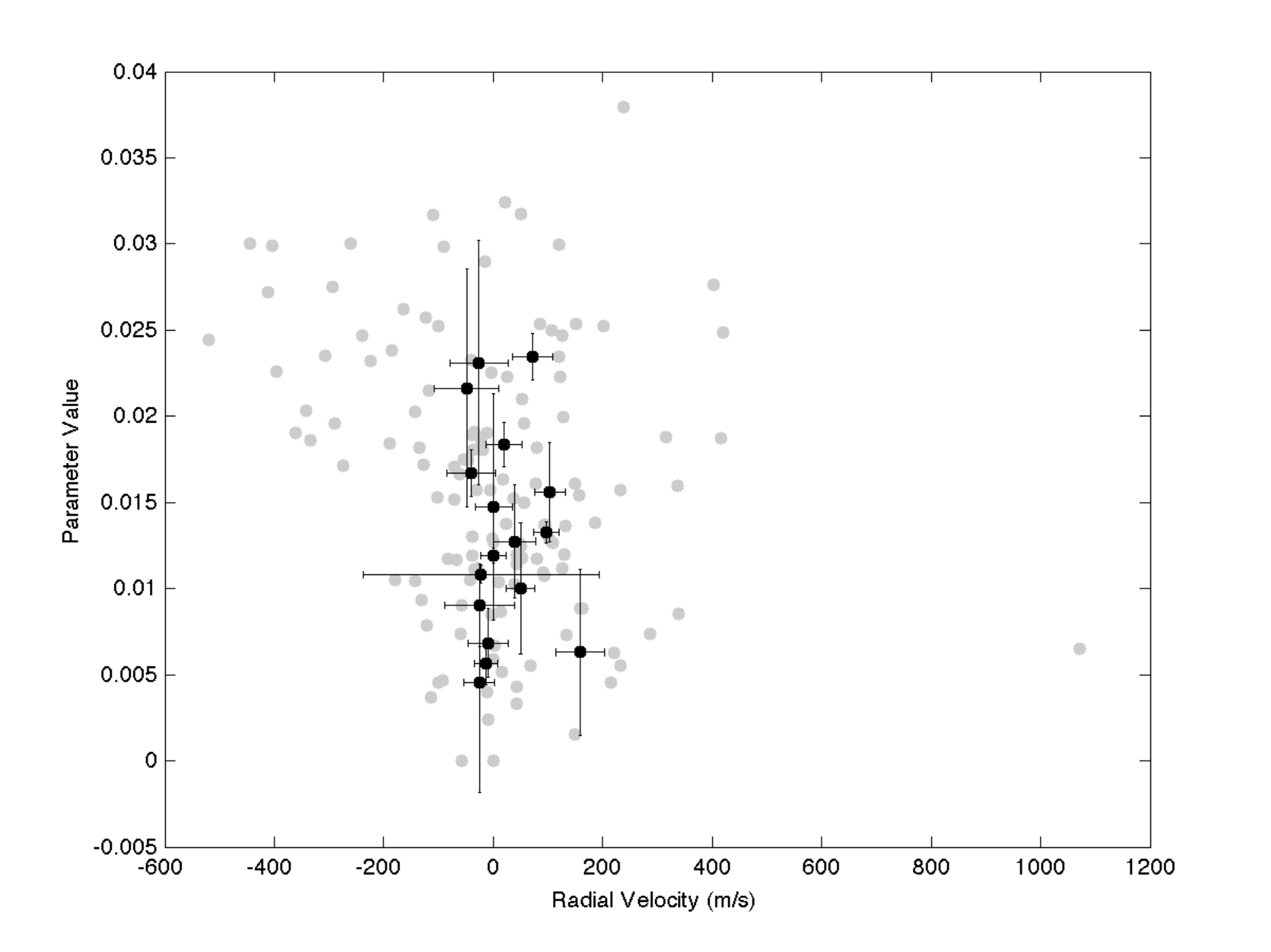}    
    \includegraphics[width=0.3\textwidth]{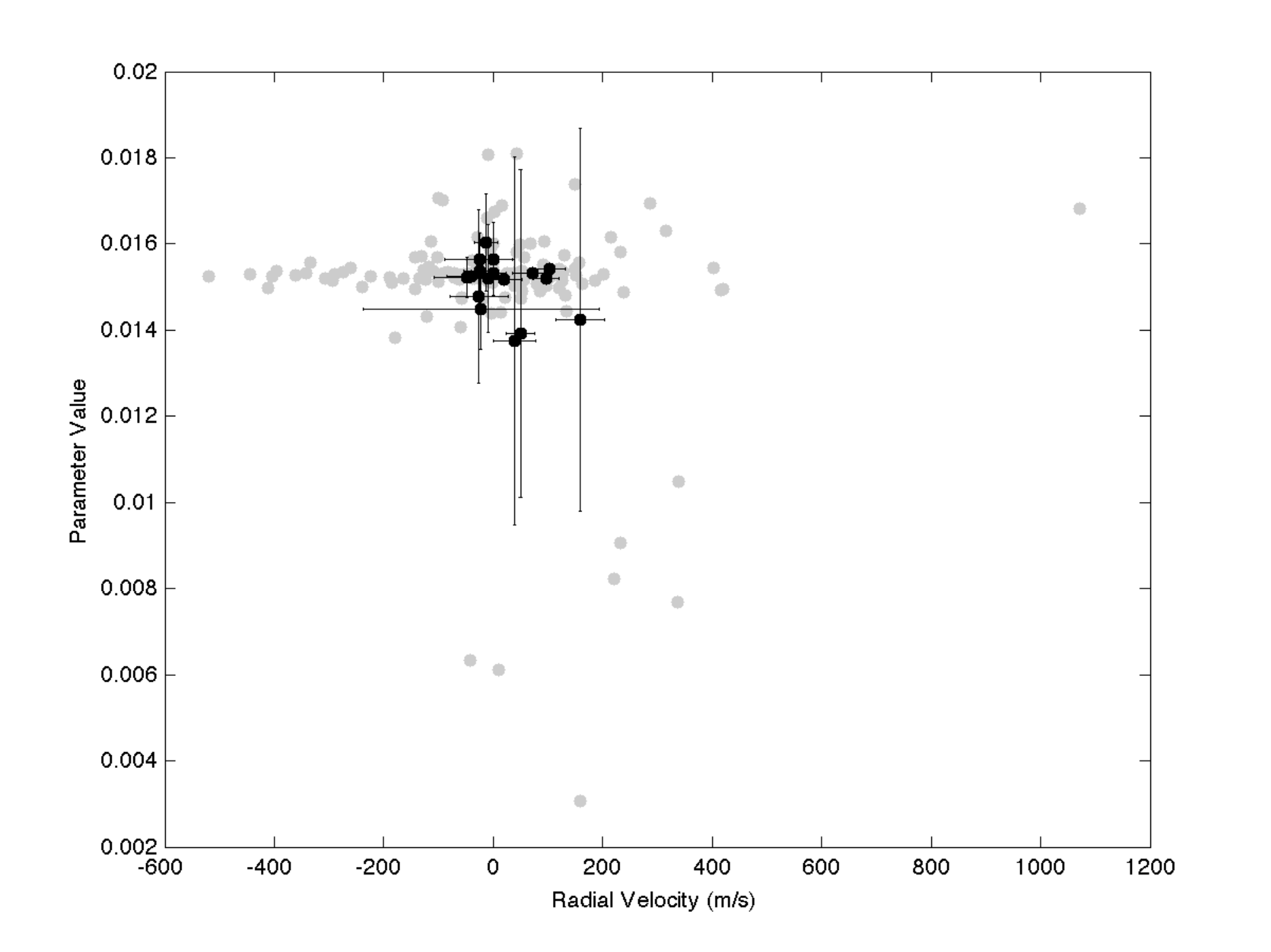}    
    \includegraphics[width=0.3\textwidth]{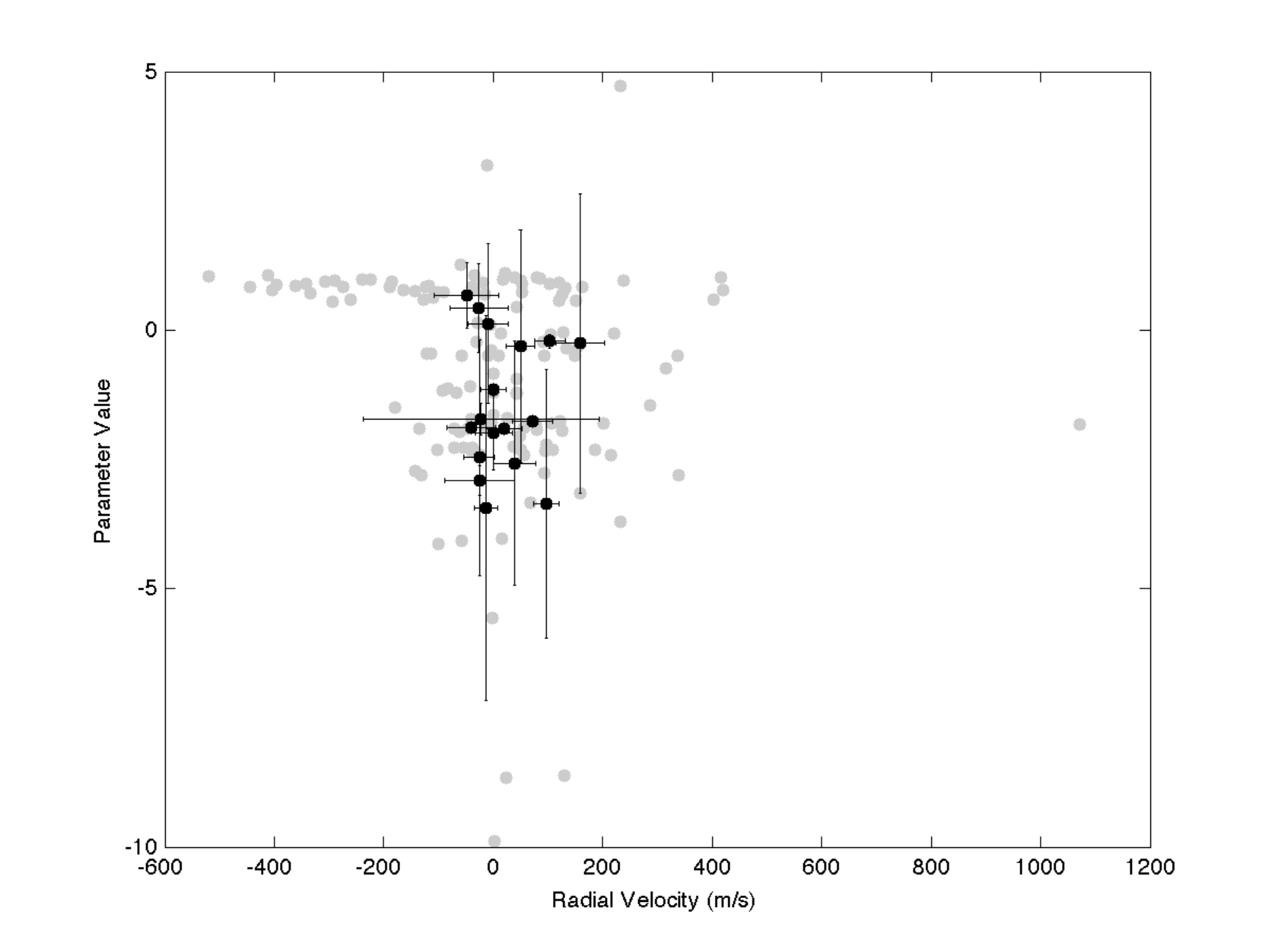}    
    \includegraphics[width=0.3\textwidth]{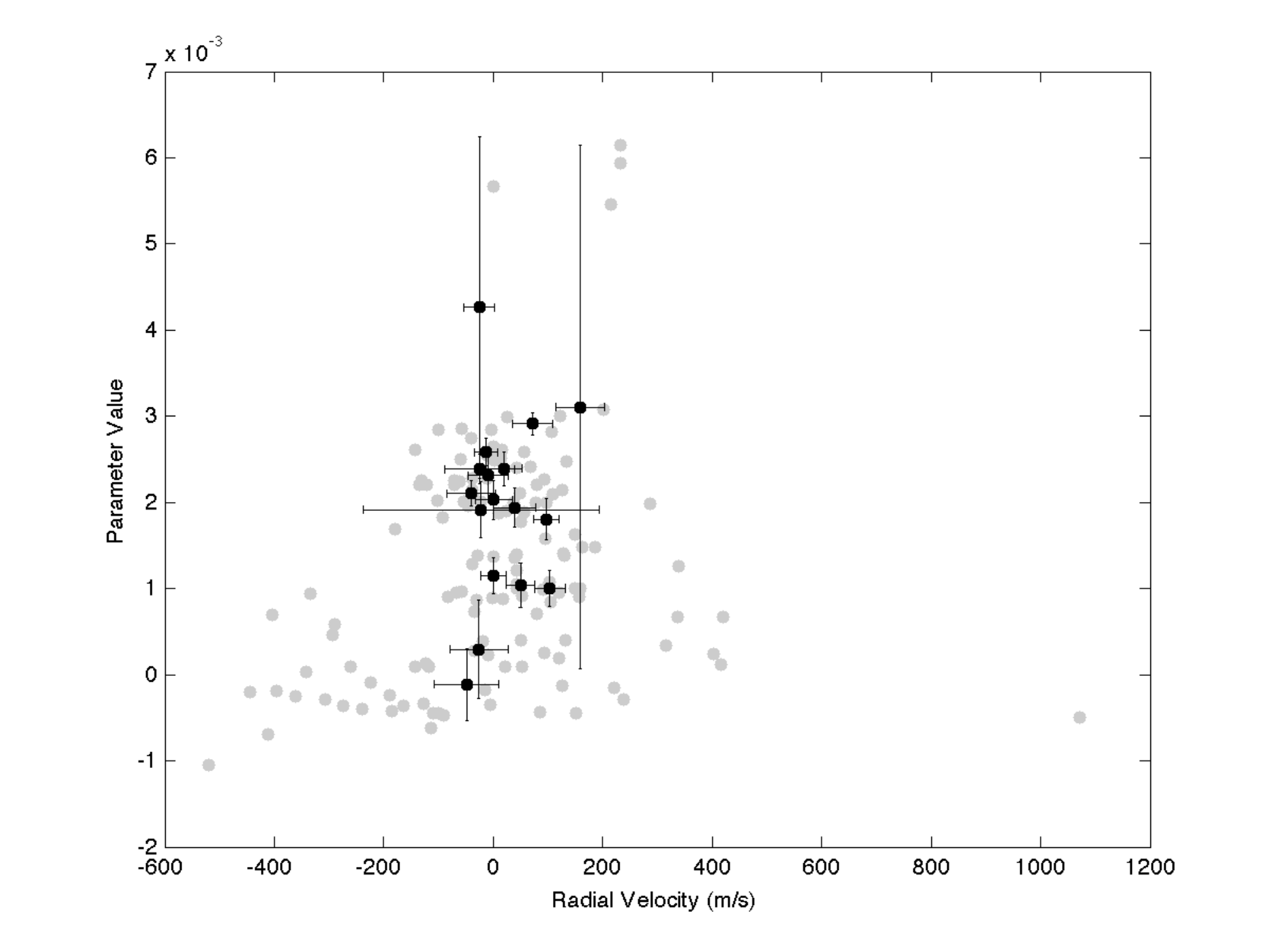}    
    \includegraphics[width=0.3\textwidth]{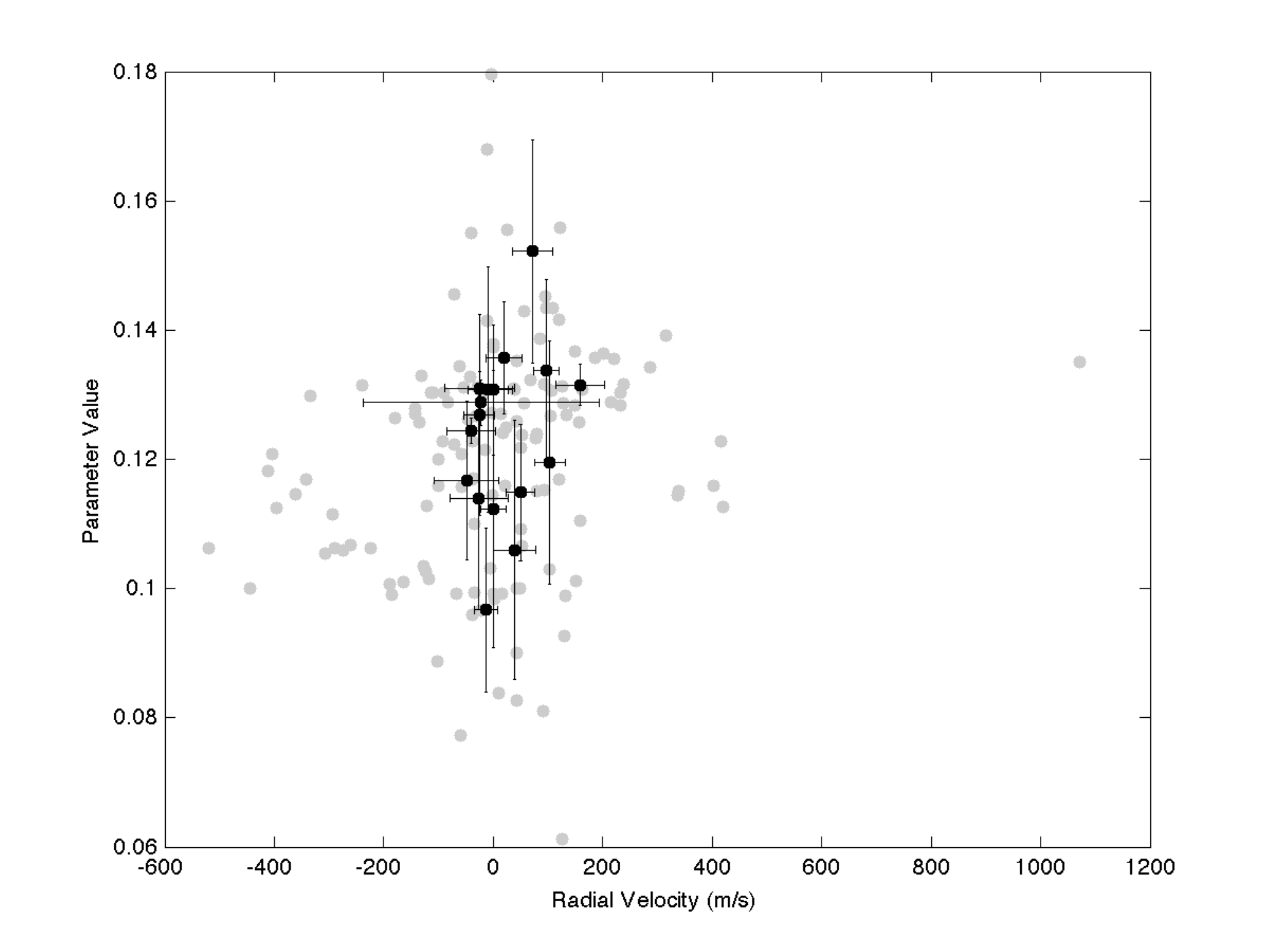}    
    \includegraphics[width=0.3\textwidth]{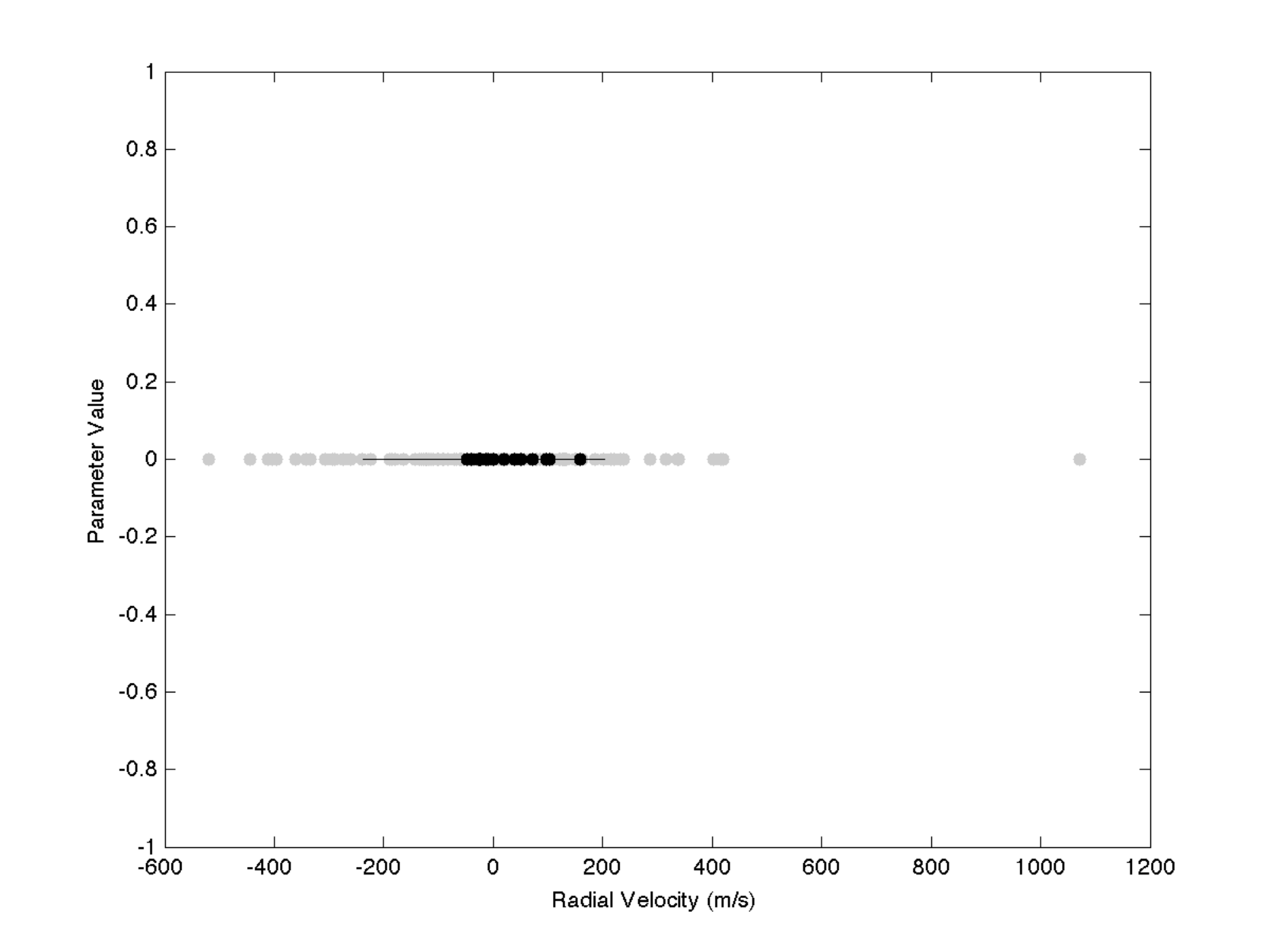}    
    \includegraphics[width=0.3\textwidth]{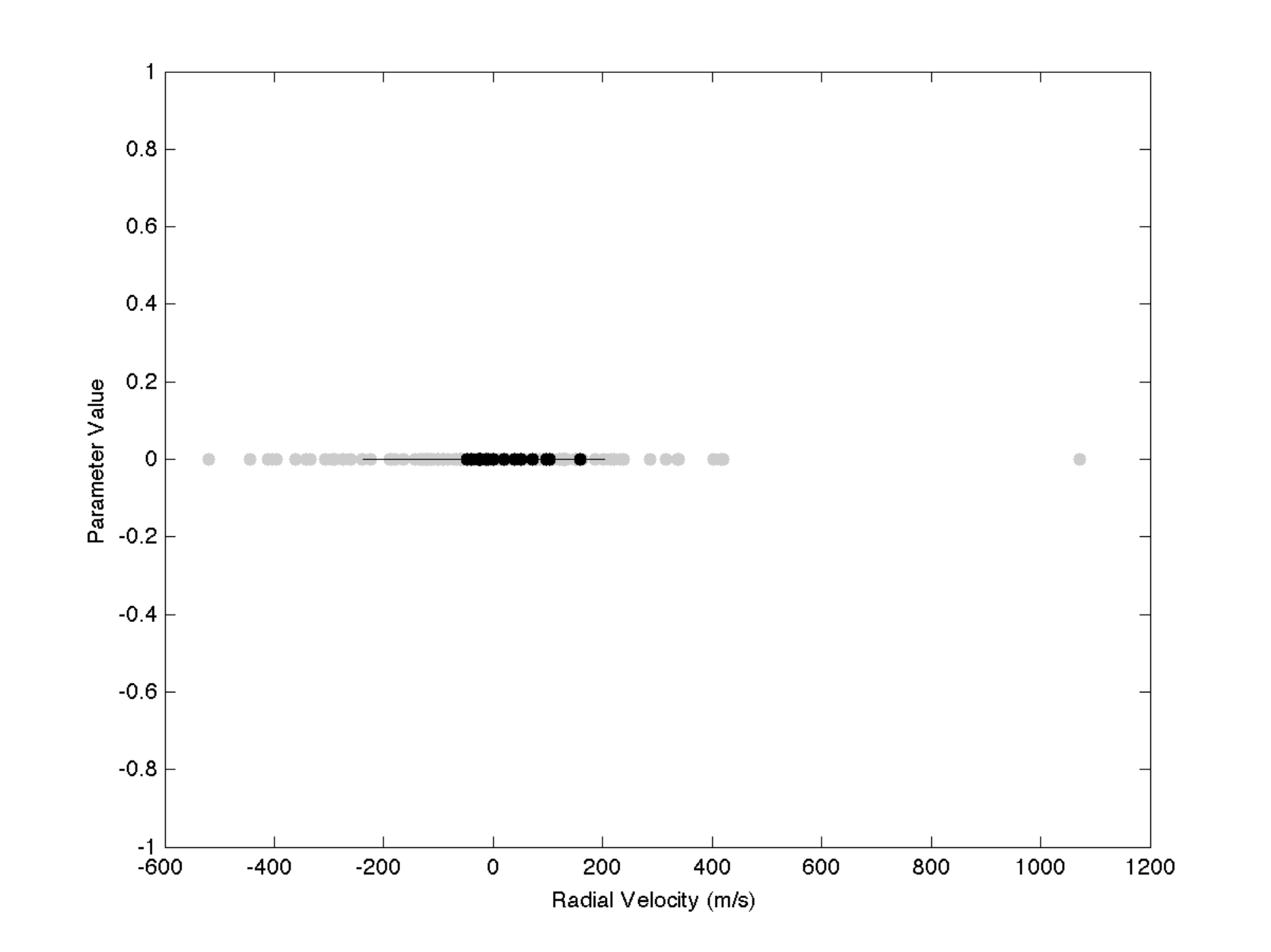}    
    \includegraphics[width=0.3\textwidth]{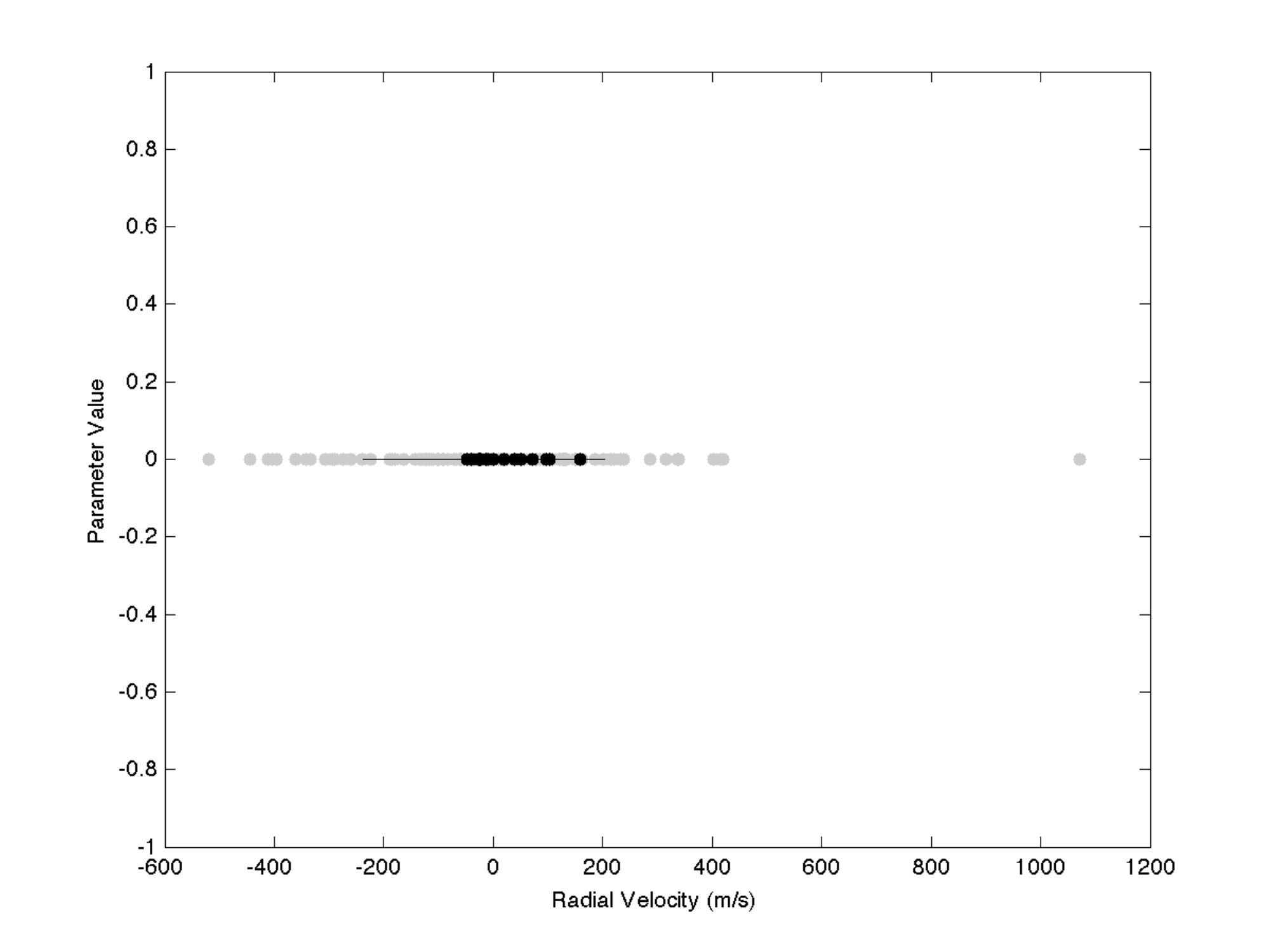}    
    \includegraphics[width=0.3\textwidth]{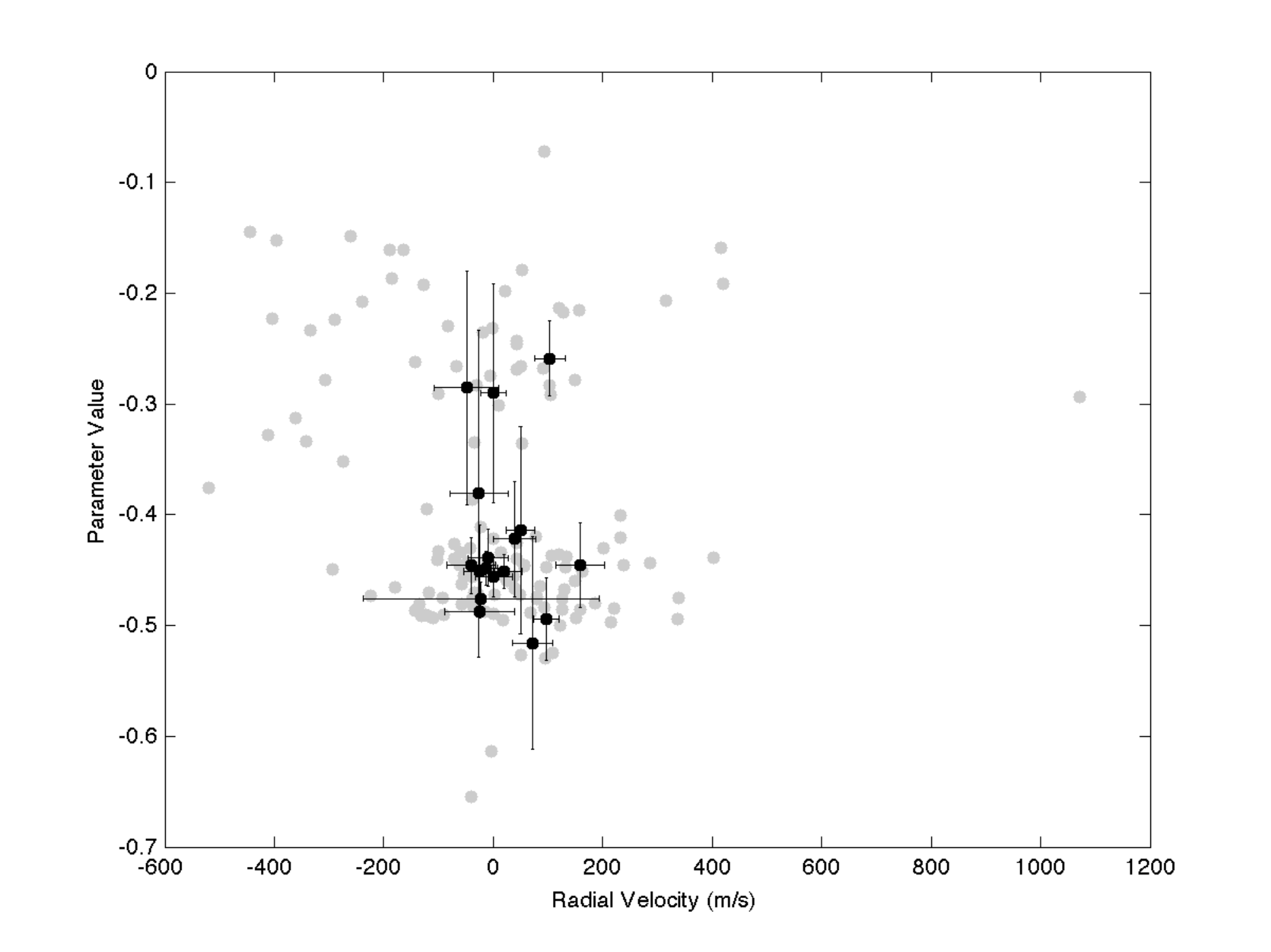}    
    \includegraphics[width=0.3\textwidth]{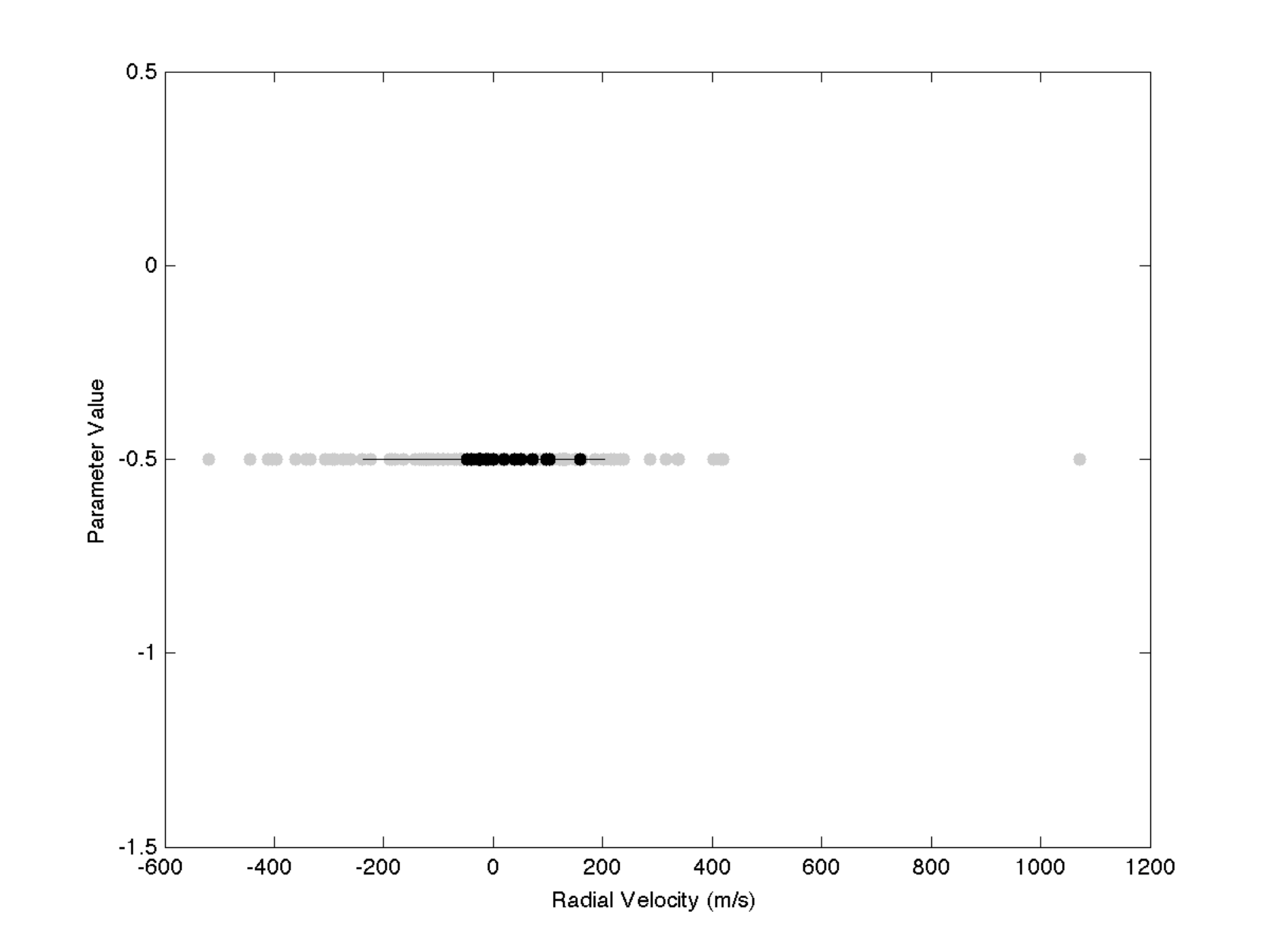}    
      \end{center}      
 \caption{The dependence of the radial velocity measurements for GJ 15A on parameters 16-29 as listed in Table 1, and run in the same order from top left to top right to the second row left to the second row right, etc. Correspondence of the radial velocities with parameters such as the telescope hour angle, target airmass and barycenter correction are not shown, but do not exhibit any correlations.   See Figure 13.   \label{fig:f13}}
\end{figure}

\subsection{Other case studies -- GJ 876 and AU Mic}

The radial velocity time-series for AU Mic is shown in Figure 12.  We obtain a long-term precision of 126 m/s, equal to the 125 m/s precision obtained for AU Mic with NIRSPEC on Keck with telluric line wavelength calibration, spanning over thirty times the spectral grasp and 10 times the light gathering power compared to our observations with CSHELL and IRTF.  Our $\chi^2$=2.87, and our median single-night uncertainty is 61 m/s.  In Bailey et al. 2012, AU Mic is suggested to be a radial velocity variable source due to its youth and activity.  With our performance for GJ 15A, we support the conclusion that AU Mic is a radial velocity variable, but more work is needed on the rest of our sample to confirm this status.

In Figure 15 we present the radial velocities for GJ 876.  We confirm GJ 876 as a radial velocity variable consistent with the planetary system detected in the visible with HIRES and the Hamspec spectrograph at Lick Observatory, demonstrating the ability of our instrumentation and analysis to detect astrophysical radial velocity variations \cite{laughlin05}.  We are the first alternate wavelength confirmation of the detection of the two most massive planets in this system.

\begin{figure}[tb]
  \begin{center}
    \includegraphics[width=0.8\textwidth,clip=true,trim=0cm 0cm 0cm 0cm]{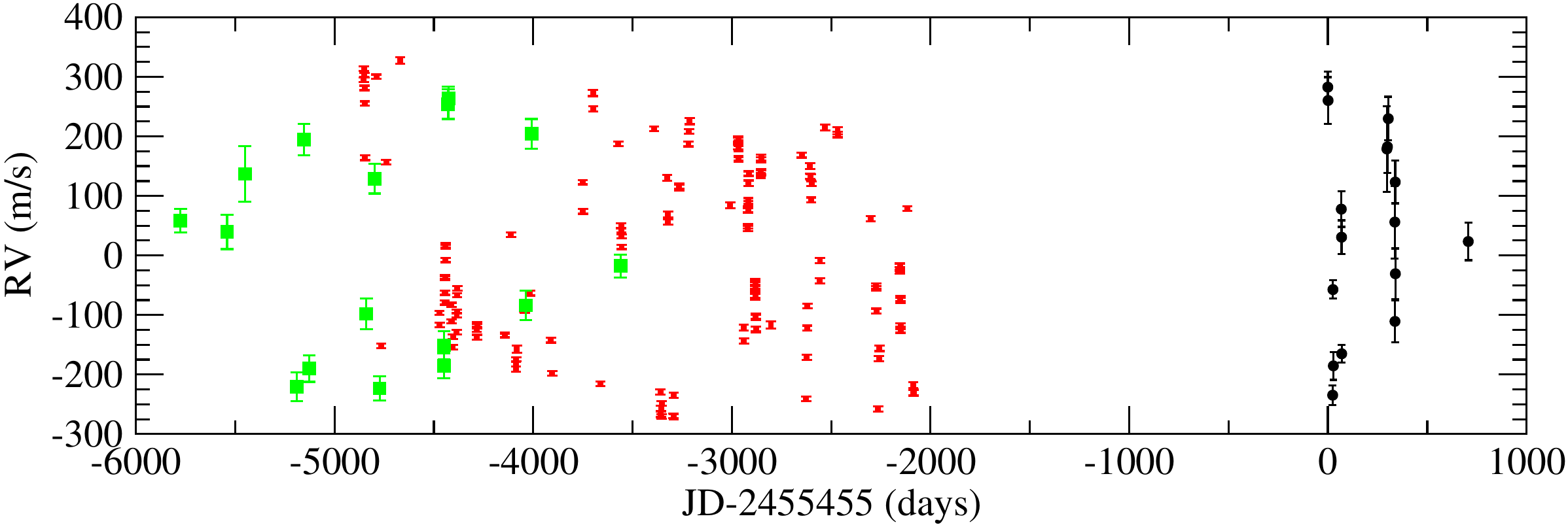}    
    \includegraphics[width=0.8\textwidth,clip=true,trim=0cm 0cm 0cm 0cm]{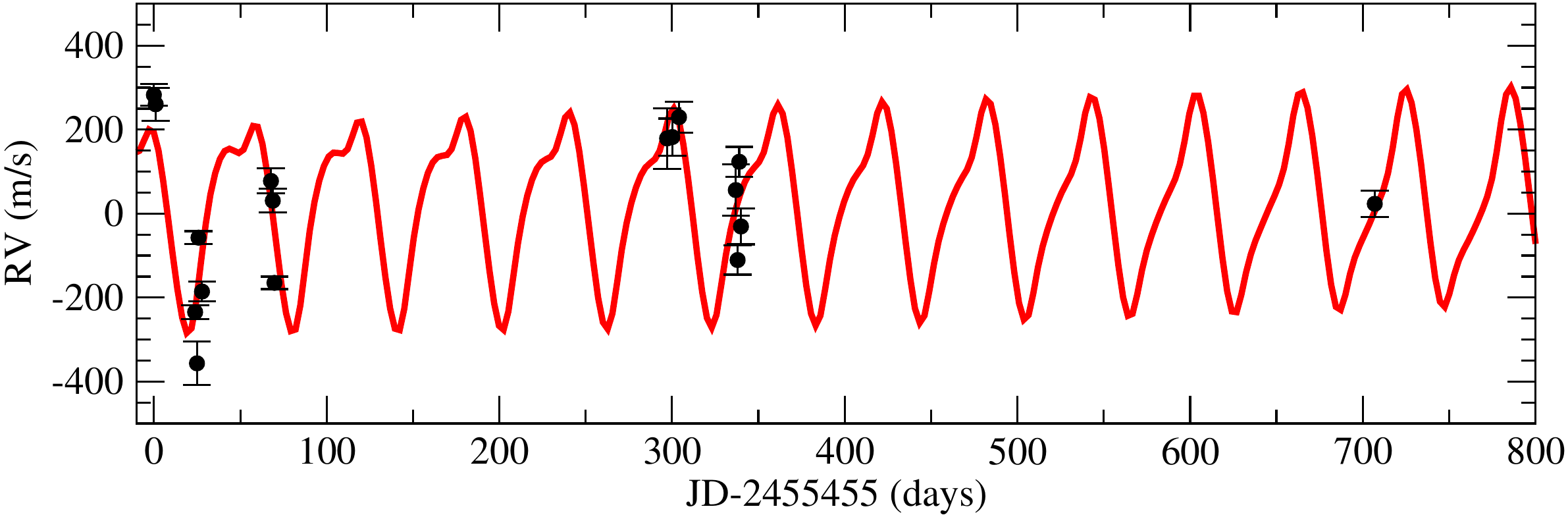}    
      \end{center}      
 \caption{Top: Lick, Keck and IRTF radial velocity time-series for GJ 876. The IRTF RV variability is consistent with the observed visible wavelength radial velocity variability, the first alternate wavelength confirmation of the detection of this system.  Bottom: Zoom of preliminary IRTF RVs for GJ 876 spanning 800 days,
overlaid with a model Systemic Console RV curve fit to the Keck+Lick RVs and extrapolated to cover our data. A 42 day shift is applied to the model, consistent with the reported periastron longitude precession of -41 degrees/year identified for GJ 876bc that is not accounted for by Systemic.\cite{laughlin05}.  \label{fig:f14}}
\end{figure}

\section{Conclusions and Future Work}

We have built and commissioned three gas cells -- isotopic and deuterated methane, and ammonia -- to work with the CSHELL spectrograph at IRTF.  We have demonstrated a short time-scale RV noise-floor for suitably bright targets of $\sim$7 m/s, and a long term precision of $<$60 m/s.   We are now applying the computationally intensive pipeline to the rest of our survey data.  Our survey results and a more detailed paper on the pipeline performance will be presented in future publications.  We have demonstrated that the gas cell technique can be extended from the visible to the infrared, and holds promise for use on planned future near-infrared spectrographs with more modern detectors and cross-dispersion for larger spectral grasp, including iSHELL on IRTF, an upgraded NIRSPEC on Keck, iGRINS, the Habitable Zone Planet Finder, SPIRou and others.

\acknowledgments     
 
Peter Plavchan would like acknowledge Wes Traub and Stephen Unwin for seed funding provided by the JPL Center for Exoplanet Science and NASA Exoplanet Science Institute, as well as JPL Research and Technology Development grant in FY13.  G. Anglada-Escud\'e would like to acknowledge the Carnegie Postdoctoral Fellowship Program and the support provided by the NASA Astrobiology Institute grant NNA09DA81A.  Part of the research at the Jet Propulsion Laboratory (JPL) and California Institute of Technology was performed under contracts with National Aeronautics and Space Administration. The stellar synthetic spectra were graciously provided by Peter Hauschildt (U. of Hamburg) and the PHOENIX group. We also thank Keeyoon Sung, Linda Brown and Pin Chen from JPL's Laboratory Studies and Astrobiology Group for their advice and support using the FTIR spectrometer. We would like to thank Paul Butler (Carnegie Institution of Washington) and Gilian Nave (NIST) for their advice in gas optimization parameters and molecular spectroscopy in general. We would like to thank Steve Osterman (U. of Colorado) for their valuable discussions. We also thank John Rayner, Morgan Bonnet, George Koenig, Kars Bergknut and Alan Tokunaga from IfA/Hawaii for their support during the CSHELL/IRTF cell design review, integration and commissioning.  We thank Rick Gerhart (Caltech), Scot Howell (Mindrum Precision) and Thurston Levy (Glass Instruments, Inc.) for their work in helping construct and fill the gas cells, and Joeff Zolkower (Caltech) for mechanical engineering advise.


\bibliography{report}   
\bibliographystyle{spiebib}   

\end{document}